\documentclass[longauth]{aa}  
\usepackage{comment}
\usepackage{threeparttable}
\usepackage{caption}
\usepackage{booktabs}
\usepackage{subcaption}
\usepackage{tikz}
\usepackage{natbib}
\usepackage{amssymb}
\usepackage{graphicx}
\usepackage{hyperref}
\usepackage{float}
\usepackage{tabularx}
\usepackage{array}
\usepackage{siunitx}
\usepackage{txfonts}
\begin{document}

   \title{Revisiting the multi-planetary system of the nearby star HD 20794 \thanks{Program IDs for the ESPRESSO observations we used in the analysis are: 106.21M2.003, 106.21M2.004, 108.2254.001, 108.2254.006, and 110.24CD.001, 110.24CD.003, 1102.C-0744, 1102.C-0744, 1102.C-0744, 1102.C-0744, 1104.C-0350, 1104.C-0350, 1104.C-0350, 1104.C-0350, 1104.C-0350, and 1104.C-0350.}
   \thanks{Programs IDs for the HARPS observations we used in the analysis are: 060.A-9709, 072.C-0488, 083.C-1001, 084.C-0229, 086.C-0230, 087.C-0990, 088.C-0011, 089.C-0050, 090.C-0849, 091.C-0936, 092.C-0579, 093.C-0062,094.C-0797, 095.C-0040, 096.C-0053, 
60.A-9036, 60.A-9709, and 110.24BB.001.}}
\subtitle{Confirmation of a low-mass planet in the habitable zone of a nearby G-dwarf}

   \author{N. Nari \inst{\ref{lbridges},\ref{iac},\ref{ull}}
\and 
X. Dumusque \inst{\ref{unige}}
\and
N. C. Hara \inst{\ref{unige},\ref{unimar}}
\and
A. Su\'arez Mascare\~no \inst{\ref{iac},\ref{ull}}
\and
M. Cretignier \inst{\ref{unioxf}}
\and
J. I. Gonz\'alez Hern\'andez \inst{\ref{iac},\ref{ull}}
\and
A. K. Stefanov \inst{\ref{iac},\ref{ull}}
\and
V. M. Passegger \inst{\ref{subaru},\ref{iac},\ref{ull},\ref{sw}}
\and
R. Rebolo \inst{\ref{iac},\ref{ull},\ref{csic}}
\and
F. Pepe \inst{\ref{unige}}
\and
N. C. Santos \inst{\ref{IAporto},\ref{uniporto}}
\and
S. Cristiani \inst{\ref{inaf-trieste},\ref{unitrieste}}
\and
J. P. Faria \inst{\ref{unige},\ref{IAporto},\ref{uniporto}}
\and
P. Figueira \inst{\ref{unige},\ref{IAporto}}
\and
A. Sozzetti \inst{\ref{inaf-torino}}
\and
M. R. Zapatero Osorio \inst{\ref{cab}}
\and
V. Adibekyan \inst{\ref{IAporto},\ref{uniporto}}
\and
Y. Alibert \inst{\ref{bern},\ref{bern2}}
\and
C. Allende Prieto \inst{\ref{iac},\ref{ull}}
\and
F. Bouchy \inst{\ref{unige}}
\and
S. Benatti \inst{\ref{inaf-palermo}}
\and
A. Castro-Gonz\'alez \inst{\ref{cab}}
\and
V. D'Odorico \inst{\ref{inaf-trieste},\ref{unitrieste}}
\and
M. Damasso \inst{\ref{inaf-torino}}
\and
J.B. Delisle \inst{\ref{unige}}
\and
P. Di Marcantonio \inst{\ref{inaf-trieste}}
\and
D. Ehrenreich \inst{\ref{unige}}
\and
R. G\'enova-Santos \inst{\ref{iac},\ref{ull}}
\and
M. J. Hobson \inst{\ref{unige}}
\and 
B. Lavie \inst{\ref{unige}}
\and
J. Lillo-Box \inst{\ref{cab}}
\and
G. Lo Curto \inst{\ref{eso}}
\and 
C. Lovis \inst{\ref{unige}}
\and
C. J. A. P. Martins \inst{\ref{IAporto},\ref{caup}}
\and
A. Mehner \inst{\ref{eso}}
\and
G. Micela \inst{\ref{inaf-palermo}}
\and
P. Molaro \inst{\ref{inaf-trieste}}
\and
C. Mordasini \inst{\ref{bern2}}
\and
N. Nunes \inst{\ref{lisbon}}
\and
E. Palle \inst{\ref{iac},\ref{ull}}
\and
S.P. Quanz \inst{\ref{eth},\ref{eth_des}}
\and
D. S\'egransan \inst{\ref{unige}}
\and
A.M. Silva \inst{\ref{IAporto},\ref{uniporto}}
\and
S. G. Sousa \inst{\ref{IAporto},\ref{uniporto}}
\and
S. Udry \inst{\ref{unige}}
\and 
N. Unger \inst{\ref{unige}}
\and 
J. Venturini \inst{\ref{unige}}
}

   \institute{
Light Bridges S.L., Observatorio del Teide, Carretera del Observatorio, s/n Guimar, 38500, Tenerife, Canarias, Spain \label{lbridges}
\and
Instituto de Astrof{\'i}sica de Canarias, E-38205 La Laguna, Tenerife, Spain \label{iac}\\
\email{nicola.nari@lightbridges.es}
\and
Departamento de Astrof{\'i}sica, Universidad de La Laguna, E-38206 La Laguna, Tenerife, Spain \label{ull}
\and
Consejo Superior de Investigaciones Cient\'{i}ficas, Spain \label{csic}
\and
Observatoire de Gen\`eve, D\'epartement d'Astronomie, Universit\'e de Genève, Chemin Pegasi 51b, 1290 Versoix, Switzerland \label{unige}
\and
Aix Marseille Univ, CNRS, CNES, Institut Origines, LAM, Marseille, France \label{unimar}
\and 
Department of Physics, University of Oxford, Oxford OX13RH, UK \label{unioxf}
\and
Instituto de Astrof\'{i}sica e Ci\^encias do Espa\c{c}o, CAUP, Universidade do Porto, Rua das Estrelas, 4150-762, Porto, Portugal \label{IAporto}
\and
Departamento de F\'{i}sica e Astronomia, Faculdade de Ci\^encias, Universidade do Porto, Rua do Campo Alegre, 4169-007, Porto, Portugal \label{uniporto}
\and
Centro de Astrof\'{\i}sica da Universidade do Porto, Rua das Estrelas, 4150-762 Porto, Portugal \label{caup}
\and
Centro de Astrobiolog\'{i}a (CAB), CSIC-INTA, ESAC campus, Camino Bajo del Castillo s/n, E-28692, Villanueva de la Ca\~nada (Madrid), Spain \label{cab}
\and
INAF - Osservatorio Astronomico di Trieste, via G. B. Tiepolo 11, I-34143, Trieste, Italy \label{inaf-trieste}
\and
IFPU--Institute for Fundamental Physics of the Universe, via Beirut 2, I-34151 Trieste, Italy \label{unitrieste}
\and
INAF - Osservatorio Astrofisico di Torino, Strada Osservatorio 20, I-10025, Pino Torinese (TO), Italy \label{inaf-torino}
\and
ESO - European Southern Observatory, Av. Alonso de Cordova 3107, Vitacura, Santiago, Chile \label{eso}
\and
INAF - Osservatorio Astronomico di Palermo, Piazza del Parlamento 1, I-90134 Palermo, Italy \label{inaf-palermo}
\and
Instituto de Astrof\'{i}sica e Ci\^{e}ncias do Espa\c{C}o, Faculdade de Ci\^{e}ncias da Universidade de Lisboa, Campo Grande, 1749-016, Lisboa, Portugal \label{lisbon}
\and
Center for Space and Habitability, University of Bern, Gesellschaftsstrasse 6, 3012 Bern, Switzerland \label{bern}
\and
Weltraumforschung und Planetologie, Physikalisches Institut, University of Bern, Gesellschaftsstrasse 6, 3012 Bern, Switzerland \label{bern2}
\and ETH Zurich, Institute for Particle Physics \& Astrophysics, Wolfgang-Pauli-Str. 27, 8093 Zurich, Switzerland \label{eth}
\and 
ETH Zurich, Department of Earth Sciences, Sonneggstrasse 5, 8092 Zurich, Switzerland \label{eth_des}
\and
Subaru Telescope, National Astronomical Observatory of Japan, 650 N Aohoku Place, Hilo, HI 96720, USA \label{subaru}
\and
Hamburger Sternwarte, Gojenbergsweg 112, 21029 Hamburg, Germany \label{sw}
}
% \abstract{}{}{}{}{} 
% 5 {} token are mandatory
 
  \abstract
  % context heading (optional)
  % {} leave it empty if necessary  
   {Close-by Earth analogs and super-Earths are of primary importance because they will be preferential targets for the next generation of direct imaging instruments. Bright and close-by G-to-M type stars are preferential targets in radial velocity surveys to find Earth analogs. Their brightness allows us to achieve the best precision on RV measurements and search for signals with amplitudes of less than 1 \si{\meter\per\second}.}
  % aims heading (mandatory)
   {We present an analysis of the RV data of the star HD 20794, a target whose planetary system has been extensively debated in the literature. The broad time span of the observations makes it possible to find planets with signal semi-amplitudes below 1 \si{\meter\per\second} in the habitable zone.
   }
  % methods heading (mandatory)
   {We analyzed RV datasets spanning more than 20 years. We monitored the system with ESPRESSO. We joined ESPRESSO data with the HARPS data, including archival data and new measurements from a recent program.
   %We exploit the information encoded in the spectra to derive and study activity indicators, which are useful to disentangle signals of stellar origin from real Keplerian signals. 
   We applied the post-processing pipeline YARARA to HARPS data to correct systematics, improve the quality of RV measurements, and mitigate the impact of stellar activity.}
  % results heading (mandatory)
   {We confirm the presence of three planets, with periods of 18.3142 $\pm$ 0.0022 \si{\day}, 89.68 $\pm$ 0.10 \si{\day}, and 647.6$_{-2.7}^{+2.5}$ \si{\day  }, along with masses of 2.15 $\pm$ 0.17 M$\oplus$, 2.98 $\pm$ 0.29 M$\oplus$, and 5.82 $\pm$ 0.57  M$\oplus$ respectively. For the outer planet, we find an eccentricity of 0.45$_{-0.11}^{+0.10}$, whereas the inner planets are compatible with circular orbits. The latter is likely to be a rocky planet in the habitable zone of HD 20794. From the analysis of activity indicators, we find evidence of a magnetic cycle with a period of $\sim$ 3000 \si{\day}, along with evidence pointing to a rotation period of $\sim$ 39 \si{\day}.}
  % conclusion
  {We have determined the presence of a system of three planets orbiting the solar-type star HD 20794. This star is bright (V=4.34 mag) and close (d = 6.04 pc), and HD 20794 d resides in the stellar habitable zone, making this system a high-priority target for future atmospheric characterization with direct imaging facilities.} 


   \maketitle
%
%-------------------------------------------------------------------

\section{Introduction}

    Exoplanetary science has experienced a rapid and successful development since the discovery of the first planet orbiting around a main sequence star other than the Sun \citep{mayor}. This detection was obtained through the radial velocity (RV) technique, %\citep{struve_1952,hatzes_book}
    which is still one of the most successful techniques for discovering and characterizing exoplanets. This seminal discovery has opened the field to more than 5600 confirmed planets up to July 2024, according to the 
    \texttt{NASA Exoplanet Archive}\footnote{\url{https://exoplanetarchive.ipac.caltech.edu/}} \citep{2013_nasa_exoplanet_archive}. 
    
   % Through the years other detection techniques were refined and became commonly used in the exoplanets research. The transit method, in particular, exploits the dimming of starlight caused by the planet passing in front of the star, blocking its light \citep{sackett_1999_transit,winn_transit}. This method is the most successful technique for discovering new planets, especially owing to the \textit{Kepler} mission \citep{kepler_borucki_2010} and the TESS mission \citep{tess_ricker_2014}. However, even if outnumbered in detections, the RV method remains fundamental for constraining planetary mass, a basic parameter to characterize the demographics of planetary systems. In addition, mass is an important parameter that sheds light on the formation and evolution of exoplanets. 
    
    %The field of exoplanets is intimately linked with the technological development of the instrumentation necessary for their detection.
    At the onset of the exoplanet field, the main issue for the detection of new planets was related to the precision of available instruments. 
    %(citation about hot Jupiters). no need for citation because 51 peg b is already an hot Jupiter.
    The High Accuracy Radial velocity Planet Searcher (HARPS) instrument \citep{harps} represents a ground-breaking advancement in the field. It is installed at the 3.6 m Telescope in La Silla, Chile and it has been the first instrument able to reach an RV precision of the level of 1 \si{\meter\per\second}. To reach this goal, the instrument is temperature-controlled and pressure-stabilized. Over the years, this instrument has achieved many results of high scientific interest, such as the discovery of the first super-Earth \citep{nuno_searth}. This instrument remained the state-of-the-art RV instrument until 2018 when the Echelle SPectrograph for Rocky Exoplanets and Stable Spectroscopic Observations (ESPRESSO) \citep{espresso_2014} began operations. The latter is installed at the 8.2 m Very Large Telescope (VLT) in Paranal, Chile. Again, ESPRESSO is temperature-controlled and pressure-stabilized. All these factors contribute to reaching a precision on a single measurement better than 10 \si{\centi\meter\per\second} \citep{pepe_espresso_2021}.
    ESPRESSO was the basis for some of the most exciting discoveries in the field of exoplanets, for example, the confirmation of Proxima d, a small, sub-Earth planet orbiting around Proxima,  \citep{suarez_proxima_2020,faria_proxima_2022}. It has been involved in other ground-breaking discoveries in the field of exoplanetary atmospheres \citep{iron_rain_ehrenreich} and the mass determination of sub-Earth radius planets detected by transits \citep{2021_transit_demangeon}.
    Combining the high-accuracy measurements of HARPS and ESPRESSO, while exploiting the long temporal baseline of the former, we can search for signals with amplitudes of below 1 \si{\meter\per\second} at long orbital periods.
    According to the
    \texttt{NASA Exoplanet Archive} only 25 planets out of more than 5600 have a measured mass of less than 10 M$_{\oplus}$ and an orbital period of > 50 \si{\day}.

    HD 20794 has been part of long RV surveys dedicated to the search for low-amplitude long-period signals around solar-type stars, with hundreds of nights of observations with HARPS and ESPRESSO spanning more than 20 years available.
    The recent publication of a candidate at $\sim$ 640 \si{\day} period \citep{cretignier_yarara_2023} gave rise to a new campaign for a dense monitoring of the star. The low stellar activity level and the brightness of HD 20794 has made this target one of the most well-suited candidates for this purpose. Furthermore, HD 20794 is only 6.04 pc away, and long-period planets orbiting this star can become targets for future facilities dedicated to the characterization of exoplanetary atmosphere, both from the ground and space as ANDES at ELT\citep{2023_palle_andes}, LIFE \citep{2022_life_quanz}, and HWO \citep{2024_mamajek_hwo}. The work is structured as follows. In Sect. \ref{sec_obs}, we report the observations obtained for HD 20794. In Sect. \ref{sec_stel_par}, we report stellar characteristics and state-of-the-art knowledge on the system. In Sect. \ref{sec_analysis}, we report the modeling of the RV time-series. In Sect. \ref{sec_dis}, we discuss the results and implications of our work. In Sect. \ref{sec_con}, we summarize our analysis and findings.
    %Other additional effects that come directly from stellar physics, such as granulation or oscillations caused by pressure waves, can interfere with our capability to detect exoplanets and create variability on short timescale. The variation related to oscillations is in the order of minutes for G-type stars and in the order of hours to days for granulation (for an overview of the strategies to mitigate oscillation and granulation see \citet{dumusque_2011_gran_strategy} and references therein). 
\\

%--------------------------------------------------------------------

\section{Observations}
\label{sec_obs}

\subsection{ESPRESSO}
\label{espresso_description}

HD 20794 has been observed as part of the ESPRESSO Guaranteed Time Observations (GTO), within working group 1 (WG1), from October 2018 to March 2023. The main objective of WG1 is the research of Earth-like planets around G-to-M type stars within the habitable zone (HZ).

ESPRESSO \citep{pepe_espresso_2021} is a fiber-fed spectrograph installed at VLT in Paranal, Chile, and it can be connected to each of the 8.2 m telescope units. It spans a wavelength range between 378.2 \si{\nano\meter} and 788.7 \si{\nano\meter} and has a resolving power of $\sim$ 140 000 in high-resolution mode. Designed to achieve 10 \si{\centi\meter\per\second} precision in RV for solar-type stars, ESPRESSO operates within a temperature- and pressure-controlled vacuum vessel for best instrumental stability. The simultaneous reference technique \citep{1996_baranne_ccf} using a Fabry-Pérot etalon \citep{wildi_2010_fp,pepe_espresso_2021} allows us to correct for any residual instrumental drift up to 10 \si{\centi\meter\per\second}. 

The ESPRESSO data reduction software (DRS)
%,which is a development of the HARPS pipeline, 
provides extracted and wavelength-calibrated spectra, with different byproducts, such as the cross-correlation function (CCF), CCF-derived RV, and telemetry data. The CCF technique measures RV by convolving stellar spectra with a mask from a similar spectral type template star \citep{fellgett_1953,1979_baranne_ccf,1996_baranne_ccf}.
For this study, we derived RVs with sBART \citep{2022_silva_sbart}, a template matching tool within a semi-Bayesian framework. sBART derives RVs by comparing each spectrum to a template spectrum built from observations. 
%We correct the velocities for secular acceleration following \citet{secular_acceleration}. pensa se togliere o lasciare

In June 2019, an intervention on the ESPRESSO fiber link improved the efficiency by 50 \%, introducing an offset in the data. We treated observations before BJD 2458721 as a separate dataset. 
%This date is the date of the first observation made after the fiber link update for HD 20794.
An interruption of operations due to COVID-19 and a lamp change introduced an additional offset, corrected in DRS version 3.0.0, eliminating the need for additional dataset divisions. We obtained 34 observations over 6 nights before the intervention and 661 observations over 65 nights after. 
We discarded pre-intervention data due to its limited usefulness. 

After nightly binning and excluding outliers, that is, nights with an error on the measurement exceeding three times the median error or nights with a difference from the median RV larger than three times the standard deviation of the measurements, we retained 63 nights, covering 1307.66 days from BJD 2458721.89 to BJD 2460029.49.
The root mean square (RMS) of the post-intervention dataset, referred to as E19, is 0.84 \si{\meter\per\second}, with a mean photonic error of 0.10 \si{\meter\per\second} per measurement. Nightly binning reduced the mean photonic error to 0.03 \si{\meter\per\second}. This is attributed to binning multiple measurements per night, usually ten or more. However, nightly stacking does not reduce the instrumental noise on a timescale beyond the intra-night timescale.  Due to the brightness of HD 20794, we used short exposure times of 60 seconds.

\subsection{HARPS}
\label{harps_section}
HARPS \citep{harps} is a fiber-fed spectrograph installed at the Observatory of La Silla, in Chile. It started operations in 2003 and is now one of the most successful ground-based facilities for the discovery of exoplanets. Designed to obtain long-term stability, the instrument has shown a precision on a single measurement better than 50 \si{\centi\meter\per\second} for bright stars and a long-term stability of 1 \si{\meter\per\second} for quiet targets such as HD 20794 \citep{cretignier_yarara_2021,cretignier_yarara_2023}. The spectrograph is equipped with two fibers and is mechanically stabilized. The instrument operates within a temperature- and pressure-controlled vacuum vessel to improve instrumental stability. One of the two fibers provides starlight to the instrument, while the other one is used for simultaneous calibrations with a Thorium-Argon lamp or Fabry-Pérot etalon as reference. It covers a spectral range of 378-691 nm and has a spectral resolving power of R $\sim$ 115000. 

The instrument underwent a fiber update in 2015, which introduced an offset in the RV \citep{lo_curto_2015}. For this reason, we consider the data taken before and after the update as independent datasets (henceforth, H03 and H15) and consider them as if the data were acquired by different instruments. We have applied the ESPRESSO pipeline version 3.2.0, which has been adapted to HARPS to reduce HARPS spectra. The pipeline provides extracted and wavelength-corrected spectra. The new pipeline automatically corrects for drifts due to lamp changes and aging and for a linear trend visible in the H15 dataset after BJD 59915 in the previous pipeline. HD 20794 has been part of an intensive observational campaign to confirm the planetary nature of the long-period candidate announced in \citet{cretignier_yarara_2023}. In our analysis, we have considered HARPS spectra available on \texttt{DACE}
\footnote{\url{https://dace.unige.ch}} \citep{2015_dace}.
We collected 765 visits of HD 20794 made by HARPS, 531 nights before the fiber link update, and 234 nights after the fiber link update. 
For our analysis we extracted the RV with the line-by-line (LBL) \citep{dumusque_lbl_2018} post-processing pipeline YARARA \citep{cretignier_yarara_2021,cretignier_yarara_2023}. 
The YARARA pipeline automatically rejects some nights due to anomalous CCF and S/N anomalies. We removed those nights from our analysis and remained with 514 nights binned from 6224 single spectra for H03 and 232 nights binned from 1409 spectra for H15.

YARARA improves the precision of RV by applying a correction to the 1-D spectra
derived from the pipeline of the instrument and normalized by RASSINE \citep{cretignier_2021_rassine}. 
RASSINE is a Python code for the normalization of merged 1D spectra. YARARA corrects systematics such as ghosts, telluric lines, stitching effects, cosmic rays, and more. The outcome from this first correction is referred to as YV1, following the definition of \citet{cretignier_yarara_2023}. 
A second correction, named YV2, is then performed in the time domain. This correction consists of linearly decorrelating the RVs from the principal component coefficients obtained from a PCA performed on the SHELL spectral representation \citep{2022_cretignier_shell}, in addition to the principal component coefficients obtained from a PCA performed on the LBL RVs \citep{cretignier_yarara_2023}. 
The SHELL spectral representation is a representation for the spectra where we consider the flux and its gradient $\left( \frac{d f_0}{d\lambda}, f \right),$ instead of the flux and the wavelength  $(\lambda, f)$. 

\begin{figure}[!h]
    \begin{minipage}{0.45\textwidth}
        \includegraphics[width=\linewidth]{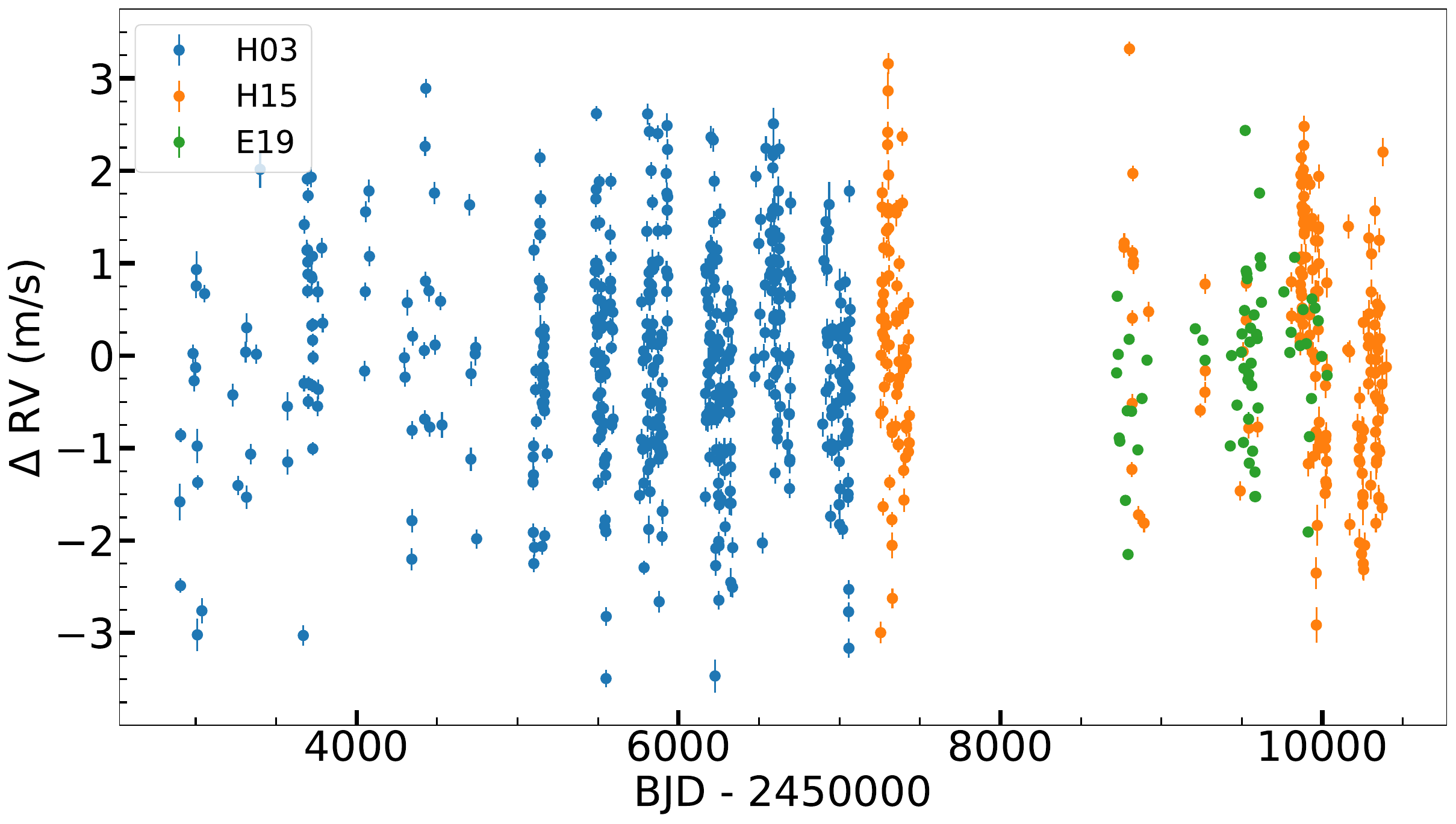}
    \end{minipage}
    \caption{RV dataset of HD20794 for the different datasets. The HARPS RVs have been extracted using YARARA \citep{cretignier_yarara_2021,cretignier_yarara_2023}. For ESPRESSO, RVs have been extracted with sBART \citep{2022_silva_sbart}.}
    \label{RV_data}
\end{figure}

For our analysis, we used YV2 RVs for HARPS before fiber update (H03), and YV1 RVs for HARPS after fiber update (H15). This decision comes from the fact that YV2 is less effective when applied to H15 due to fewer observations and inhomogeneous sampling. After nightly binning observations and applying a 3$\sigma$ cut to remove outliers, with the same procedure we followed for ESPRESSO, we retain 512 and 231 nights available for H03 and H15, respectively. Thanks to the YARARA correction we can derive HARPS RVs with a root mean square of the binned data of 1.14 \si{\meter\per\second} and 1.21 \si{\meter\per\second} for H03 and H15, respectively. The mean error in the measurement of binned points is  0.11 \si{\meter\per\second} for H03 and 0.12 \si{\meter\per\second} for H15. Here, we are considering only the photonic error on the binned measurements. We show the RVs time series in Fig. \ref{RV_data}.

\subsection{TESS}

\begin{figure}[!h]
    \centering
    \begin{subfigure}[]{0.45\textwidth}
        \begin{tikzpicture}
            \node[inner sep=0] (image) at (0,0) {\includegraphics[width=\textwidth]{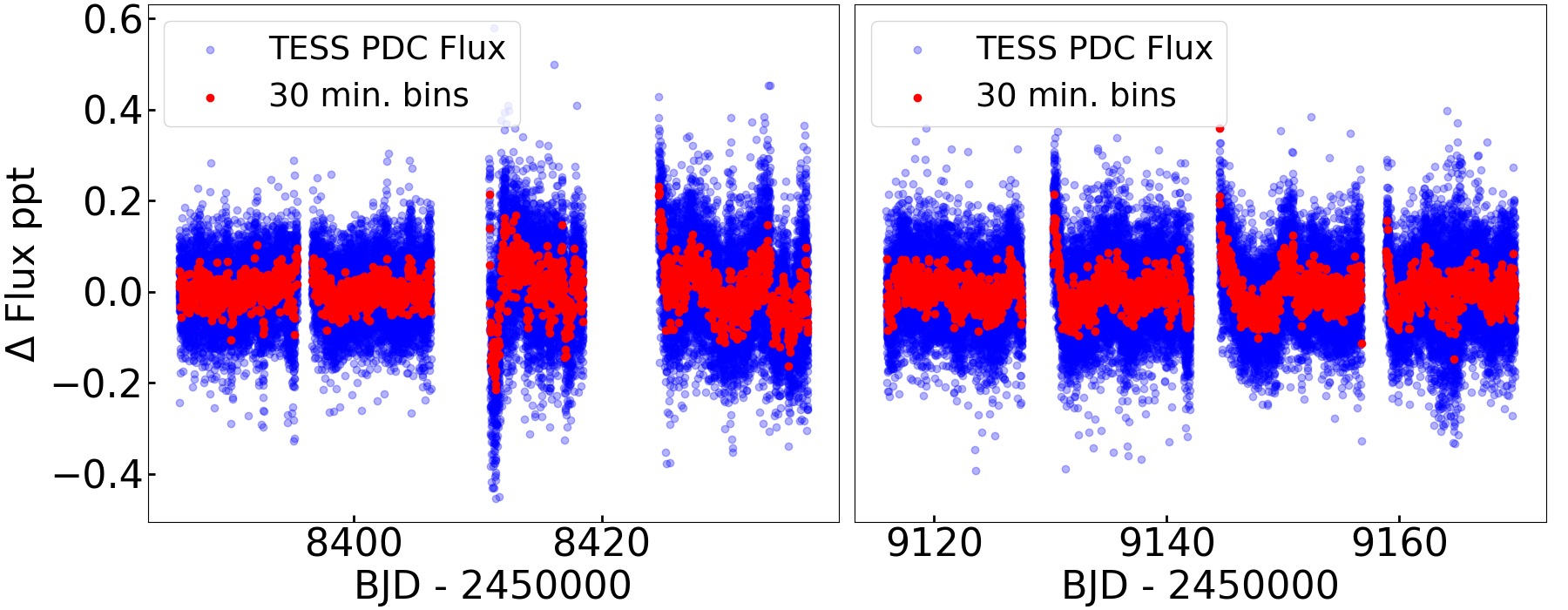}};
            \node[anchor=north west] at ([xshift=-0.05cm, yshift=-0.05cm]image.north west) {\textbf{a)}};
        \end{tikzpicture}
    \phantom{}
    \end{subfigure}
    \hfill
    \begin{subfigure}[b]{0.45\textwidth}
        \begin{tikzpicture}
            \node[inner sep=0] (image) at (0,0) {\includegraphics[width=\textwidth]{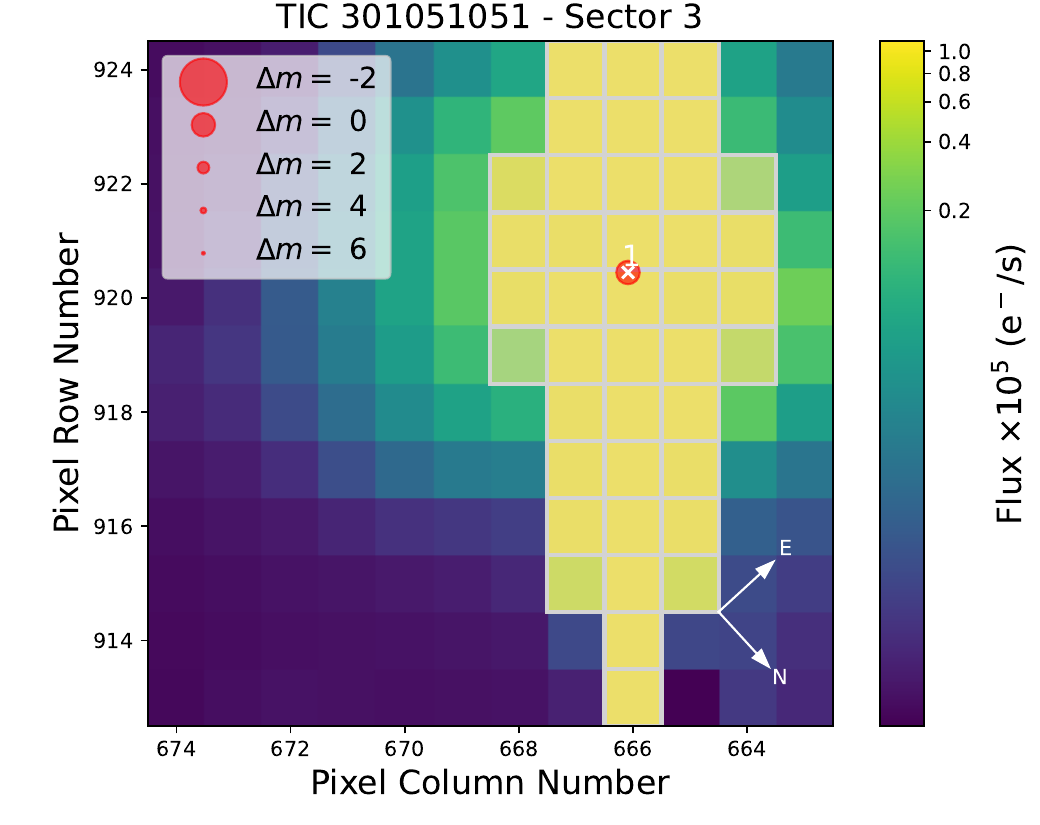}};
            \node[anchor=north west] at ([xshift=0.1cm, yshift=-0.3cm]image.north west) {\textbf{b)}};
        \end{tikzpicture}
        \label{target_pixel_file}
    \end{subfigure}
    \caption{Tess photometry and TESS target pixel file for HD 20794. Panel (a): TESS photometry for HD 20794. We do not find any evidence of transits in the dataset. Panel (b): TESS target pixel file for HD 20794. No nearby sources are detected by the \textit{Gaia} DR3 catalog up to a contrast magnitude of +6 within the TESS aperture.}
    \label{tess_hd20794}
\end{figure}

The Transiting Exoplanet Survey Satellite \citep[TESS;][]{tess_ricker_2014} observed HD 20794 with a 2 min cadence in sectors 3, 4, 30, and 31. The data were processed with the SPOC (Science Processing Operations Center) pipeline \citep{spoc_pipeline_2016} and the preliminary research for transiting planets was done with an adaptive, wavelet-dependent algorithm for transit \citep{jenkins_2010_transit}.
We analyzed the four time series independently and all together. We excluded flux measurements from sector 4 taken between BJD 2458420 and 2458424 for a systematic present in the data. 
We applied a BLS periodogram \citep{2002_kovacs_bls,hartman_bakos_bls} to the time series,  considering the four sectors independently and together, but we did not find any evidence of a transit. The RMS of TESS measurements is 0.08 ppt and it goes down to 0.044 ppt when we bin the data in 30 min bins. TESS photometry is visible in Fig. \ref{tess_hd20794}a. We have used the publicly available code \texttt{tpfplotter}\footnote{\url{https://github.com/jlillo/tpfplotter}} \citep{2020_tpf_plotter} to plot the target pixel file. The target pixel file is visible in Fig. \ref{tess_hd20794}b. Any other sources in the field are fainter by at least six magnitudes.

\section{HD 20794}
\label{sec_stel_par}
\subsection{HD 20794: stellar parameters}

HD 20794 ($\alpha$ = 03:19:55.669, $\delta$ = -43:04:11.29) is a bright star (V-mag $\sim$ 4.34) of the spectral type G6V, located at a distance of 6.04 pc from the Sun \citep{gaia_dr3_2020}. We give a list of stellar parameters in Table \ref{tab_stel_par}. Surface gravity $\log_{10} g$, the effective temperature T$eff$, and the metallicity $[Fe/H]$ were derived directly from the analysis of the ESPRESSO spectra with the ARES + MOOG method described in \citet{2015_sousa_stellar}. This method combines the ARES code \citep{sousa_ares_2007} for the derivation of the equivalent width of lines, with the MOOG model for the atmospheric derivation of abundances \citep{sneden_moog_1973}. ARES + MOOG is a specific realization of the equivalent width method for deriving stellar parameters. The equivalent width of a selected number of stellar lines is derived directly from the spectrum and it is used together with an atmospheric model to calculate the individual line abundances. The stellar parameters are found when the ionization and excitation balances of all lines are achieved; otherwise, the atmospheric model is changed to refine the fit. The ARES + MOOG method is designed to be automatically applied to a large sample of stars, so the line list was set to be wide and stable at the same time \citep{2008_sousa_lines}.
%The oscillator strength associated with every line is not known precisely, so the value is derived in a differential analysis where the solar spectra are taken as a benchmark. 
This method is more effective when the star analyzed is similar to the Sun. The ARES code derives the equivalent width and MOOG is used to calculate line abundances in the context of local thermodynamic equilibrium (LTE) through an interpolation with a grid of Kurucz Atlas-9 plane-parallel model atmosphere \citep{kurucz_1993_atlas}. A total of 14 spectra were added to reach the S/N of $\sim$ 2500 necessary to extract the parameters.
The stellar mass, M$_{\star}$, and the stellar radius, R$_{\star}$, are estimations coming from a calibration based on the derived spectroscopic parameters: $\log_{10} g$, T$eff$, and $[Fe/H]$ \citep{2010_torres_mass}.

The star has color B-V $\sim$ 0.7 \citep{2000_tycho}, and a mass of 0.79 $\pm$ 0.01 M$_{\odot}$. We measure a log$_{10}$R'$_{HK}$ = -4.98 $\pm$ 0.02.
Another source for deriving the stellar parameters is the eDR3 \textit{Gaia} catalog \citep{gaia_edr3_2021}. The luminosity of the star is 0.6869 $\pm$ 0.0026 L$_{\odot}$ \citep{gaia_dr2}. We can use the stellar mass and luminosity to derive the HZ of the system following \citet{kopparapu_habitable_zone}. We refer to the HZ limits defined in \citet{kopparapu_habitable_zone} as "recent-Venus" and "early-Mars" as the optimistic inner and outer edge of the HZ, respectively. We refer to the HZ limits defined as "runaway greenhouse" and "maximum greenhouse" as the conservative inner and outer edge of the HZ.
%The distance of the habitable zone depends on assumptions about the atmosphere of the planets and the characteristics of their guest star. 
The conservative HZ for HD20794 spans between 0.833 $\pm$ 0.003 au and 1.450 $\pm$ 0.008 au, while the optimistic HZ is comprised between 0.632 $\pm$ 0.002 au and 1.524 $\pm$ 0.009 au. The corresponding orbital period for the inner optimistic and conservative HZ is 207 $\pm$ 2 \si{\day} and 313 $\pm$ 3 \si{\day} respectively, while for the outer HZ, we obtain values of 718 $\pm$ 8 \si{\day} and 773 $\pm$ 9 \si{\day} for conservative and optimistic calculations.

\begin{table}[h!]
  \caption[]{Stellar parameters of interest for HD20794.}
  \label{tab_stel_par}
  \begin{tabular}{p{0.5\linewidth}ll}
    \hline
    \hline
    \noalign{\smallskip}
    Parameter & HD20794 & Ref\\
    \noalign{\smallskip}
    \hline
    \noalign{\smallskip}
    $\alpha$ & 3:19:55.67 & 1\\
    $\delta$ & -43:04:11.29 & 1\\
    Parallax (mas) & 165.524 $\pm$ 0.078 & 1 \\
    d (pc) & 6.0414 $\pm$ 0.0028 & 1\\
    $\mu_{\alpha}$ cos $\delta$ (mas yr$^{-1}$) & 3035.017 $\pm$ 0.081 & 1\\
    $\mu_{\delta}$ (mas yr$^{-1}$) & 726.964 $\pm$ 0.101 & 1\\
    $T_\text{eff}$ (K) & $5368 \pm 64$ & 0 \\
    $\log_{10} g$ (cgs) & $4.39 \pm 0.03$ & 0 \\
    Spectral Type & G6V & 4\\
    RV (\si{\kilo\meter\per\second}) & 87.8820 $\pm$ 0.0001 & 1\\
    $[Fe/H]$ (dex) & -0.42 $\pm$ 0.02 & 0\\
    $M_{\star}$ (M$_{\odot}$) & 0.79 $\pm$ 0.01 & 0\\
    $R_{\star}$ (R$_{\odot}$) & 0.93 $\pm$ 0.03 & 0\\
    log$_{10}$R'$_{HK}$ & -4.98 $\pm$ 0.02 & 3 \\
    B (mag) & 5.130 $\pm$ 0.014 & 2 \\
    V (mag) & 4.336 $\pm$ 0.009 & 2 \\
    L (L$_{\odot}$) & 0.6869 $\pm$ 0.0026 & 5 \\
    $\xi$ (\si{\kilo\meter\per\second}) & 0.555 $\pm$ 0.050 & 0 \\
    Inner optimistic HZ (au) &  0.6322 $\pm$ 0.0020& 0 \\
    Inner conservative HZ (au) & 0.8329 $\pm$ 0.0033 & 0 \\
    Outer conservative HZ (au) & 1.4502 $\pm$ 0.0086 & 0 \\
    Outer optimistic HZ (au) & 1.5236 $\pm$ 0.0091 & 0 \\
    \noalign{\smallskip}
    \hline
  \end{tabular}
  \medskip % Spazio verticale tra la tabella e la lista di riferimenti
    
    \begin{minipage}{0.5\textwidth}
        \raggedright
        References: 0 - This work, 1 - \citep{gaia_dr3_2020}, 2 -\citep{2000_tycho}, 3 - \citep{2019_dace}, 4 - \citep{1989_perkins}, 5 - \citep{gaia_dr2}
    \end{minipage}
\end{table}
\defcitealias{pepe}{Pepe}
\defcitealias{feng_hd20794}{Feng}
\defcitealias{cretignier_yarara_2023}{Cretignier}
\subsection{Planetary system}
HD 20794 is known for hosting a multi-planetary system \citep{pepe,feng_hd20794,2022_ritvik_hd20794,cretignier_yarara_2023}. Through the years, multiple analyses have been conducted on this star. Different analyses use non-identical datasets and different methods for red-noise correction. As a result, they bring on outcomes that are not fully compatible. Considering the periods of the planets detected in the different works, we refer to the planets as the 18-\si{\day}, 40-\si{\day}, 89-\si{\day}, 147-\si{\day}, and 640-\si{\day} planet. 

\citet{pepe} reports the discovery of three planets: HD 20794 b with a period of 18.315 $\pm$ 0.008 \si{\day} and amplitude of 0.83 $\pm$ 0.09 \si{\meter\per\second}; HD 20794 c with a period of 40.114 $\pm$ 0.053 \si{\day} and an amplitude of 0.56 $\pm$ 0.10 \si{\meter\per\second}; and HD 20794 d with a period of 90.31 $\pm$ 0.18 \si{\day} and an amplitude of 0.85 $\pm$ 0.10 \si{\meter\per\second}. \citet{feng_hd20794} considered a larger dataset of HARPS observations, as compared to \citet{pepe}. They used the template matching tool TERRA \citep{2012_terra} to extract the RVs. A wavelength-dependent noise model was applied to the time series to correct spurious signals in the RVs. For the correction details, we refer to the original paper. \citet{feng_hd20794} confirmed the detection of planets at 18-\si{\day} and 89-\si{\day,} but did not find evidence of the planet at 40-\si{\day}. Furthermore, they found an additional candidate at orbital period of 147.0 $\pm$ 1.1 \si{\day}, with an amplitude of 0.69 $\pm$ 0.14 \si{\meter\per\second}. \citet{feng_hd20794} reported a significant signal at P = 331.4$_{-3.0}^{+5.1}$ \si{\day,} but since this is so very close to a year and there is the presence of signals with the same periodicity in activity indicators, it does not allow for the RV signal to be claimed as a safe detection. 
Recently, \citet{cretignier_yarara_2023} revisited the system and considered the historical time series of HARPS before the fiber intervention the instrument underwent in 2015. The main change in this analysis was the usage of the YARARA tool to extract velocities \citep{cretignier_yarara_2021,cretignier_yarara_2023}. The signals at 18 \si{\day} and 89 \si{\day} have been recovered with a high level of confidence. A new candidate is also recovered with a period of 644.6 $\pm$ 8.8 \si{\day} and an amplitude of 0.61 $\pm$ 0.06 \si{\meter\per\second}. \citet{cretignier_yarara_2023} did not find evidence of any additional signals. 
The possible presence of an additional planet with an orbital period between 549 \si{\day} and 733 \si{\day} is also reported in \citet{2022_ritvik_hd20794}. 
We report a summary of the results from the literature in Table \ref{table_literature}.
%We are using the YARARA correction for the HARPS observations also in this work and we will consider \citet{cretignier_yarara_2023} as the reference work for our analysis, taking as planets b, c, and d the planets at 18.3, 89.6, 644 days.
\section{Analysis}
\label{sec_analysis}
Here, we report the analysis of the HARPS and ESPRESSO datasets. We refer to Appendix \ref{methods_sec} for details about the methods and tools used in our analysis to conduct parameter estimation and model comparison. 
\subsection{Stellar activity}
\label{sec_stel_act}

We analyzed different proxies for activity, such as the full width at half maximum (FWHM), bisector timespan (BIS), S-index, H$_{\alpha}$, Na I D, and the contrast of the CCF to retrieve the stellar characteristics. The periodicities we found in the activity proxies (if they were also found in RVs) would require specific attention to be paid to determine their nature, so that we could avoid misunderstanding stellar-related signals from signals that are Keplerian in origin. 

When we modeled the activity, we considered both activity indicators derived from the spectra before and after the YARARA correction. YARARA can correct signatures of activity and we are interested in the inference of stellar parameters, such  as the period of an eventual magnetic cycle and the period of stellar rotation. We corrected the activity indicators for changes in temperature of the echelle gratings of the instruments with an approach similar to \citet{suarez_gj1002}. We found evidence of a magnetic cycle in FWHM with a period of 3020$_{-50}^{+111}$ \si{\day}. We were also able to find evidence of a long-term magnetic cycle  in the analysis of other activity indicators, such as the S-index, BIS, and contrast. 
%\LEt{ \textcolor{red}{Please make sure the specific steps taken in your study are in the past tense, other than general steps/approaches/procedures.***}}

We recovered the rotation period from FWHM and BIS, for the latter in the YARARA-corrected dataset. We found a rotation period of 35.0$_{-2.5}^{+3.2}$ \si{\day} derived from FWHM and a rotation period of 38.8 $_{-2.6}^{+2.4}$ \si{\day} derived from BIS. The latter is particularly important because this period is compatible with the period of the 40-\si{\day} planet detected in \citet{pepe}. We conclude this signal is related to stellar activity. Considering the main focus of the analysis of HD 20794 is the analysis of RVs, we refer to Appendix \ref{stellar_activity_appendix} for a detailed description of the impact of stellar activity on the detection of exoplanets and the full analysis of the different activity indicators. Figure \ref{stellar_activity_full} shows time series and generalized Lomb-Scargle (GLS) periodograms for different activity indicators extracted from spectra that were not corrected by YARARA. Figure \ref{stellar_activity_full_yarara} shows the time series and GLS periodograms for different activity indicators from spectra corrected by YARARA.

\subsection{Radial velocities analysis}
\label{sec_rv}
Figure \ref{RV_data} shows the RVs of HD20794 used in this analysis. The full RV dataset spans 7496 \si{\day}, roughly corresponding to 20 years. 
% pensa di fare una sezione con il metodo che hai utilizzato
After removing outliers and binning data as explained in Appendix \ref{methods_sec} we have a total of 806 nights of observations, divided into 512 nights for HARPS before the fiber update (H03), 231 for HARPS after the fiber update (H15), and 63 nights for ESPRESSO after the intervention on the fiber link (E19). First, we conduct searches for planets in the ESPRESSO dataset alone. Then, we included the HARPS dataset directly corrected for YARARA to make a full analysis of the system.

\subsection{ESPRESSO}

\begin{figure}[!h]
    \centering
    % Figura 1 (a)
    \begin{subfigure}[b]{0.45\textwidth}
        \centering
        \begin{tikzpicture}
            \node[inner sep=0] (image) at (0,0) {\includegraphics[width=\textwidth]{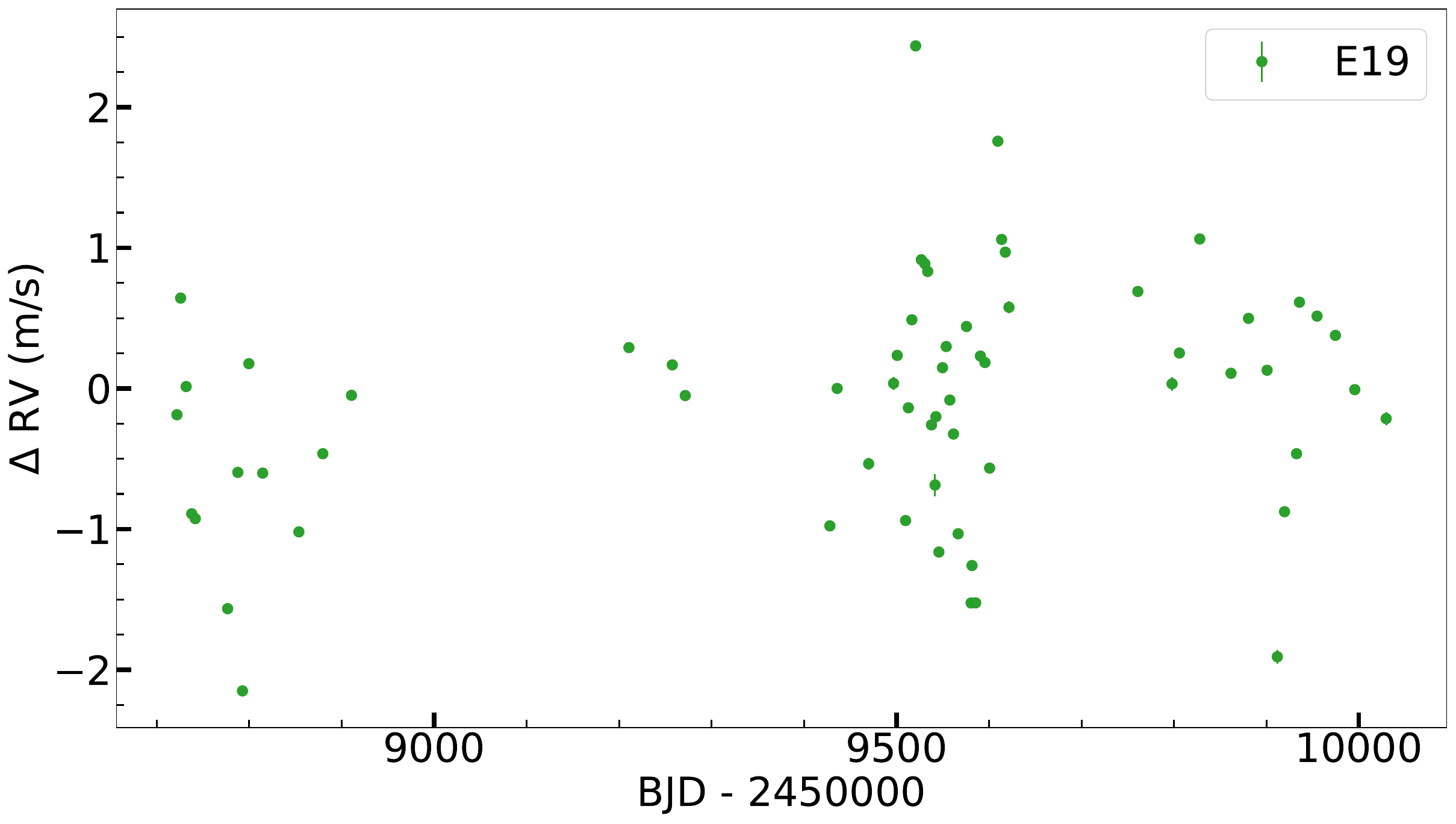}};
            \node[anchor=north west] at ([xshift=0.8cm, yshift=-0.25cm]image.north west) {\textbf{a)}};
        \end{tikzpicture}
    \end{subfigure}
    \hfill
    % Figura 2 (b)
    \begin{subfigure}[b]{0.45\textwidth}
        \centering
        \begin{tikzpicture}
            \node[inner sep=0] (image) at (0,0) {\includegraphics[width=\textwidth]{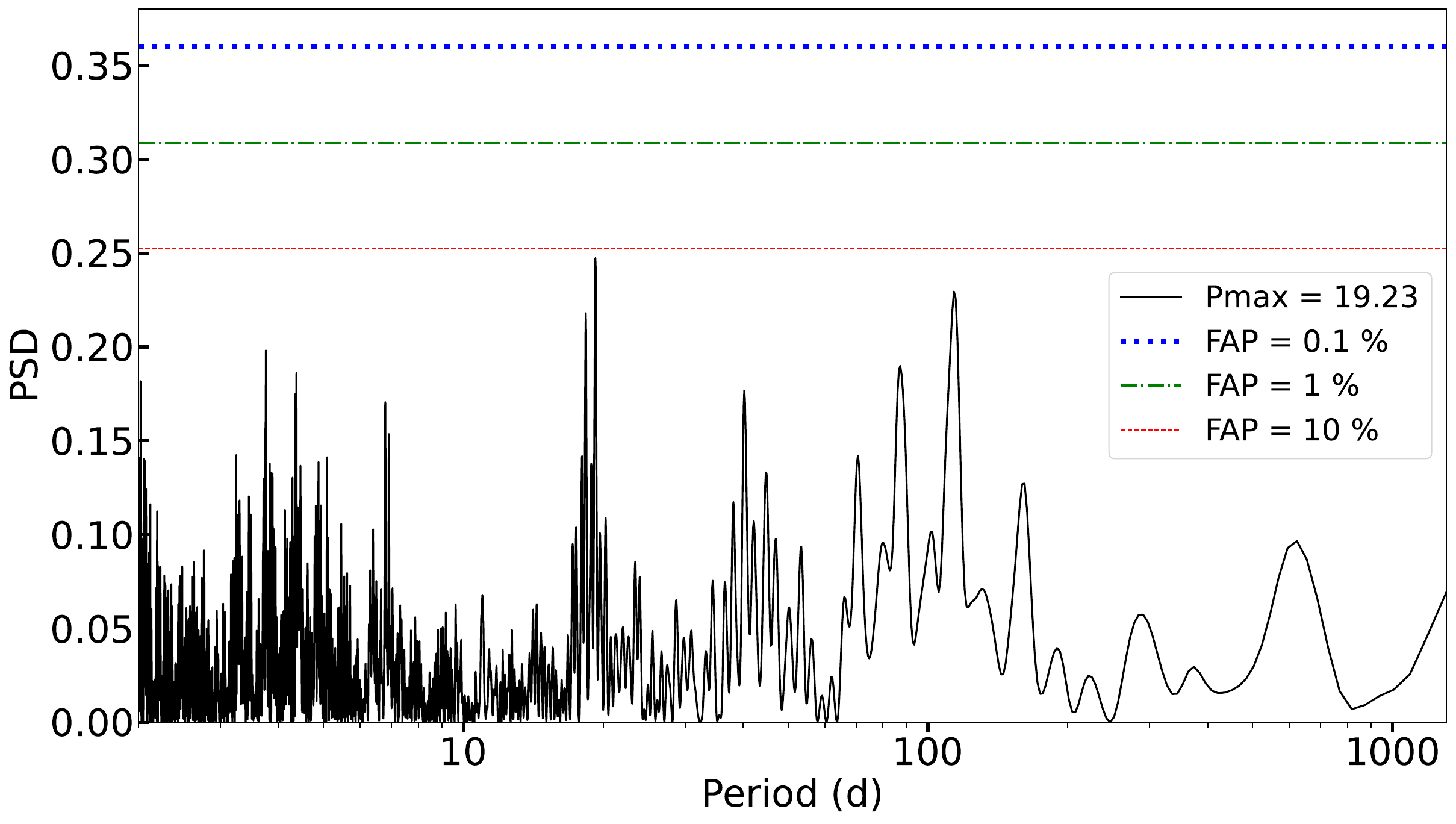}};
            \node[anchor=north west] at ([xshift=1.00cm, yshift=-0.32cm]image.north west) {\textbf{b)}};
        \end{tikzpicture}
    \end{subfigure}
    \caption{ESPRESSO observations of HD 20794. Panel (a): E19 RVs for HD20794. Panel (b): GLS periodogram of the ESPRESSO RVs for HD20794. The main peak is around 19.23 \si{\day}, a year alias of 18.3 \si{\day} signal, and a second peak at 115 \si{\day}, a year alias of the peak at $\sim$ 87 \si{\day}.}
    \label{espresso_data}
\end{figure}

We first consider the analysis of the system with the ESPRESSO dataset alone. We refer to Sect. \ref{espresso_description} for a description of this dataset. We show the ESPRESSO dataset in Fig. \ref{espresso_data}a and the ESPRESSO dataset GLS periodogram in Fig. \ref{espresso_data}b.
%The standard deviation on the measurements is E19 = 0.84 \si{\meter\per\second}.

%\subsubsection{One-planet model}

%\citet{cretignier_yarara_2023} claims the presence of 3 planets in the system, with periods of 18.3 \si{\day}, 89.6 \si{\day}, and $\sim$ 650 \si{\day}. The planets with orbital periods of 18.3 \si{\day} and 89.6 \si{\day} are claimed in all the literature works that dealt with the system \citep{pepe,feng_hd20794}. The timespan of ESPRESSO observations is not large enough to recover the signal at a longer period, and also the inner planets are not trivial to recover due to their low-amplitude signal, expected on the order of $\sim$ 50 \si{\centi\meter\per\second}. 
The GLS periodogram of Fig. \ref{espresso_data}b shows the most prominent peak at 19.23 \si{\day}. This peak does not show a false alarm probability (FAP) of < 10 \%, but it is still remarkable that it stands as precisely the 1 year alias of 18-\si{\day} planet seen in all the literature works. We can see a double peak at 87.5 \si{\day} and 115 \si{\day}, with one being the 1 year alias of the other. Again, even if the peak has a FAP of > 10 \%, the signal at 87.5 \si{\day} has a period comparable with the 89-\si{\day} signal found in previous literature works. 

The first model we considered is a model only concerning a zero-point of the velocity. Then, we tried a blind search for the planets but we do not find a convergent result. The following step is conducted as an informed search for a sinusoidal signal, with a uniform prior on the amplitude between 0 and 3 \si{\meter\per\second}. We consider a normal prior on the period centered at 18.3 \si{\day}, the period of planet b, and a width of 0.2 \si{\day}. This is necessary to avoid a conflict between the signal and its aliases in the determination of the period. We find an amplitude for the signal of 0.51 $\pm$ 0.15 \si{\meter\per\second}, with a period of 18.33 $^{+0.03}_{-0.04}$ \si{\day}. 

\begin{figure}[!h]
    \centering
    \begin{subfigure}[b]{0.45\textwidth}
        \begin{tikzpicture}
            \node[inner sep=0] (image) at (0,0) {\includegraphics[width=\textwidth]{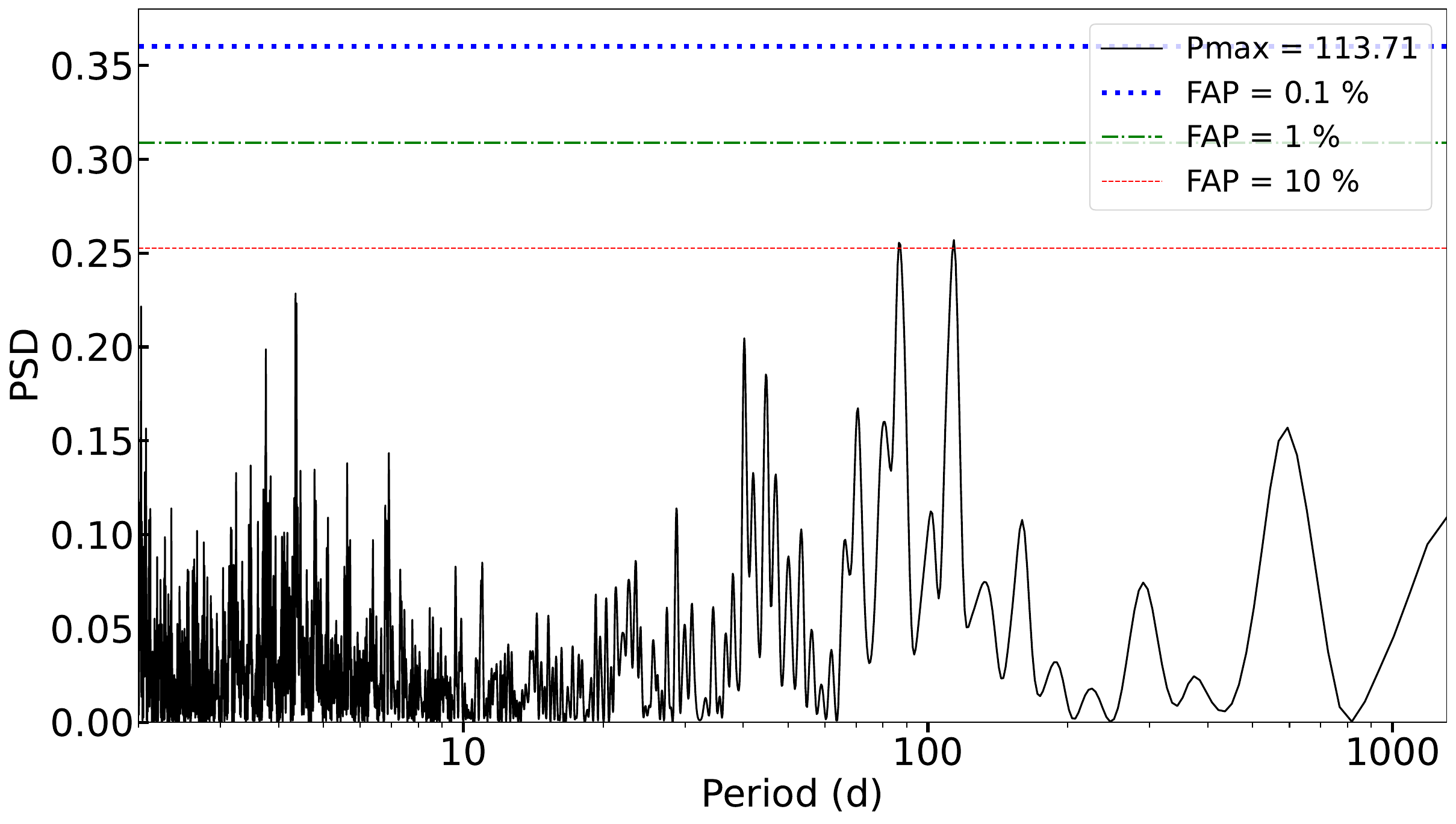}};
            \node[anchor=north west] at ([xshift=1.00cm, yshift=-0.32cm]image.north west) {\textbf{a)}};
        \end{tikzpicture}
    \end{subfigure}
    \hfill
    \begin{subfigure}[b]{0.45\textwidth}
        \begin{tikzpicture}
            \node[inner sep=0] (image) at (0,0) {\includegraphics[width=\textwidth]{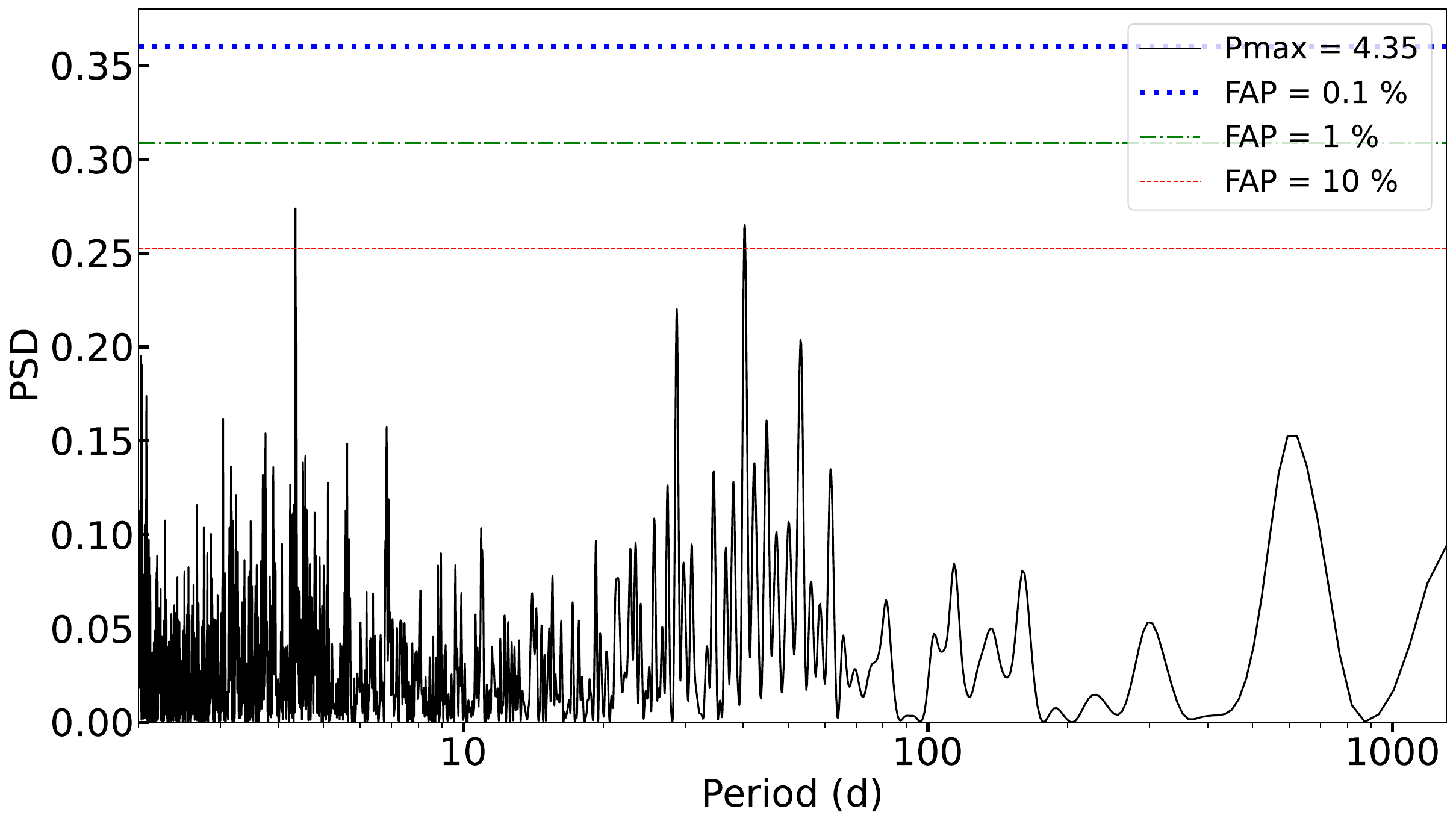}};
            \node[anchor=north west] at ([xshift=1.00cm, yshift=-0.32cm]image.north west) {\textbf{b)}};
        \end{tikzpicture}
    \label{espresso19_periodogram_residuals_2p}
    \end{subfigure}
    \caption{Analysis of the ESPRESSO residual time series. Panel (a): GLS periodogram of the residuals of ESPRESSO RVs for HD20794 once we subtract the signal at 18.3 \si{\day}. We can see a double peak at 87 \si{\day} and 113 \si{\day}, one peak being the 1 year alias of the other. Panel (b): GLS periodogram of the residuals of ESPRESSO RVs for HD20794 after subtracting the signals at 18.3 \si{\day} and 87 \si{\day}. We can see some peaks at 4 \si{\day} and 40 \si{\day}. The peak at 4 \si{\day} probably comes from the sampling, while the peak at 40 \si{\day} could be a signature of the stellar rotation period.}
    \label{espresso19_periodogram_residuals_1p}
\end{figure}

To compare the different models, we used the criterion of the natural logarithm of the evidence  (lnZ) associated with a model (Appendix \ref{methods_sec}). The difference in evidence with the no-planet model is $\Delta$ lnZ = +1.7. This is not sufficient to claim detection.
%We have tried to model the signal with a Keplerian instead of a sinusoidal. The small number of observations of the dataset does not allow us to recover a significant value for the eccentricity, with only an upper limit for eccentricity of 0.67 (84th percentile).
Once we subtract the sinusoidal we have fitted in the one-planet model, we obtain the GLS periodogram of the residuals shown in Fig. \ref{espresso19_periodogram_residuals_1p}a.
In the GLS periodogram, we can see a weak peak at the period of $\sim$ 87 \si{\day} and its 1-year alias at 113 \si{\day}. These two peaks have a FAP that is slightly lower than 10 \%.
%\subsubsubsection{Two-planet model}

When we modeled the second signal, we considered an informed search on the period. We followed the same strategy of the one-planet model, performing an informed search with an additional sinusoidal. For this second sinusoidal we consider a normal prior on the period centered at 89.5 \si{\day}, with a width of 5 \si{\day}. We kept the same prior as before for the signal at 18.3 \si{\day}. The model with two sinusoidals has a $\Delta$ lnZ = +4.1 \textcolor{red}{with respect to the model} with no planets and a $\Delta$ lnZ = +2.4 to the model with one sinusoid. The model is slightly favored both on the one-sinusoidal model and the no-planet model. For the additional signal we find an amplitude of K2 = 0.53 $\pm$ 0.12 \si{\meter\per\second} and a period of P2 = 87.0 $^{+1.0}_{-0.9}$ \si{\day}. We plot the GLS periodogram of the residuals once we also subtract  the signal of the second planet in Fig. \ref{espresso19_periodogram_residuals_1p}b.

\begin{figure}[!h]
    \centering
    \begin{subfigure}[b]{0.45\textwidth}
        \begin{tikzpicture}
            \node[inner sep=0] (image) at (0,0) {\includegraphics[width=\textwidth]{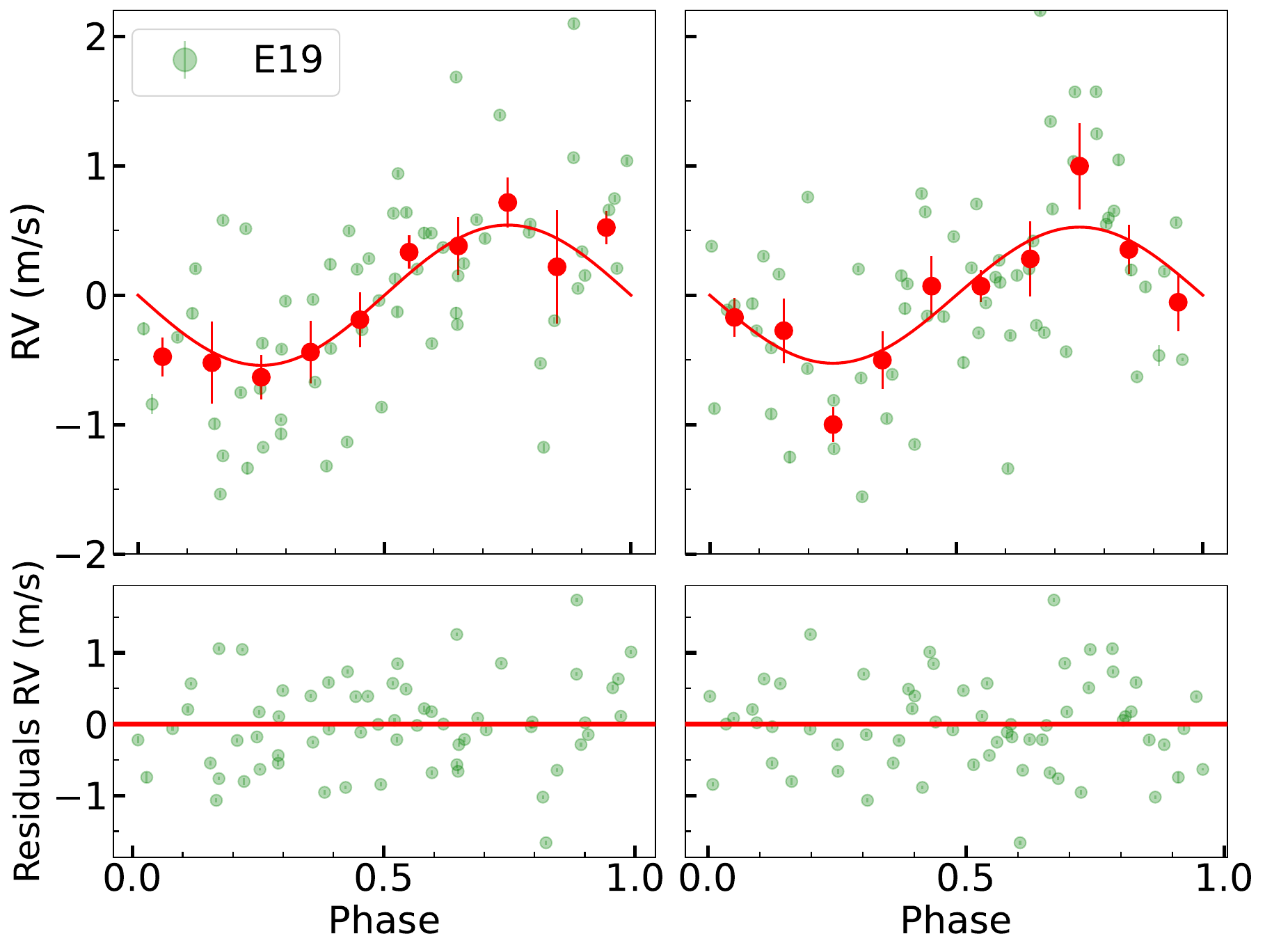}};
            \node[anchor=north west] at ([xshift=1.00cm, yshift=-0.52cm]image.north west) {\textbf{a)}};
        \end{tikzpicture}
    \end{subfigure}
    \hfill
    \begin{subfigure}[b]{0.45\textwidth}
        \begin{tikzpicture}
            \node[inner sep=0] (image) at (0,0) {\includegraphics[width=\textwidth]{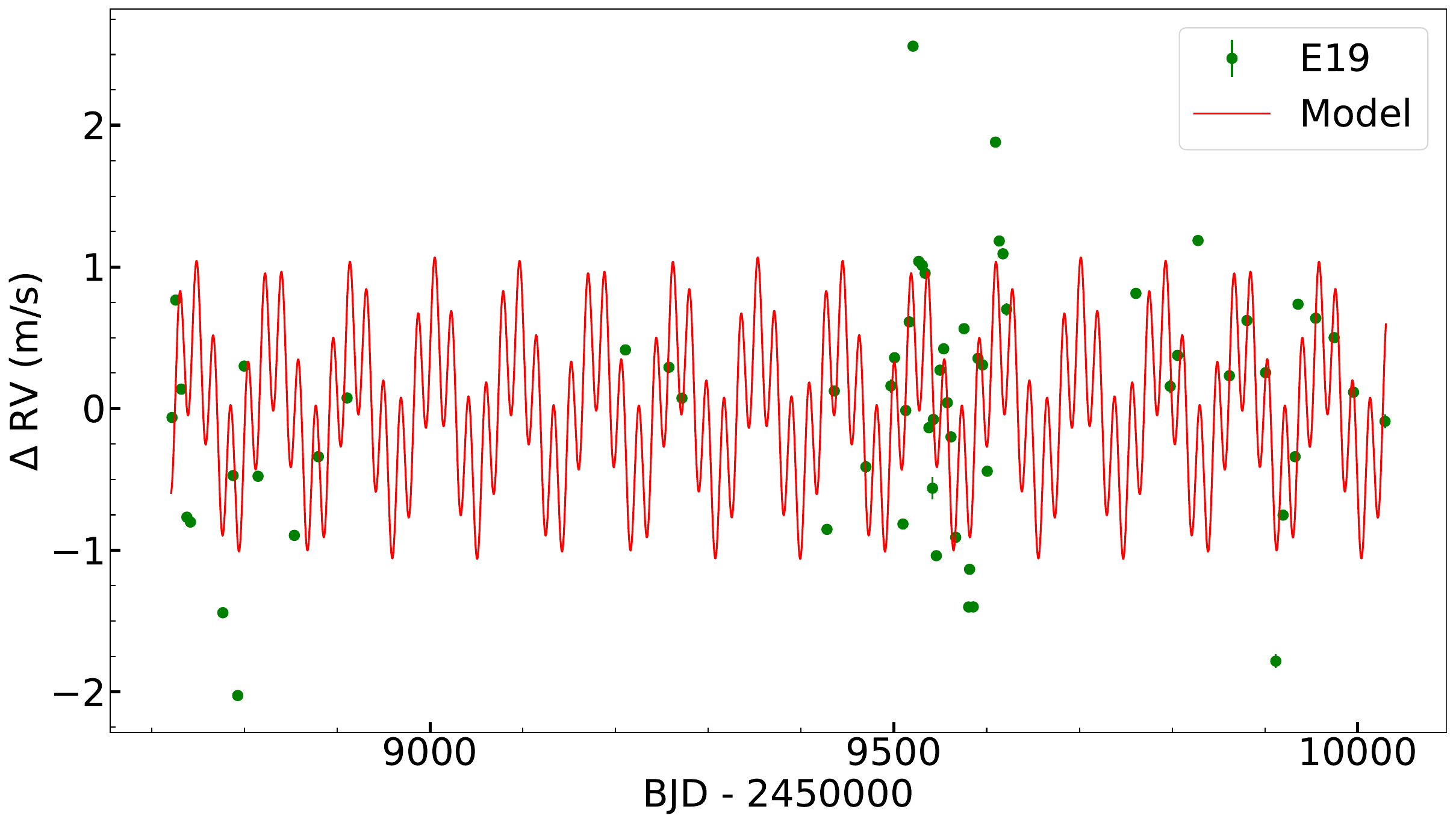}};
            \node[anchor=north west] at ([xshift=1.00cm, yshift=-0.32cm]image.north west) {\textbf{b)}};
        \end{tikzpicture}
        \label{model_espresso_2_planet}
    \end{subfigure}
    \caption{Phase-folded plots and model of planets recovered in the ESPRESSO dataset. (a): Phase-folded plot of 18-\si{\day} planet (on the left) and 87-\si{\day} planet (on the right). Panel (b): Two-planet model for the ESPRESSO dataset alone. We subtracted a zero-point velocity from the dataset.}
    \label{espresso19_phase_folded_2}
\end{figure}

In the GLS periodogram, we can see two peaks with a FAP < 10 \%, at 4.35 \si{\day} and $\sim$ 40 \si{\day}. The period at 40 \si{\day} is of particular interest because it corresponds to the orbital period of the 40-\si{\day} planet found in \citet{pepe}, and, following our analysis of the activity, it is likely caused by stellar rotation. We  followed the same approach as before:\ fitting a sinusoidal with a normal prior on the period of each of the two signals in addition to the previous ones. In the case of the 4.35-day signal, we find worse evidence in comparison to the two-sinusoidal model, with a $\Delta$ lnZ = -0.3 in disfavor of the three-sinusoidal model. We discarded this model. In the case of the 40-\si{\day} signal, we find evidence of a $\Delta$ lnZ = +0.8 compared to the two-planet model, which means a slight improvement. The amplitude related to the third signal is K = 0.46 $^{+0.13}_{-0.15}$ \si{\meter\per\second}, with a period of 40.31 $^{+0.24}_{-0.18}$ \si{\day}. Considering that the stellar origin is the most likely origin of the 40-\si{\day} signal, we tried to model the signal at 40 \si{\day} with a GP, but the fit did not converge, so we  discarded that model. We  tried different kernels as SHO and MEP (see Appendix \ref{stellar_activity_appendix} for details on the definition of the two kernels). Even if we were not able to model this signal with a GP, we considered the signal as likely to be coming from stellar rotation, because we do not see any signature of it in the full analysis. Rotation-related signals are not stable over long timespan and this is the case for the 40-\si{\day} signal. The paucity of measurements in the ESPRESSO dataset alone can also make the GP fit a non-trivial task, considering the low amplitude of activity signals in HD 20794. Taking into account what we know from the analysis of activity indicators and the slight improvement in terms of lnZ once we added a third signal, we considered the best model for the ESPRESSO dataset alone to be the two-planet model with planetary periods of 18.33 $^{+0.03}_{-0.04}$ \si{\day} and 87.0 $^{+1.0}_{-0.9}$ \si{\day}.
For this model, we found a minimum mass for the planet at 18.33 \si{\day} of 1.91 $^{+0.45}_{-0.47}$ M$\oplus$ and a minimum mass for the planet at 87.0 \si{\day} of 3.11 $\pm$ 0.72 M$\oplus$.
We show a phase-folded plot of the planets in Fig.\ref{espresso19_phase_folded_2}a.

A plot of the two-planet model imposed on the ESPRESSO dataset, where a zero-point velocity was subtracted, is shown in Fig.\ref{espresso19_phase_folded_2}b.
A full analysis of the system based on the ESPRESSO dataset alone would require a larger dataset and a longer timespan of observations considering we are not able to  properly recover the planets, even if the photon noise we reach is at the level of 10 \si{\centi\meter\per\second} for a single exposure. 

\subsection{HARPS + ESPRESSO}

\begin{figure}[!h]
    \includegraphics[width=\linewidth]{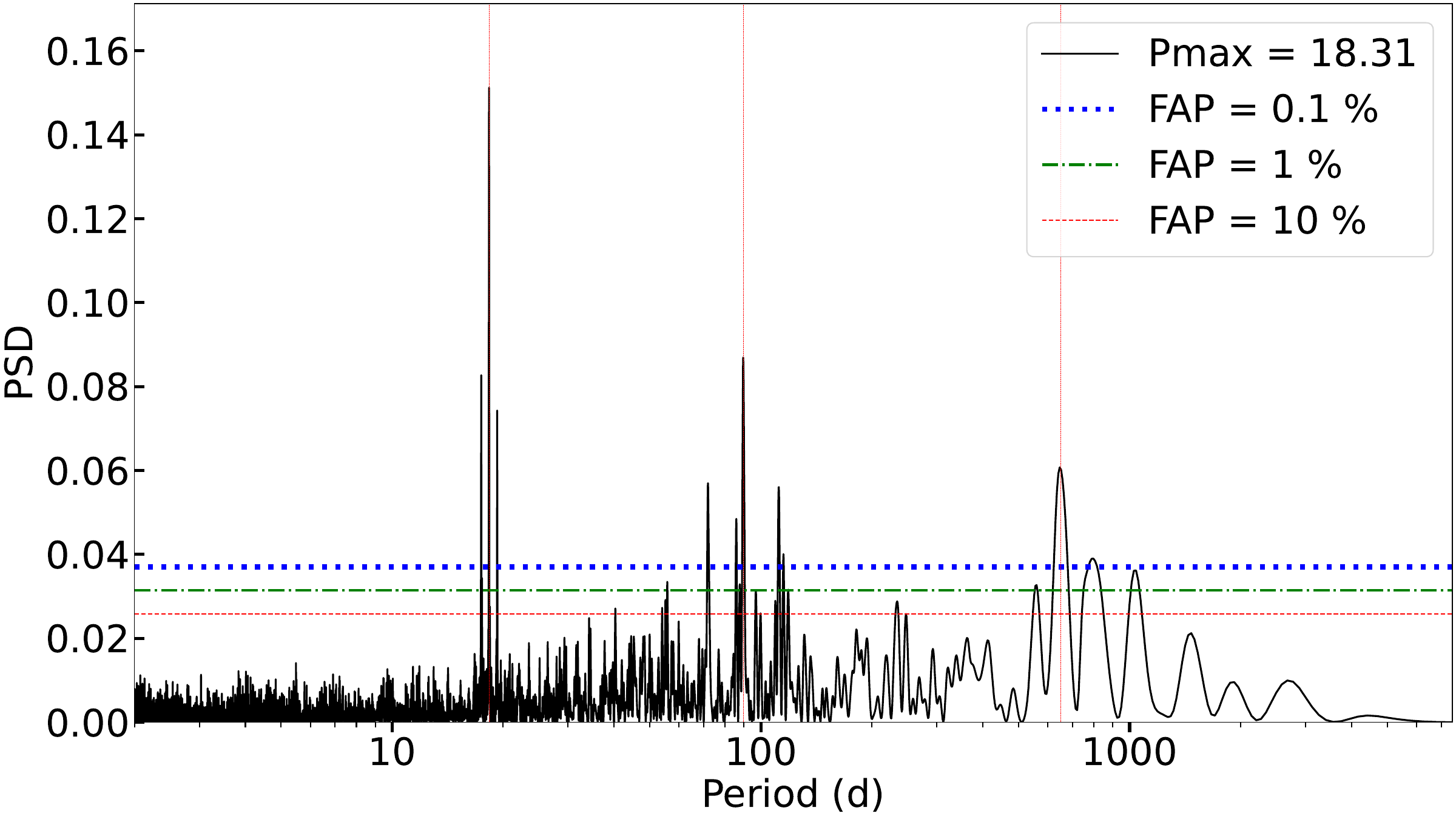}
    \caption{GLS periodogram of the full dataset of RVs for HD20794. The strongest peak is at 18.3 \si{\day}. Also, peaks at 89.6 \si{\day} and 650 \si{\day} are visible. The periods found in \citet{cretignier_yarara_2023} are highlighted by red vertical lines.}
    \label{hd20794_full_dataset_periodogram}
\end{figure}

Once we had explored the possibilities of the ESPRESSO dataset alone, we added the HARPS dataset corrected by YARARA to our analysis. The full dataset is visible in Fig. \ref{RV_data}. The offset between H03 and H15 was calculated for this target at 17.0 $\pm$ 1.7 \si{\meter\per\second} \citep{lo_curto_2015}. 
%We have a dispersion on the measurements for H03 = 1.07 \si{\meter\per\second}, with a dispersion for H15 = 1.56 \si{\meter\per\second}. 
In our analysis, we  fit for the offset as a free parameter.
The instrument efficiency has improved by 33-40 \% after the change of fibers, while the resolution remained constant \citep{lo_curto_2015}. The GLS periodogram of the full dataset is visible in Fig. \ref{hd20794_full_dataset_periodogram}.
A peak at 18.3 \si{\day} is visible. Also, peaks at 89.6 \si{\day} and 650 \si{\day} are visible in the GLS periodogram, all with FAP  < 0.1 \%. 

\subsubsection{One-planet model}

For the analysis of the presence of planetary signals in the data, we followed the same approach we  followed for the
ESPRESSO dataset alone, namely: adding the signals one by one. Considering the larger significance of the peaks and the large number of observations, we go first for a blind search, considering a model with a single sinusoidal with a uniform prior for the period in log space between 2 and 3000 \si{\day}. The blind search in log space for the period of the signal avoids sampling with the same density regions of the parameter space at short and long orbital periods.
We have a uniform prior on the amplitude between 0 \si{\meter\per\second} and 3 \si{\meter\per\second}. We ran our analysis multiple times to investigate the stability and robustness of the result.

We find that the result on lnZ in the blind search is not constant within the error associated with the measurement by our script. This issue with the calculation of the evidence in nested sampling was reported in \citet{nested_issue_2020}. To mitigate this effect, we ran each model five times. We discuss the mean of the best three runs, the best value and the standard deviation of the measurements of lnZ to infer the best model. The decision to consider only the best three runs comes from the model being intrinsically degenerate trying to fit a multiple planetary system with a single planet model that for construction does not have any preference among the signals. Through this procedure, we mitigate the impact of lnZ outliers in the calculation of the mean lnZ.
In the multiple runs, we found two different solutions, the fit converging to both 18.3 \si{\day} and 89.6 \si{\day}. We consider as a reference in evidence the evidence associated with a model with no planets. When we consider the mean of the evidence of the one-planet model compared to the no-planet model we find a $\Delta$ lnZ = +38.0 in favor of the model with one planet. If we compare the best result of the one-planet model with the best result of the no-planet model, the first is favored by a $\Delta$ lnZ = +40.9. The standard deviation of the repeated measurement of lnZ in the one-planet model is 3.0, while for the no-planet model, the standard deviation is 0.2. The difference in lnZ is more than 30.  We summarize in Table \ref{table_lnz} the different models in the analysis with their significance. We consider the best solution for our one-planet model the solution with the best lnZ. We find a planet with an amplitude of K = 0.61 $\pm$ 0.05 \si{\meter\per\second} and a period of P = 18.314 $\pm$ 0.002 \si{\day}. The GLS periodogram of the residuals after subtracting the signal at 18.313 \si{\day} is visible in Fig. \ref{hd20794_final_1p_residuals}a.

In the GLS periodogram of the residuals, it is possible to see a strong peak at 89.66 \si{\day}, which is the period of a planet found in all previous works on the system. There is also a second prominent peak at $\sim$ 650 \si{\day} and a peak at $\sim$ 111 \si{\day}.

\subsubsection{Two-planet model}

We continue in the analysis and test a model comprising a second sinusoidal signal. In this model, we carried out a blind search for both the periods of the two planets, with overlapping priors in log space between 2 and 3000 \si{\day}. We keep the same setup for the other parameters.
The two-planet model is favored in terms of the mean evidence of repeated runs by a $\Delta$ lnZ = +32.2 compared to the one-planet model, and by a $\Delta$ lnZ = +70.2 compared to the no-planet model. If we compare the best models, we have a $\Delta$ lnZ = +33.3 with respect to the one-planet model and a $\Delta$ lnZ = +74.2 with respect to the no-planet model in favor of the two-planet model. The standard deviation on the evidence of the repeated measurement is 3.7. 
In the multiple runs, we find results converging to three different planets, at 18.3, 89.6, and 650 \si{\day}.
Considering the best run in terms of lnZ we find a short-term signal with an amplitude K= 0.60 $\pm$ 0.05 \si{\meter\per\second}, and a period of P = 18.313 $\pm$ 0.002 \si{\day}, and a second signal recovered with an amplitude K2 = 0.46 $\pm$ 0.05 \si{\meter\per\second} and a period P2 = 89.65$^{+0.10}_{-0.09}$\si{\day}. We show the GLS periodogram of the residuals in Fig. \ref{hd20794_final_1p_residuals}b, where we can see a strong peak at 651.79 \si{\day} and two additional peaks at $\sim$ 85 \si{\day} and $\sim$ 111 \si{\day}, one the 1-year alias of the other. The long-period peak is compatible in period with the long-term planet found in \citet{cretignier_yarara_2023}.  

\subsubsection{Three-planet model}

\begin{figure}[!h]
    \centering
    \begin{subfigure}[b]{0.45\textwidth}
        \begin{tikzpicture}
            \node[inner sep=0] (image) at (0,0) {\includegraphics[width=\textwidth]{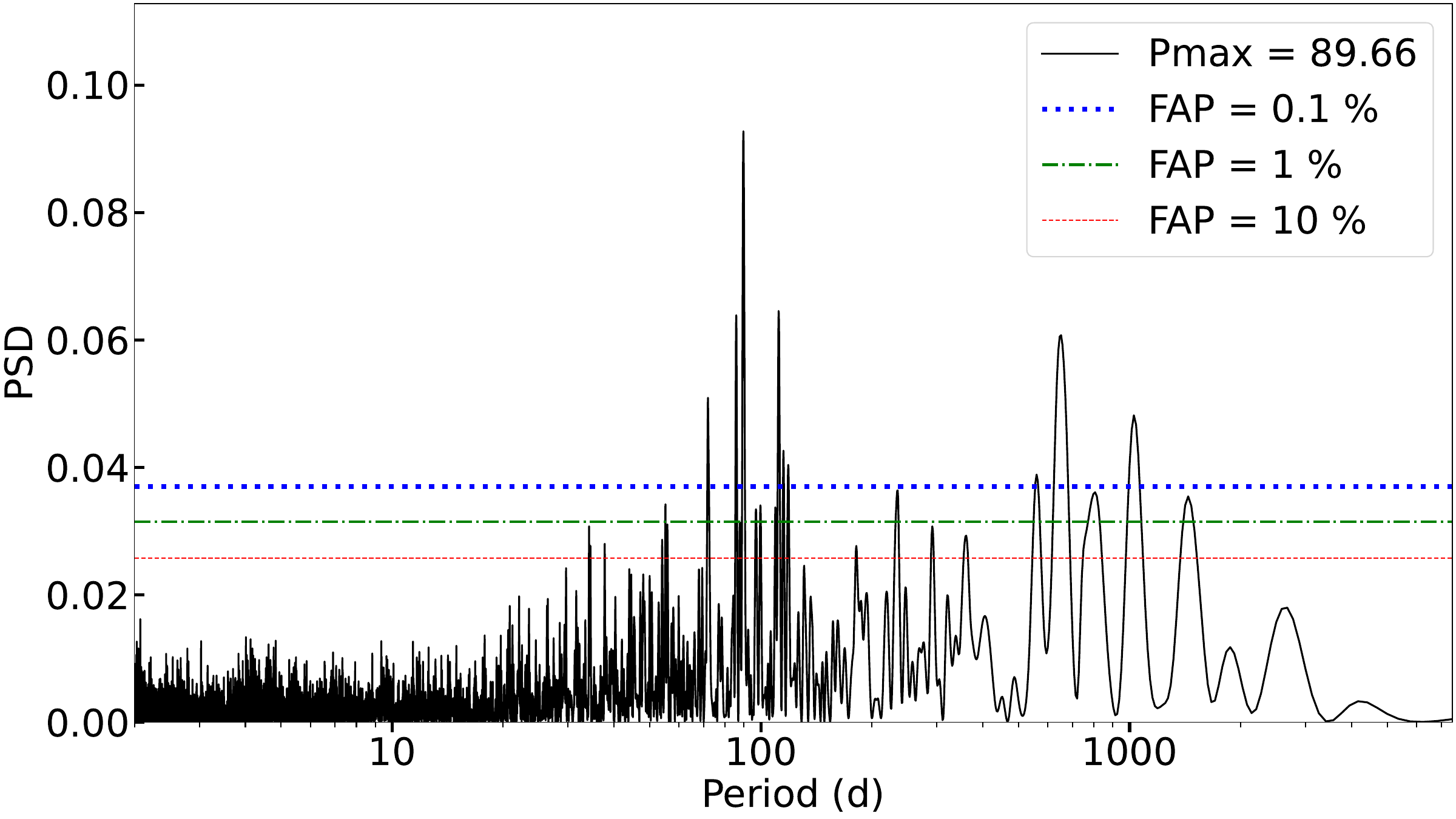}};
            \node[anchor=north west] at ([xshift=0.8cm, yshift=-0.32cm]image.north west) {\textbf{a)}};
        \end{tikzpicture}
    \end{subfigure}
    \hfill
    \begin{subfigure}[b]{0.45\textwidth}
        \begin{tikzpicture}
            \node[inner sep=0] (image) at (0,0) {\includegraphics[width=\textwidth]{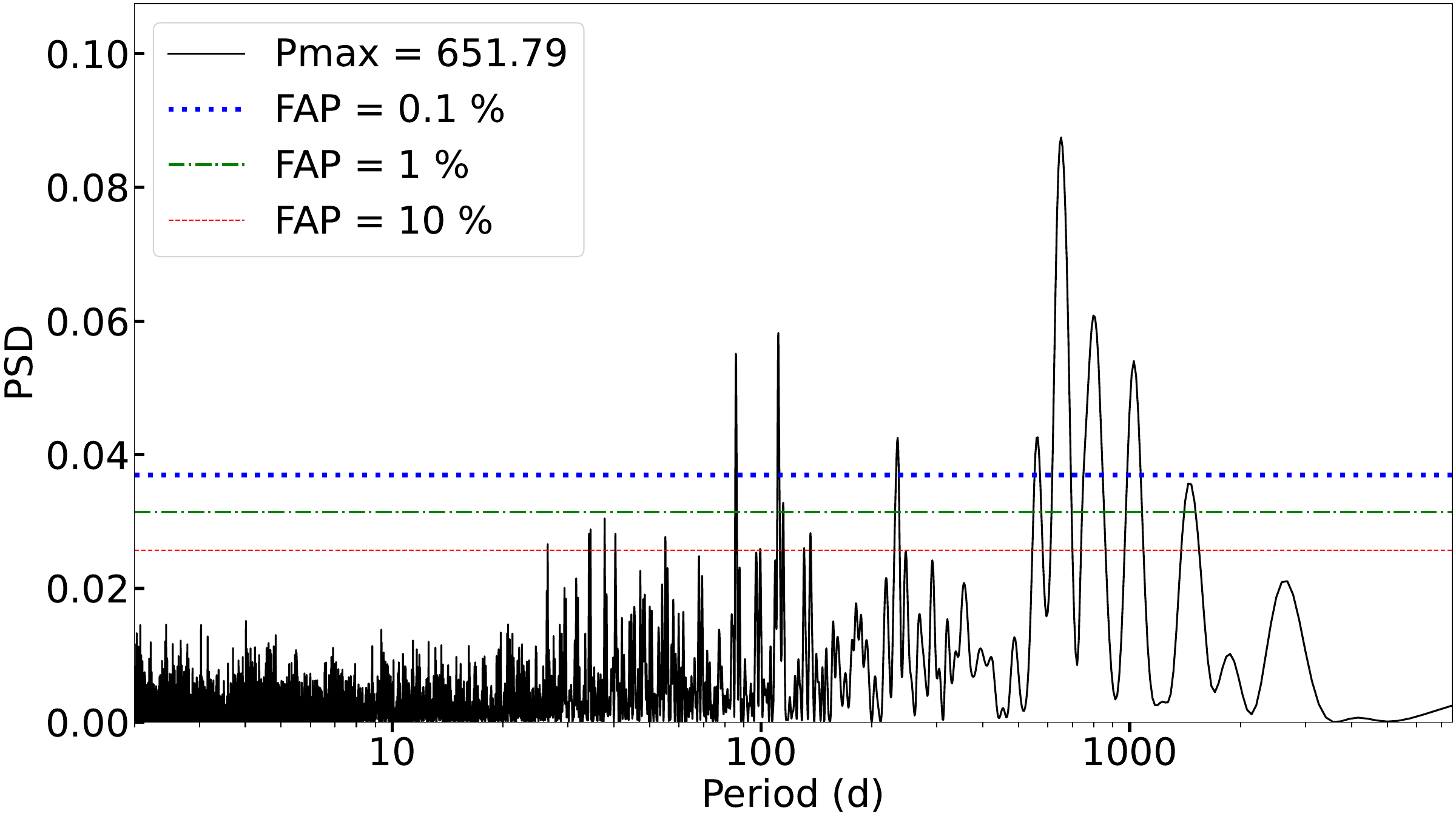}};
            \node[anchor=north east] at ([xshift=-0.2cm, yshift=-0.32cm]image.north east) {\textbf{b)}};
        \end{tikzpicture}
        \label{hd20794_final_2p_residuals}
    \end{subfigure}
    \hfill
    \begin{subfigure}[b]{0.45\textwidth}
        \begin{tikzpicture}
            \node[inner sep=0] (image) at (0,0) {\includegraphics[width=\textwidth]{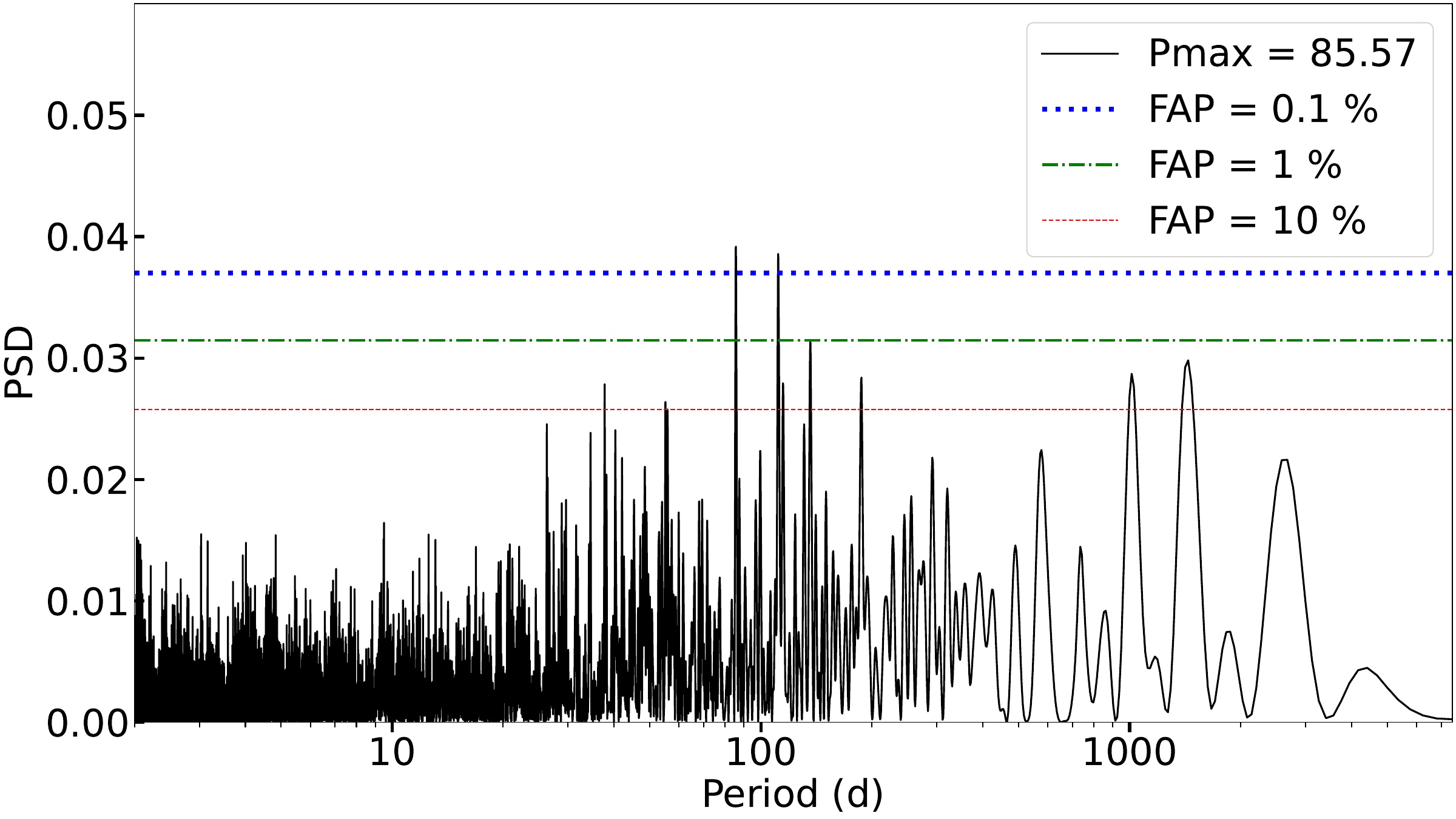}};
            \node[anchor=north west] at ([xshift=0.8cm, yshift=-0.32cm]image.north west) {\textbf{c)}};
        \end{tikzpicture}
        \label{hd20794_final_3p_residuals}
    \end{subfigure}
    \caption{Analysis of the residuals time series after planetary signals subtraction. Panel (a): GLS periodogram of the residuals after subtracting the 18.3 \si{\day} signal. A peak at $\sim$ 89 \si{\day} and another one at $\sim$ 650 \si{\day} are visible. Panel (b): GLS periodogram of the residuals after subtracting the 18.3 \si{\day} and 89.6 \si{\day} signals. A peak at $\sim$ 650 \si{\day} is visible. Panel (c): GLS periodogram of the residuals after subtracting the 18.314 \si{\day}, 650.9 \si{\day}, and 89.65 \si{\day} signals. We see some additional peaks in the periodogram of the residuals with FAP < 0.1 \% at periods of $\sim$ 85 \si{\day} and 111 \si{\day}.}
    \label{hd20794_final_1p_residuals}
\end{figure}
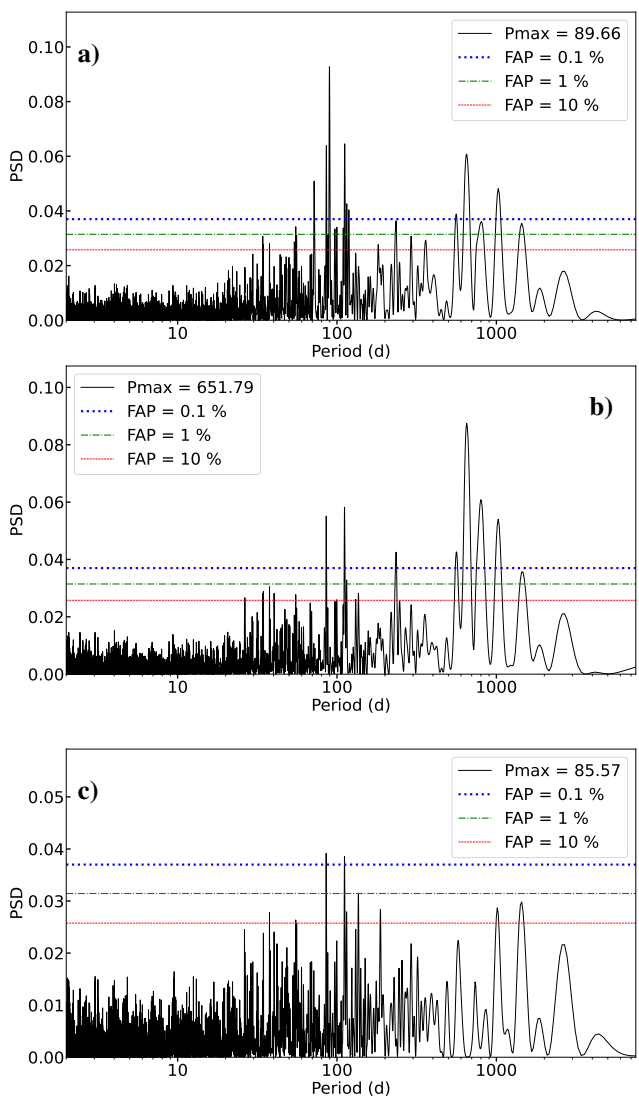

The GLS periodogram of the residuals in the two-planet configuration indicates the presence of an additional planet at an orbital period of 651 \si{\day}. This peak corresponds to the candidate planet detected in \citet{cretignier_yarara_2023}. To model this third signal, we followed the same procedure as we did for the others. We considered three sinusoids with the same blind priors on the period, spanning in log space all periods between 2 \si{\day} and 3000 \si{\day}, and a uniform prior on the amplitudes between 0 and 3 \si{\meter\per\second}. The average evidence of the three-planet model is favored, compared to the two-planet model in terms of lnZ by $\Delta$ lnZ = +29.4, while the best models differ by $\Delta$ lnZ = +27.4, in favor of the three-planet model. The repeated runs have a standard deviation of 2.0. The large difference in the evidence compared to the two-planet model points toward the direction of this model as the best to describe the dataset. The periods of the planets are always recovered with the same three periods, even if the order those are recovered can differ from run to run. We refer to the planets at 18.3 \si{\day}, 89.6 \si{\day}, and 650 \si{\day} as HD 20794 b, HD 20794 c, and HD 20794 d, respectively. For HD 20794 b, we find an amplitude of K$_{b}$ = 0.60 $\pm$ 0.05 \si{\meter\per\second} and a period of P$_b$ = 18.314 $\pm$ 0.002 \si{\day}. For HD 20794 c, we find an amplitude of K$_{c}$ = 0.52 $\pm$ 0.05 \si{\meter\per\second} and a period of P$_c$ = 89.65 $\pm$ 0.08 \si{\day}. For HD 20794 d, we find an amplitude of K$_{d}$ = 0.46 $\pm$ 0.05 \si{\meter\per\second} and a period of P$_d$ = 650.9 $^{+5.0}_{-4.9}$ \si{\day}. Once we subtract this model, we have the GLS periodogram of the residuals shown in Fig. \ref{hd20794_final_1p_residuals}c.

\begin{figure}[!h]
    \includegraphics[width=\linewidth]{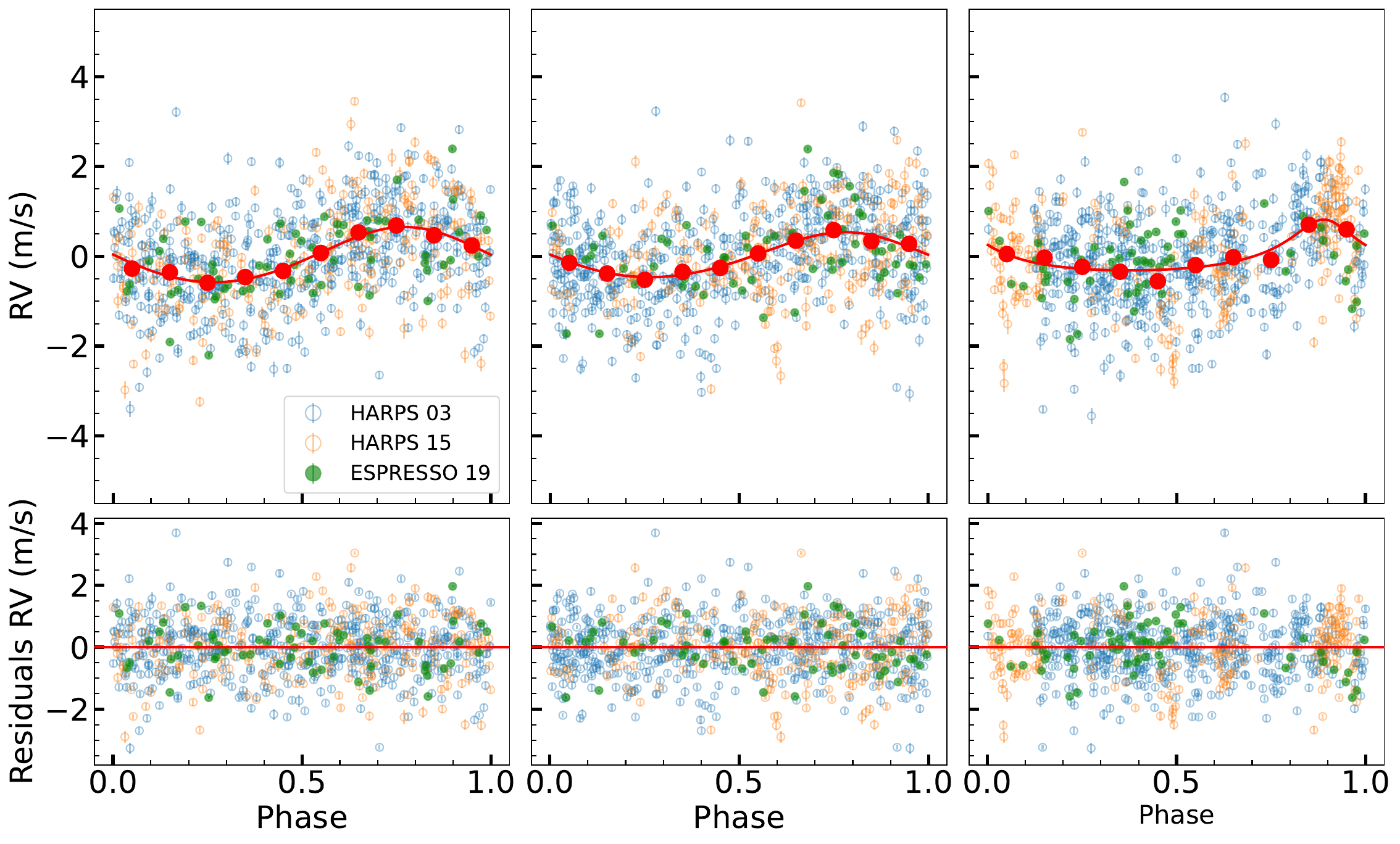}
    \caption{Phase-folded plots: HD 20794 b (left);  HD 20794 c (middle); and\ HD 20794 d (right).}
    \label{hd20794_final_3p_phase_folded}
\end{figure}

\begin{figure}[!h]
    \includegraphics[width=\linewidth]{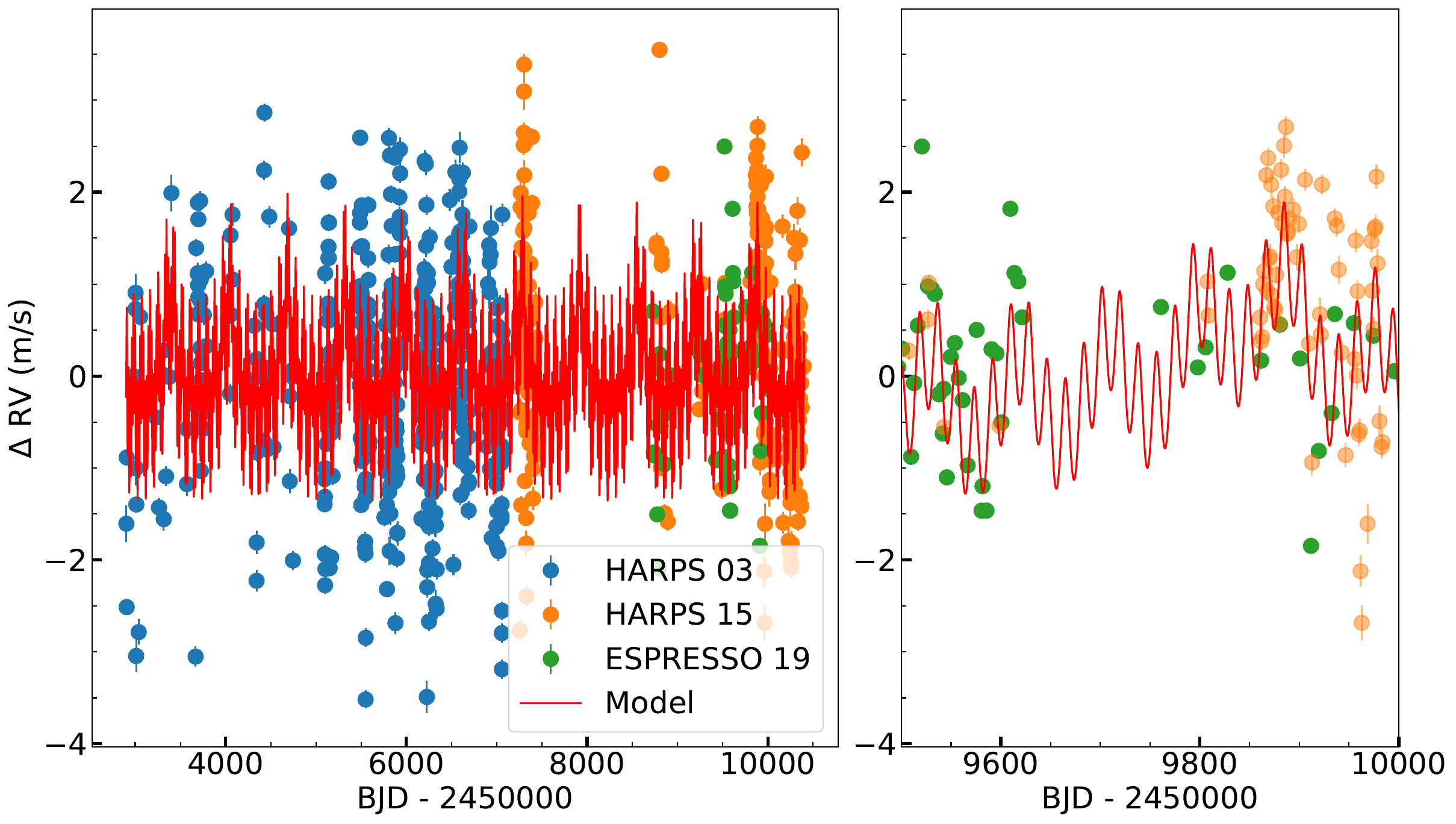}
    \caption{Three-planet model for the full dataset (left) and zoom on the model on the ESPRESSO dataset (right).}
    \label{hd20794_3p_model}
\end{figure}

We see two peaks with FAP < 0.1 \% at 85.6 and $\sim$ 111.7 \si{\day}. These two signals are one the one-year aliases of the other.
We tested a model with three Keplerians instead of three sinusoids to compare with \citet{cretignier_yarara_2023}. To sample for eccentricity and argument of the pericenter, we did not  directly use the parameters, but  combinations of them:
$\sqrt{e}$ $\cos$($\omega$) and $\sqrt{e}$ $\sin$($\omega$). We used a normal prior centered at 0 with a width of 0.3. We imposed the condition e $\leq$ 1. The mean evidence of the Keplerian model was favored, compared to the model with circular orbits by a $\Delta$ lnZ = +8.7, with a $\Delta$ ln Z = +7.1 for the best model evidence. HD 20794 b has an amplitude K$_b$ = 0.62 $\pm$ 0.05 \si{\meter\per\second} and a period P$_b$ = 18.314 $\pm$ 0.002 \si{\day}; HD 20794 c has an amplitude K$_c$ = 0.50 \si{\meter\per\second} and a period P$_c$ = 89.67 $_{-0.10}^{+0.11}$ \si{\day}; HD 20794 d has an amplitude K$_d$ = 0.57 $\pm$ 0.07 \si{\meter\per\second} and a period P$_d$ = 647.5 $_{-2.8}^{+2.6}$ \si{\day}. We can only put upper limits on e$_b$ < 0.13 (84th percentile) for HD 20794 b, and e$_c$ < 0.16 (84th percentile) for HD 20794 c; whereas for HD 20794 d, we find a value for the eccentricity from the best run e$_d$ = 0.45 $_{-0.11}^{+0.10}$. The eccentricity of HD 20794 is significantly different from zero, according to \citet{cretignier_yarara_2023}. We see peaks at $\sim$ 85 \si{\day} and 111 \si{\day} in the periodogram of the residuals again.
In Fig. \ref{hd20794_final_3p_phase_folded} we show the phase-folded plots of the three planets from the Keplerian model. 
In Fig. \ref{hd20794_3p_model} we show the three-Keplerian model. A zoom on the model on the ESPRESSO dataset is also shown.

\subsubsection{Four-planet model}

We applied a blind search for a model with four Keplerians to search for additional signals in the dataset. We considered a Keplerian model because we found a significative eccentricity for HD 20794 d. This model has a $\Delta$ lnZ = +2.0 compared to the three-planet model, with a standard deviation of 4.6. The best solution for the four-planet model has a $\Delta$ lnZ = +6.9 compared to the three-planet model. We do not see the fourth signal converging clearly to any of the peaks we see in Fig. \ref{hd20794_final_1p_residuals}c. The posterior distribution of the additional signal is shared between the signal at $\sim$ 85 \si{\day}, the signal at $\sim$ 111 \si{\day}, and additional long-term signals at $\sim$ 1000 \si{\day} and $\sim$ 1400 \si{\day}. The amplitude of this hypothetical signal would be at $\sim$ 30 \si{\centi\meter\per\second}. Furthermore, the improvement in lnZ is not significant enough to claim a new detection. Investigating the presence of additional planetary candidates, at the level of 30 \si{\centi\meter\per\second}, requires additional observations to be thoroughly investigated.

\captionsetup{justification=raggedright, singlelinecheck=false} % Allinea a sinistra

\begin{table}[h!]
  \centering
  \begin{threeparttable}
    \caption{Evidence of different models for HD 20794.} % Caption allineata a sinistra
    \label{table_lnz}
    \begin{tabular}{p{0.5\linewidth}ll}
      \hline
      \hline
      \noalign{\smallskip}
      Model & $\Delta$ lnZ-mean &  $\Delta$ lnZ-best \\
      \noalign{\smallskip}
      \hline
      \noalign{\smallskip}
      No-planet & -108.3 & -108.6 \\
      One-sinusoidal & -69.3 & -60.2 \\
      Two-sinusoidal & -38.1 & -34.5 \\
      Three-sinusoidal & -8.7 & -7.1 \\
      Three-Keplerian & 0.0 & 0.0 \\
      Four-Keplerian & +2.0 & +6.9 \\
      \noalign{\smallskip}
      \hline
    \end{tabular}
    \medskip % Spazio verticale tra la tabella e la lista di riferimenti
    
    \begin{minipage}{0.5\textwidth}
        \raggedright
        Notes: We consider as the zero-point of each column separately the lnZ associated with a three-Keplerian model. The values in the table reflect the comparison between different models for the planetary system around HD 20794.
    \end{minipage}
  \end{threeparttable}
\end{table}

\subsubsection{False inclusion probability}

\begin{figure}[!h]
    \includegraphics[width=\linewidth]{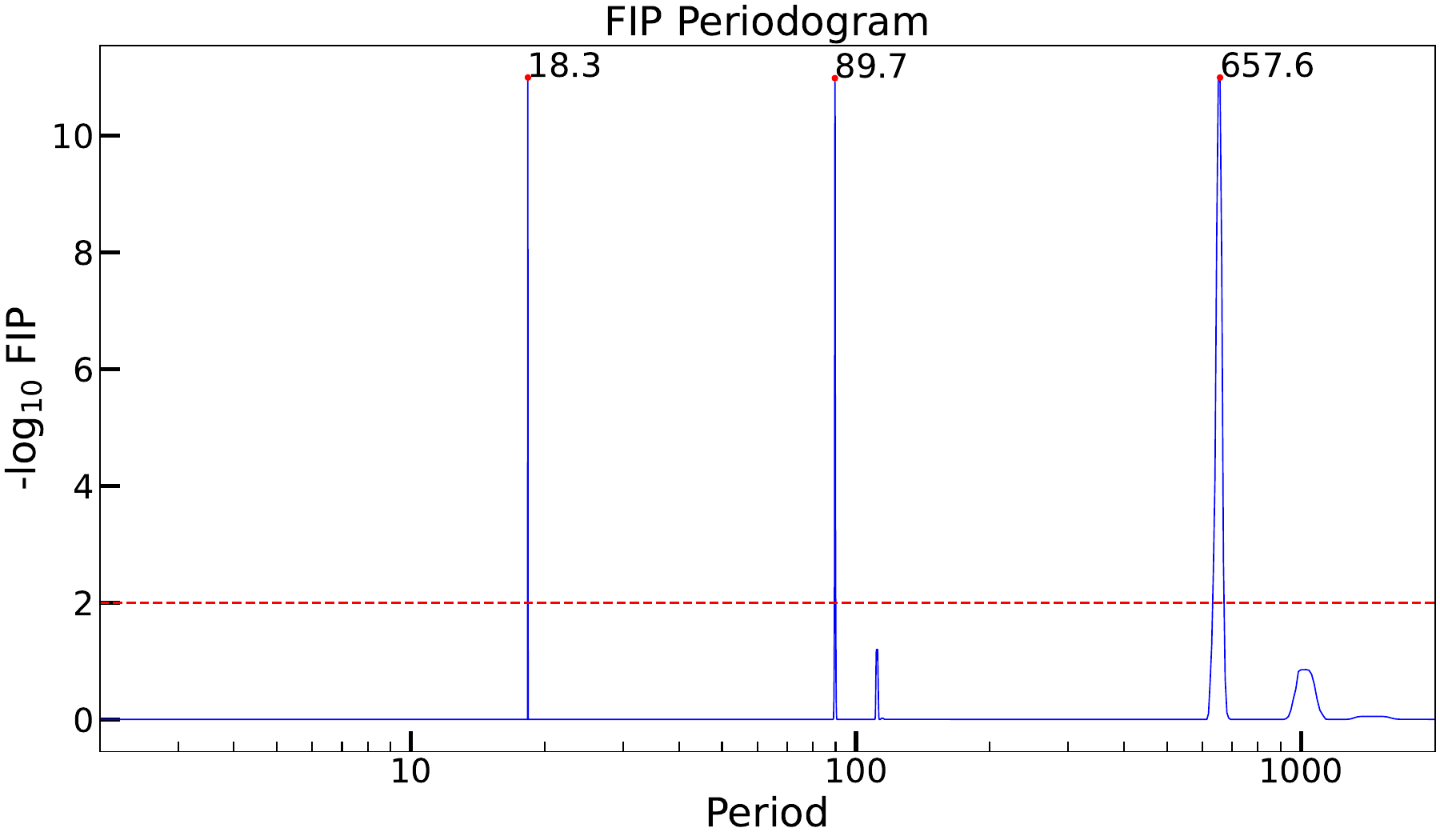}
    \caption{FIP periodogram for HD20794. We recover three signals at the same periodicities we derived in our blind search. The FIP levels for the planets at 18.3 \si{\day}, 89.7 \si{\day}, and 657.6 \si{\day} are cut because the numerical calculation gave us back an infinite value.}
    \label{fip_3_planets}
\end{figure}

The analysis of the ESPRESSO dataset and the HARPS dataset corrected by YARARA revealed the presence of at least three planets in the system. Statistical tools are available to assess the robustness of the result. We analyzed the false inclusion probability (FIP) \citep{fip_hara_2022} to assess the number and significance of planetary signals in the data. FIP has been demonstrated to be optimal to maximize true detections for a certain tolerance to false positives \citep{fip_hara_2024}. We take a grid of frequency intervals with a fixed length. The \textit{k} element of the grid is defined as [$\omega_{\textit{k}}$ - $\Delta$$\omega$/2,$\omega_{\textit{k}}$ + $\Delta$$\omega$/2], where $\omega_{\textit{k}}$ is defined as $\omega_{\textit{k}}$ = \textit{k}$\Delta$$\omega$/$N_{\mathrm{oversampling}}$ and $\Delta$$\omega$ is defined as $\Delta$$\omega$=2$\pi$/T$_{obs}$. Here, T$_{obs}$ is the total time of the observations, and we take a $N_{\mathrm{oversampling}}$ = 5. We considered a Keplerian model for the planets.
In Fig. \ref{fip_3_planets}, we show the result of our FIP analysis. In the plot, we cut the levels of FIP for the 18.3 \si{\day}, the 89.6 \si{\day}, and the 657.6 \si{\day} signals because the numerical result of the FIP is infinite (this points toward a strong confirmation for these two signals). 
We see some additional peaks in the FIP at $\sim$ 111 \si{\day} and $\sim$ 1000 \si{\day}, but not exceeding the 1 \% FIP, that translates in a confidence in the detection lower than 1 \%.

Considering the issue with the stability of the evidence calculation of the nested sampling we reported in our previous analysis we repeated the FIP calculation up to five times to check the consistency through the runs. We always recover each of the three signals with confidence > 99.999 \% in every run. The peaks related to the planets at 111 \si{\day} and 1000 \si{\day} exceed the 1 \% FIP limit just once over multiple runs. The hypothetical characteristics of the 111-\si{\day} signal are discussed in Appendix \ref{111_day}.
%We do not find a peak in the FIP corresponding to the period of the possible fourth signal previously reported. 

\subsubsection{Final results}
We conducted multiple runs to marginalize the intrinsic instability of our sampling method. The dispersion on lnZ due to the sampler is one order of magnitude less than the difference of lnZ we obtain every time we add planets b,c, and d, respectively. The FIP indicates the presence of the same three planets we have found in the blind search. An eventual additional candidate arises once we subtract the signal of the three known planets from our dataset, but both a blind search and the FIP analysis do not corroborate its presence solidly. Considering these elements we take as our best model the one with three Keplerians. Once we have assessed the significance of the three signals in our analysis framework, we can relax the constraints on searching the periods blindly. We do this to find the best parameters for the model. We want to define disjoint regions of the parameter space where to search for different signals, to avoid the nested sampling method to struggle between competing signals. We considered a normal prior on the periods based on our previous knowledge derived from the blind-search and the FIP analysis. 
In principle, the decision on the width of the prior can influence the result of the analysis. To study this possible source of noise, we performed different analyses with a width of the normal prior equal to 5 \%, 10 \%, and 20 \% the period of the planet. We chose, as the starting periods for the planets, 18.3 \si{\day}, 89.6 \si{\day}, 650 \si{\day} for HD 20794 b, c, and d, respectively. We determined our results on the periods of the planets different at less than 0.1 $\sigma$.
We concluded that the prior definition, on the limits of a prior larger than 5 \% of the period and narrower than 20 \%, does not influence the result of our analysis. We refer to the results obtained with a prior width of 10 \% of the period value. We obtain for HD 20794 b an amplitude of K$_{b}$ = 0.614 $\pm$ 0.048 \si{\meter\per\second}, with a period P$_{b}$ = 18.3140 $\pm$ 0.0022 \si{\day}, and un upper limit on eccentricity, e$_b$ < 0.13 . This corresponds to a planet with a minimum mass of M$_{p}$sini = 2.15 $\pm$ 0.17 M$\oplus$ orbiting at a distance of 0.12570$_{-0.00052}^{+0.00053}$ au. For HD 20794 c we obtain an amplitude of K$_{c}$ = 0.502 $_{-0.049}^{+0.048}$ \si{\meter\per\second}, with a period of P$_{c}$ = 89.68 $\pm$ 0.10 \si{\day}, and un upper limit on eccentricity e$_c$ < 0.16. This corresponds to a planet with a minimum mass of M$_{p}$sini = 2.98 $\pm$ 0.29 M$\oplus$ orbiting at a distance of 0.3625 $_{-0.0016}^{+0.0015}$ au from the star. HD 20794 d has an amplitude of K$_{d}$ = 0.567$_{+0.067}^{-0.064}$ \si{\meter\per\second}, a period of P$_{d}$ = 647.6$_{-2.7}^{+2.5}$ \si{\day}, an eccentricity of e$_d$ = 0.45 $_{-0.11}^{+0.10}$, a minimum mass of M$_{p}$sini = 5.82 $\pm$ 0.57 M$\oplus$, and an orbital distance of 1.3541 $\pm$ 0.0068 au.
Once we subtract the model with three Keplerian, the RMS of the dataset goes from 1.14 \si{\meter\per\second} to 0.93 \si{\meter\per\second}. The RMS for H03 goes from 1.14 \si{\meter\per\second} to 0.93 \si{\meter\per\second}, the RMS for H15 goes from 1.21 \si{\meter\per\second} to 0.99 \si{\meter\per\second}, and the RMS for E19 goes from 0.84 \si{\meter\per\second} to 0.72 \si{\meter\per\second}.
In Table \ref{tab:parameters}, we summarize the main results obtained in our analysis.
\captionsetup{justification=centering, singlelinecheck=true}
\renewcommand{\arraystretch}{1.4}
\begin{table*}[h!]

    \centering
    \caption{Parameters for planets b,c,d derived in a model with norm priors.}
    \begin{tabular}{cccc} 
        \hline 
        \hline
        Parameter & HD 20794 b & HD 20794 c & HD 20794 d \\ 
        \hline
        T0 - 2 450 000 [d] & 10383.26 $\pm$ 0.54 & 10356.2 $_{-4.5}^{+5.2}$ & 9933$_{-17}^{+22}$ \\
        P (d) & 18.3140 $\pm$ 0.0022 & 89.68 $\pm$ 0.10 & 647.6 $_{-2.7}^{+2.5}$\\
        K [ms$^{-1}$] & 0.614 $\pm$ 0.048 & 0.502 $_{-0.049}^{+0.048}$ & $0.567_{-0.064}^{+0.067}$\\
        e & 0.064$_{-0.046}^{+0.065}$ & 0.077$_{-0.055}^{+0.084}$ & $0.45_{+0.10}^{-0.11}$\\
        m$_{p}$sini [M$_{\oplus}$] & 2.15 $\pm$ 0.17 & 2.98 $\pm$ 0.29 & 5.82 $\pm$ 0.57  \\
        \textit{a} [AU]  & 0.12570 $_{-0.00053}^{+0.00052}$ & 0.3625$_{-0.0016}^{+0.0015}$ & 1.3541 $\pm$ -0.0068\\
        \hline 
    \end{tabular}
    \medskip % Spazio verticale tra la tabella e la lista di riferimenti
    
    \begin{minipage}{0.5\textwidth}
        \raggedright
        Notes: A  Keplerian model is considered here.
    \end{minipage}
    \label{tab:parameters} % Etichetta per fare riferimento alla tabella
\end{table*}

\section{Discussion}
\label{sec_dis}
\subsection{Planetary system}
\begin{figure}[!h]
    \includegraphics[width=\linewidth]{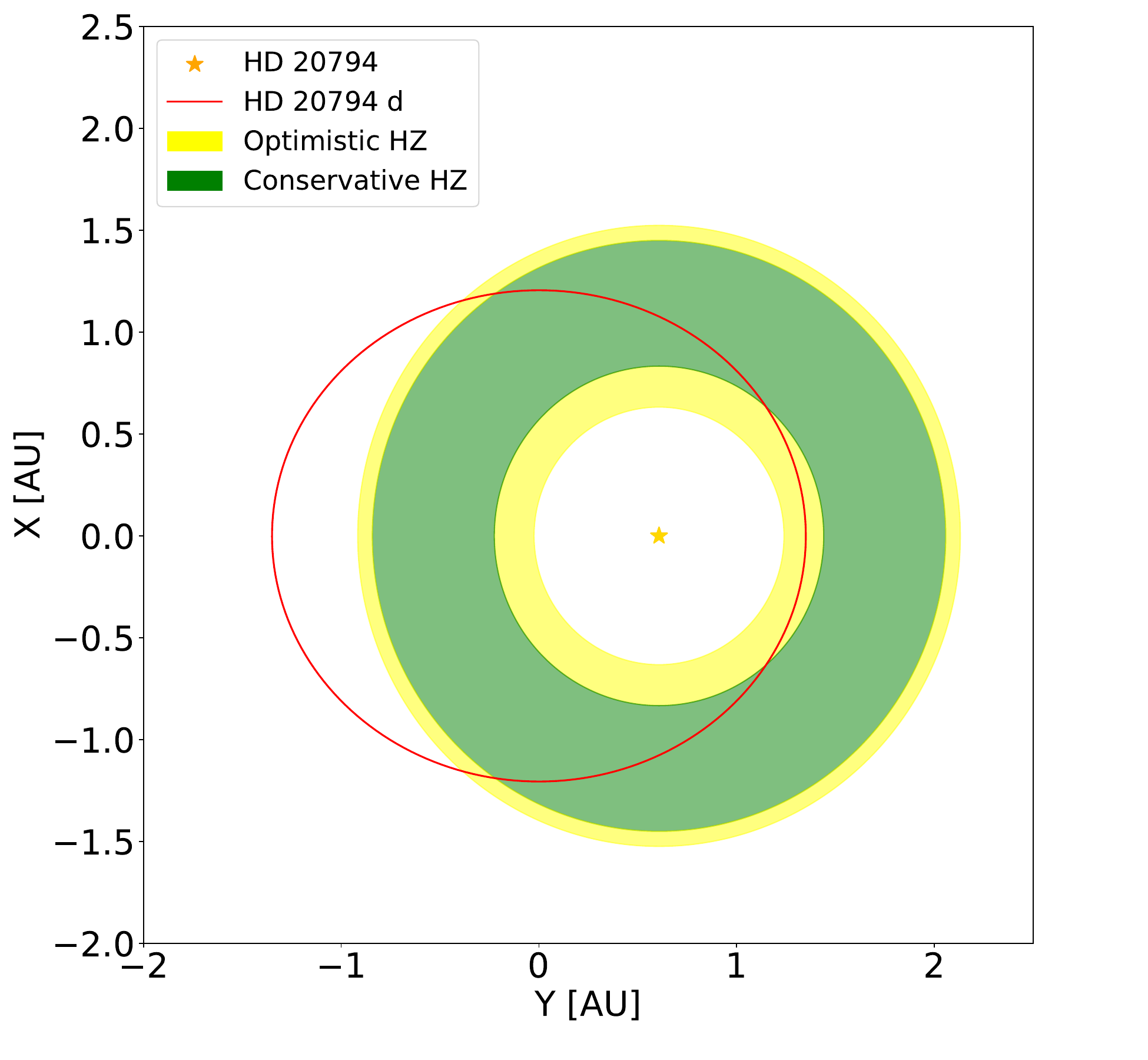}
    \caption{ Position of the HZ relative to the elliptical orbit of HD 20794 d. It is possible to see how the planet crosses the HZ both in the optimistic and conservative boundaries for a large proportion of time.}
    %\label{hd20794_final_3p_phase_folded}
    \label{cross_hz}
\end{figure}

\begin{figure*}[!h]
    \includegraphics[width=\linewidth]{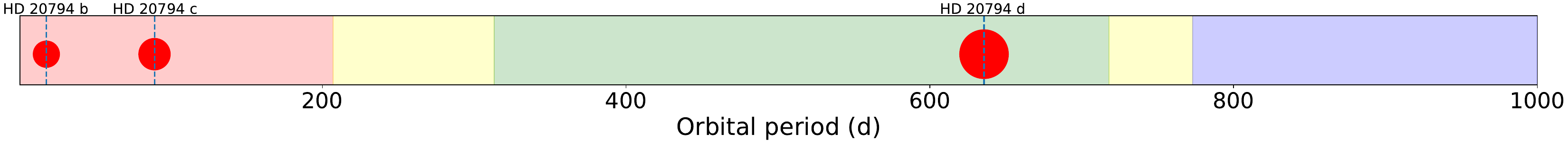}
    \caption{
    HZ for HD 20794. Following \citep{kopparapu_habitable_zone} we mark the conservative HZ in
green, while the optimistic HZ is marked in yellow. The red part of the plot considers all the periods remaining on the interior of the optimistic HZ, while the blue part represents the periods wider than the outer edge of the optimistic HZ. %Red dots represent planets b,c, and d, with radius corresponding to their minimum-mass ratio.
    }
    \label{hz_plot}
\end{figure*}

\begin{figure}[!h]
    \includegraphics[width=\linewidth]{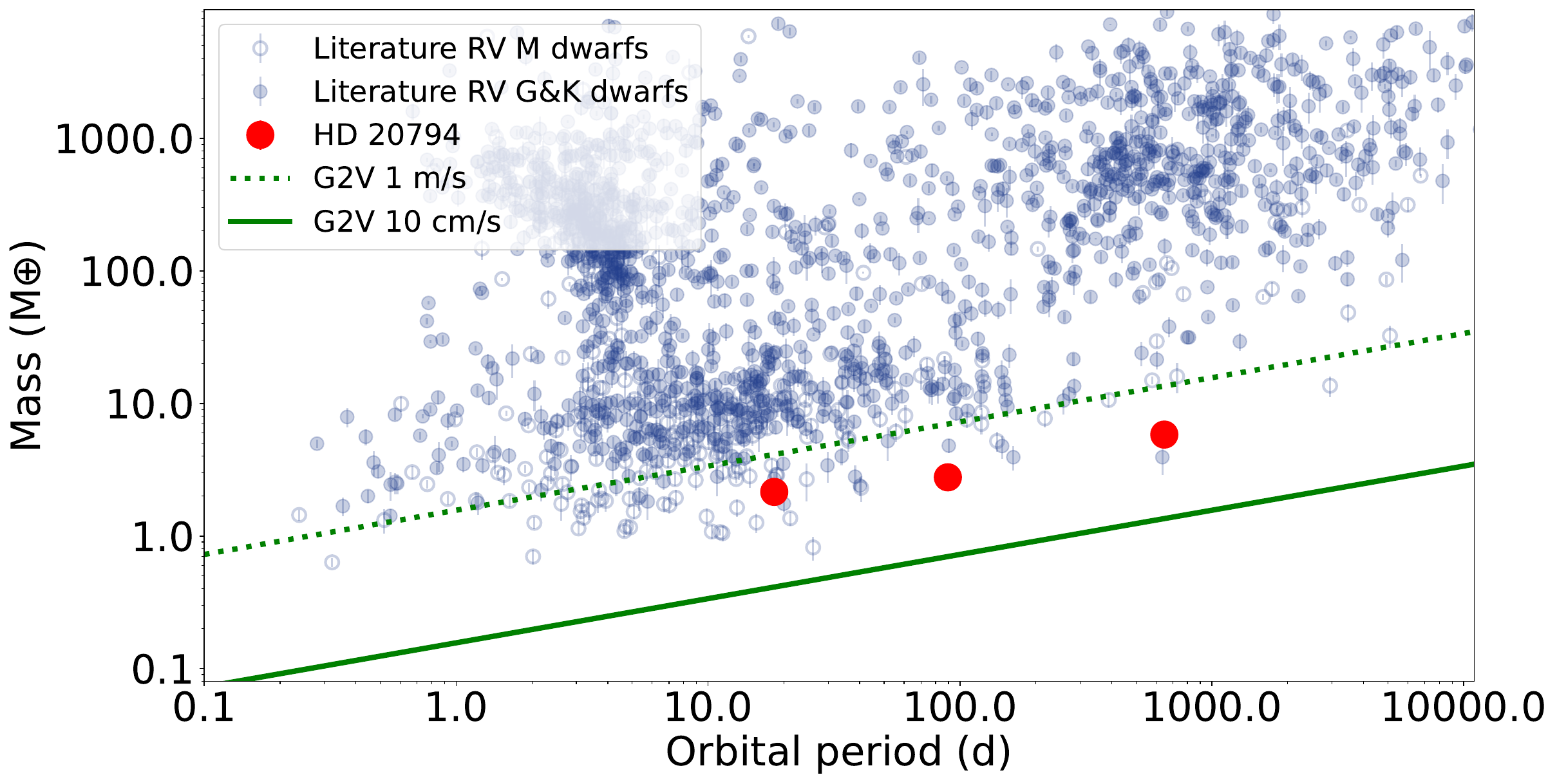}
    \caption{Mass-period relationship for planets orbiting around HD 20794 compared to the planets detected in literature with the RV method. The planets orbiting HD 20794 are among the lightest planets detected for G-type stars at their respective orbital period, well below the 1 \si{\meter\per\second} threshold. }
    \label{mass_per}
\end{figure}

Previous works from the literature assess the presence of a multi-planetary system orbiting around HD 20794.
In our analysis, we exploited a larger number of HARPS observations that were collected to investigate the 640-\si{\day} candidate reported in \citet{cretignier_yarara_2023}. Furthermore, we added the ESPRESSO dataset, and exploited a new tool for extracting velocities after the correction of activity and systematics at the spectral level, YARARA. We considered a consequential blind search for new signals to avoid any possible bias from previous works on the same subject. With this approach, we detected three significant signals, corresponding to the same planets detected in \citet{cretignier_yarara_2023}. HD 20794 b has an orbital period of 18.3140$\pm$ 0.0022 \si{\day} and its RV signal has an amplitude of 0.614 $\pm$ 0.048 \si{\meter\per\second}. These values correspond to a planet with a minimum mass of 2.15 $\pm$ 0.17 M$\oplus$, orbiting the star at a distance of 0.12570 $_{-0.00052}^{+0.00053}$ au. We can only put an upper limit on the eccentricity with e$_b$ < 0.13
HD 20794 c has an orbital period of 89.67 $\pm$ 0.10 \si{\day} and its RV signal has an amplitude of 0.502 $_{-0.049}^{+0.048}$ \si{\meter\per\second}. These values correspond to a planet with a minimum mass of 2.98 $\pm$ 0.29 M$\oplus$, orbiting the star at a distance of 0.3625 $_{-0.0016}^{+0.0015}$ au. 
Again, we can only put an upper limit on the eccentricity, with e$_c$ < 0.16.
HD 20794 d has an orbital period of 647.5 $_{-2.7}^{+2.5}$ \si{\day} and its RV signal has an amplitude of $0.567_{-0.067}^{+0.064}$ \si{\meter\per\second}. For this planet, we find an eccentricity of e$_d$ = 0.45 $_{-0.11}^{+0.10}$. These values correspond to a planet with a minimum mass of 5.82 $\pm$ 0.57 M$\oplus$, orbiting the star at a distance of 1.3541 $\pm$ 0.0068 au. The orbit of the long-period planet is compatible with an eccentric one. This result is in agreement with the result obtained in \citet{cretignier_yarara_2023}. In that work, it was derived an eccentricity for the outer planet e$_d$ = 0.40 $\pm$ 0.07. The larger dataset and the addition of ESPRESSO point toward a similar solution. 
All the planets of the system have a minimum mass compatible with a super-Earth. Still, we cannot infer the true mass of the planets without information on the inclination of the orbit with respect to the line of sight. Masses of HD 20794 b and HD 20794 c are also compatible with an Earth-like scenario.  HD 20794 d could also be a mini-Neptune with a non-negligible H/He atmosphere. The fact that we cannot access the radii of these planets does not permit us to infer their nature.
%Following \citet{kopparapu_habitable_zone} we can define the HZ for HD 20794 in different scenarios. The optimistic HZ for HD 20794 extends from 207 $\pm$ 2 \si{\day} to 773 $\pm$ 9 \si{\day}, while the conservative HZ extends from 313 $\pm$ 3 \si{\day} to 718 $\pm$ 8 \si{\day}. In both cases, HD 20794 d resides inside the HZ of the star.
The orbital period of HD 20794 d resides both in the optimistic and conservative HZ. In Fig. \ref{hz_plot} we show the position of the planets in the conservative and optimistic HZ. 
This is an interesting result because we do not have many examples of planets with M < 10 M$\oplus$ with mass measurement from RVs in the HZ of Sun-like stars.
Due to the high eccentricity of this signal we need to take into account the large variation in stellar flux received by the planet during its orbit. At the apoaster, the distance of the planet to the star is 1.96$_{-0.16}^{+0.13}$ AU, while at the periastron the planet-star distance is 0.75$_{-0.13}^{+0.15}$ AU.
This means the orbit of the HD 20794 d crosses the HZ as shown in Fig. \ref{cross_hz} for the best-fit solution.
The stellar flux at the periastron is almost seven times stronger than the stellar flux at the apoaster. HD 20794 d spends $\sim$ 59 \% of its orbit inside the optimistic HZ, and $\sim$ 38 \% of its orbit inside the conservative HZ.
\citet{2024_biasiotti_habitability} investigated the habitability of planets crossing the HZ for just a fraction of their orbit, as is the case for GJ 514 b \citep{2022_damasso_gj514}. This planet has a similar eccentricity to HD 20794 d. The level and number of tests of their analysis are beyond the scope of this work. In their work, we can see how habitability could be possible for planets on high-eccentric orbits based on stellar and planetary parameters such as stellar age, CH$_4$ abundance in the atmosphere, axis obliquity, ocean fraction, rotation period of the planet, and other properties. The possibility of maintaining habitable conditions even in a highly eccentric orbit raises the interest in investigating the HD 20794 system in  future studies.
% here let's explicit which are the parameters.
\captionsetup{justification=default, singlelinecheck=true}
 Figure \ref{mass_per} offers a plot of the mass-period relation for detected planets with masses measured from RVs. Planets orbiting HD 20794 populate the lower edge of the diagram in mass at their orbital period. This brings us to highlight the importance of long-term and high-cadence campaigns for the characterization of low-mass signals, especially in the outer regions of planetary systems. Upcoming surveys as the Terra Hunting Experiment \citep{terra_hunting_experiment}, will follow the same approach, paving the way to the characterization of habitable terrestrial planets orbiting Sun-like stars. 

In our analysis, we did not find any evidence of the presence of the planet at 147 \si{\day} reported in \citet{feng_hd20794}. We tried an informed search for the planet with priors centered around the period of the missing planet. The addition of the signal worsened the evidence of the model compared to the three-planet model. 
We can see some power excess in the GLS periodogram of the ESPRESSO dataset alone at the period of $\sim$ 40 \si{\day}. We did not recover the same signal when we consider the full dataset. We found evidence of a signal of the same period related to stellar rotation in some activity indicators (see Sect. \ref{sec_stel_act}). The possibility the 40-\si{\day} signal is related to rotation can explain the fact the signal is not persistent in the full dataset.
%The analysis of the TESS lightcurve did not reveal the presence of any transit, even if the star was observed in 4 sectors, 3, 4, 30, 31. We can see the photometric stability of the star with an RMS of 0.08 ppt in the raw photometric time series, and an RMS of 0.04 once we bin the data every 30 mins.
\subsection{Detection limits}
\label{section_detection_limits}
\begin{figure}[!h]
    \includegraphics[width=\linewidth]{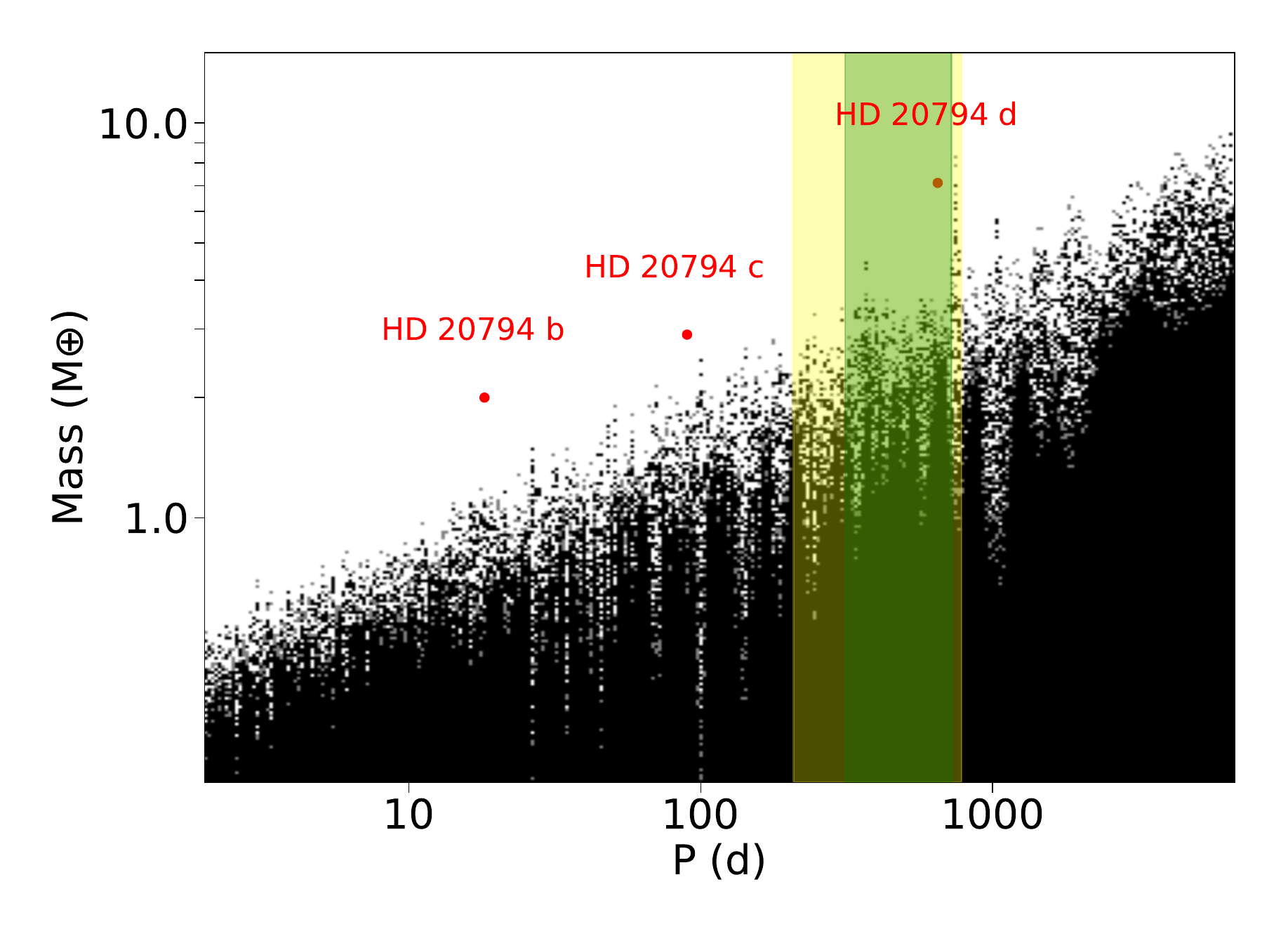}
    \caption{Sensitivity limits for HD 20794 in mass. HD 20794 b,c, and d are above our sensitivity limits. The dataset, thanks to the YARARA correction of HARPS and the extreme precision of the ESPRESSO dataset, shows a sensitivity limit in mass under 10 M$\oplus$, with the possibility to detect Earth-like planets for up to $\sim$ 20 \si{\day}.
    We have a sensitivity down to 2.5 M$\oplus$ for the inner edge of the conservative HZ and down to 4.1 M$\oplus$ for the outer edge of the HZ. We highlight in green the conservative HZ and in yellow the optimistic HZ.}
    \label{sensitivity_mass}
\end{figure}

We detected three planets orbiting HD 20794. The possibility of detecting a planet orbiting around a star depends on a list of factors: the quality of data, the quality of correction for stellar activity, and the quality of sampling. We want to investigate the regions of the parameter space where we are not sensitive to the detection of planets. To determine our sensitivity limits we do an injection-recovery test similar to the one in \citet{suarez_gj1002}. We base our knowledge of the planetary system on the result of Table \ref{tab:parameters}. We subtracted planetary signals with the same ephemerides as the ones we found for HD 20794 b, c, and d. 
We did not consider any additional correction for activity outside of the YARARA correction for HARPS. Figure \ref{sensitivity_mass}
% da cambiare con la massa
gives a plot of the results of the injection-recovery test for HD 20794, where we show the sensitivity depending on the mass and period of the hypothetical planets. To perform this test  we injected 90,000 sinusoidal signals with periods between 2 \si{\day} and 8000 \si{\day} taken from a grid of 300 different periods uniformly spaced in logarithm and a grid of 300 masses between 0.1 and 15 M$\oplus$ uniformly spaced in logarithm.
For each injection, we considered a GLS periodogram of the resulting dataset. We considered to be sensitive to injected signals when we can find a peak in the GLS periodogram with FAP < 1 \% at the injected period.
We needed to subtract an additional Keplerian signal from the dataset fitted with a norm prior on the period centered at 111 \si{\day}. This is necessary to remove this signal and its aliases from the analysis. Even if we cannot claim the planetary nature of these signals they have a FAP < 1 \%. Their inclusion would corrupt the test at their respective periodicities.
We can detect Earth or sub-Earth planets up to $\sim$ 20 \si{\day} orbital period. We have the sensitivity to detect potential candidates with a minimum mass superior to 10 M$\oplus$ for all the periods. We considered the minimum mass where we did not obtain a detection for the inner and outer conservative edges of the HZ. For the inner edge of the HZ, we are sensitive to planets with M$_{p}$sini > 2.5 M$\oplus$, while for the outer edge of the HZ, we are sensitive to M$_{p}$sini > 4.1 M$\oplus$. We can exclude the presence of giant planets in the HD 20794 system, making its architecture different from our Solar system. 
HD 20794 has a metallicity $[Fe/H]$ = -0.42 $\pm$ 0.02. The frequency of gas giants in systems known for hosting a super-Earth is derived for metal-poor stars in \citet{2024_bryan_metallicity} as P(GG|SE,$[Fe/H]$ < 0.0) = 4.5$_{-1.9}^{+2.6}$ \%. Their work defines a gas giant as a planet with a mass of > 0.5 M$_{Jup}$, while a super-Earth is a planet with a mass of between 1 and 20 M$\oplus$. \citet{2023_bonomo_frequency} search for the frequency of cold Jupiters (CJ) in the presence of short-period (SP) planets. They define as cold Jupiters planets with mass comprised between 0.3 and 13 M$_{Jup}$ and orbital distance comprised between 1 and 10 AU. They define short-period planets with masses comprised between 1 and 20 M$\oplus$ and orbital period < 100 \si{\day}. \citet{2023_bonomo_frequency}
find P(CJ|SP) = 9.3$_{-2.9}^{+7.7}$ \% for a sample of 38 stars with transiting planets from Kepler \citep{kepler_borucki_2010} and K2 \citep{2014_howell_k2}.
The absence of a cold Jupiter orbiting around HD 20794 is in line with the absence of a strong correlation between the population inner of super-Earths and the presence of an outer giant seen in statistical surveys.
%unless they are highly misaligned, in which case their mass derivation could be influenced
%The additional planets present in the system could be super-Earths, Earth-like planets, or sub-Earths, making the system different from the Solar System where there are cold giants such as Saturn and Jupiter.
We repeated the test for considering the detection limits as a function of the amplitude. We considered the same period grid as before and a grid of 300 amplitudes uniformly spaced in logarithm spanning between 0.01 and 1.5 \si{\meter\per\second}. We recovered a detection limit in the amplitude of $\sim$ 30 \si{\centi\meter\per\second} almost constant with the orbital period.

We have for HD 20794 a \textit{Gaia} DR3 renormalised unit weight error (RUWE) metric = 1.97. A value of this metric larger than one could indicate the source is not single, or problematic for the astrometric solution. Furthermore, there is evidence for astrometric acceleration due to a marginally significant \textsc{Hipparcos}-\textit{Gaia} proper motion anomaly \citep{2021_brandt_hipparcos,2022_kervella_gaia}. We are sensitive to the presence of a companion with a true mass of 50 M$\oplus$ in the range of 3-10 au. A 50 M$\oplus$ planet at 3-10 au would correspond, for such a star as HD 20794, to a signal with an amplitude of 2.9-1.6 \si{\meter\per\second} at an orbital period of $\sim$ 2000 \si{\day} and $\sim$ 13000 \si{\day,} respectively. Such a signal would be detectable as a Keplerian signal at periods shorter than our baseline and should generate a detectable linear trend at longer orbital periods. We do not see evidence of additional Keplerian signals with P > 2000. We considered a model with an acceleration term to model an eventual long-term trend present in the data and we found a result compatible with no acceleration. We conclude that an eventual long-term companion with the characteristics indicated should lie on a low-inclination orbit. 

\subsection{A candidate target for future atmospheric characterization}
The atmospheric characterization of exoplanets is fundamental to studying the composition of planets, and to infer their habitability. The atmosphere of exoplanets is studied both from the ground and space taking advantage of the footprint the atmosphere of the planet impresses to the stellar spectra once the planet transits. The most common technique used for this investigation is transmission spectroscopy. 

Future facilities such as the high-resolution spectrographs  RISTRETTO \citep{2022_lovis_ristretto} at VLT and ANDES \citep{marconi_elt_2021,2022_marconi_andes,2023_palle_andes} at ELT \citep{padovani_elt_2023} suggest an alternative path for the characterization of the atmosphere of exoplanets. The large diameter of the mirror of VLT and ELT makes it possible to spatially resolve the angular distance between stars and outer planets. Disentangling the light coming from the star from the light coming from the planet will make it possible to analyze the spectroscopy of exoplanets' atmospheres directly through high-dispersion coronagraphy (HDC).
RISTRETTO is expected to permit the characterization of the atmosphere of exoplanets at planet-star contrast of 10$^{-7}$. The angular distance limit is given by 2$\lambda$/D, where D is the diameter of the mirror. For VLT this formula transforms in an angular resolution of $\sim$ 37.5 mas for a wavelength of 760 nm \citep{2022_lovis_ristretto}. We are interested in this wavelength because it is where an oxygen band resides.  The angular distance of HD 20794 d from the star is $\sim$ 222 mas considering as reference the semi-major axis, while it spans between $\sim$ 124 mas and $\sim$ 322 mas for periastron and apoaster. The angular distance for HD 20794 c and HD 20794 b are $\sim$ 60 and 20 mas respectively considering a circular orbit. Following \citet{2020_otegi_mass_radius} we can derive a minimum radius for our planets from their minimum masses. We considered the two regimes described in this work, $\rho$ < 3.3 $g·cm^{-3}$ and $\rho$ > 3.3 $g·cm^{-3}$. In the rocky planet scenario we derived R$_b$ $\sim$ 1.3 R$\oplus$, R$_c$ $\sim$ 1.4 R$\oplus$, and R$_d$ $\sim$ 1.7 R$\oplus$. In the volatile-rich scenario, we derived R$_b$ $\sim$ 1.1 R$\oplus$, R$_c$ $\sim$ 1.4 R$\oplus$, and R$_d$ $\sim$ 2.1 R$\oplus$. To derive the planet-to-star contrast, we followed \citet{2017_lovis_ristretto}. The planet-to-star contrast ratio is $\sim$ 1.8·10$^{-8}$/1.3·10$^{-8}$ for planet b, $\sim$ 2.6·10$^{-9}$/2.2·10$^{-9}$ for planet c, and $\sim$ 2.7·10$^{-10}$/4.1·10$^{-10}$ for planet d in rocky/volatile-rich case. We are considering an Earth-like albedo. If we consider the condition for HD 20794 d at the periastron we obtain a planet-to-star contrast ratio of $\sim$ 8.9·10$^{-10}$/1.4·10$^{-9}$ for the rocky/volatile-rich case. All the targets are not feasible for RISTRETTO due to limits in contrast and/or distance from the star.
ANDES is supposed to span a range between 400 \si{\nano\meter} and 1800 \si{\nano\meter}. ELT has an angular resolution of $\sim$ 13.7 mas for a wavelength of 1300 nm \citep{suarez_gj1002}. 
This wavelength is important because it is where we can find another oxygen band.  
The maximum field of view of ANDES will be 100X100 mas. Due to the characteristics of the spectrograph HD 20794 b and HD 20794 c will be feasible, in terms of angular separation, for atmospheric characterization. HD 20794 d, even if it is interesting for its location in the HZ, resides at an angular distance from the star outside the field of view of ANDES, so the telescope should be pointed away from the guiding star. Furthermore, the low planet-to-star contrast ratio for HD 20794 d and HD 20794 c would make it necessary for long integration times to achieve a significant S/N ratio. HD 20794 b could be an interesting case of directly characterizing the atmosphere of a 2 M$\oplus$ planet orbiting close to a G-type star. 

HD 20794 is part of the Habitable World Observatory Exoplanet Exploration Program (ExEP) Precursor Science Stars \citep{2024_mamajek_hwo}, a list of amenable candidates for characterization with Habitable World Observatory (HWO). In the future, HWO will serve as a space mission with a 6-m telescope aimed at characterizing exoplanets in reflected light through direct imaging at optical/near-infrared wavelengths. The main goal of the mission will be to study the atmosphere of planets in the HZ. HWO is currently part of the Great Observatory Maturation Program (GOMAP) of NASA. The selection criteria for the target stars were based on the brightness and similarity to the Sun. HD 20794 d is a planet of particular interest for this mission because it is a light planet orbiting in the HZ of its hosting star. HWO requires targets with a separation from the star of more than 60-70 mas, while HD 20794 d is at more than 200 mas from the star, crossing the HZ. 

The Large Interferometer For Exoplanets (LIFE) is a mission aimed at characterizing habitable planets through nulling interferometry at mid-infrared wavelengths \citep{2022_life_quanz}. LIFE will investigate the thermal emission of the planets and the goal of the mission is twofold: the discovery of new planets and an in-depth atmospheric characterization of the most promising targets. The mid-infrared wavelength regime is suitable for the detection of various biosignatures, such as ozone (O$_3$), methane (CH$_4$), nitrous oxide (N$_2$O), chloromethane (CH$_3$Cl), and phosphine (PH$_3$) \citep{2018_schwieterman_biosginatures}. These biosignatures are not necessarily produced only in biological processes but combined detection of more than one of those will improve the possibility of a biological origin. Again, HD 20794 d could be a preferential candidate considering its characteristics.
\subsection{Stellar activity}
HD 20794 is known in the literature for being a quiet star. A measurement for the stellar rotational period is missing in the previous works. The RVs do not show clear signatures of the activity. We look for additional info in the activity indicators derived from spectra not corrected with YARARA. We can find a magnetic cycle in FWHM, BIS, Contrast, and S-index with a period of $\sim$ 3000 \si{\day}. Further analysis of the FWHM time series with GP points toward a rotation period of 35.0$_{-2.5}^{+3.2}$ \si{\day} for the star. When we analyzed the time series of activity indicators derived from spectra corrected by YARARA, we found in BIS a strong peak at $\sim$ 40 \si{\day}. A fit with a GP found a rotation period of 38.8$_{-2.6}^{+2.4}$ \si{\day}. The fact we can see a signal compatible with a period of $\sim$ 40 \si{\day} is of particular importance because this period is reported in \citet{pepe}
as the orbital period of a planetary signal. This work did not use a correction for the activity. Furthermore, we exploit a larger number of observations in a longer observation timespan. Activity signals are not stable over time and this can lead their signature to change in strength over time. We conclude that the possibility for the 40-\si{\day} signal to be of planetary origin is unlikely, while the possibility this periodicity is related to the rotation period is the most plausible option. We refer to Appendix \ref{stellar_activity_appendix} for more detailed activity analysis.

\subsection{Telemetry of the instrument}
%The sub \si{\meter\per\second} precision achieved by both HARPS and ESPRESSO requires a stabilized environment, to have the minimum change possible in the temperature and pressure of the instrument. A change in the instrument condition can affect the measurements at the level of precision we are working. \citet{suarez_gj1002} shows a correlation between the temperature of the echelle gratings and FWHM and bisector span. 
We investigated the variation in the temperature of echelle gratings of the spectrographs and how this impacts RVs and activity indicators \citep{suarez_gj1002}. We looked for correlations with FWHM and RVs and we found a correlation between the temperature of the Echelle gratings and the FWHM. In HARPS, we found a periodicity of the temperature variation related to the year. In ESPRESSO, we see a linear trend in the variation of temperature. We have a Pearson linear correlation of 0.67, with a p-value of 2.48·10$^{-9}$.
We see a jump in the temperature of the echelle grating of ESPRESSO in correspondence with the intervention the instrument underwent in May 2022. If we consider measurements taken before and after the intervention as separate datasets and subtract the mean value, we can see a variation related to the year with a peak-to-peak amplitude of 25 mK. For HARPS, this variation is larger, with a peak-to-peak variation of more than 300 mK. We can see the same trends in the temperature of other detector elements, such as the dichroic, the main collimator, and the field mirror. We do not see a correlation between ESPRESSO RVs, FWHM, and BIS with the pressure of the blue and red cryostat. We considered a linear detrending between indicators and temperature for all indicators in the analysis. This detrending is fundamental to revealing the signatures of magnetic cycle and rotation in our dataset.

\subsection{Intra-night scatter and possible origins of the dispersion of ESPRESSO dataset}
During the observations of HD 20794 with ESPRESSO, a sequence of short exposures was taken within the same night. This included one night with one single exposure and an additional night with two exposures. For all the remaining nights, we had between 7 and 25 exposures per night.
The sampling strategy we followed observing HD 20794 with ESPRESSO was dictated by the brightness of the target and the presence of phenomena such as granulation and oscillations, which have relatively short timescales (from minutes to several days) compared to the periodicities we are interested in the detection of exoplanets. The observing strategy, alongside the nightly binning of the observations, should permit us to mitigate the effect of p-mode oscillations. The median timespan between the first and the last exposure on the same night is 15 min. We observe a mean intranight dispersion of 49 \si{\centi\meter\per\second}, considering only the nights with more than five exposures. The maximum scatter we observe over a single night is 90 \si{\centi\meter\per\second}, while the minimum is 22 \si{\centi\meter\per\second}.
%with a standard deviation of 14 \si{\centi\meter\per\second} on the dispersion. 
 This must be compared to an error on the single measurement with a mean of 10 \si{\centi\meter\per\second}. For a star such as HD 20794, pulsations should have a timescale of the order of minutes \citep{2019_chaplin}. The most common strategy to take out the contribution of the oscillation is to integrate the stellar spectra for a time comparable to the pulsation timing of the star. In our case, this approach was impossible because a longer exposure time would have led to detector saturation. Instead of a longer exposure time, we exploit the presence of multiple exposures to mitigate the effect of pulsation \citep{2008_otoole_oscillation,dumusque_2011_oscillation}. The mitigation of granulation effects on RVs can be obtained by observing a star multiple times per night with a span of some hours between observations. We do not have multiple observations per night for HD 20794 and we cannot even follow the granulation pattern with the multiple exposures we have on a single night, because of the short baseline (15 min) of our exposures. The granulation effect on RVs is difficult to predict,
% maybe reference
but it can induce a scatter similar to the scatter we observe in the ESPRESSO dataset once we remove the planetary signals (RMS of 72 \si{\centi\meter\per\second} in the residuals). Most likely this scatter is of stellar origin, but in the time series, there is no evidence of a significant impact from the magnetic cycle of the star and rotation. Considering we are mitigating the effect of pulsations by binning, granulation remains the main effect we cannot model. Only a deep analysis of the behavior of granulation on RVs, through dedicated campaigns of consecutive exposures over a full night or longer, can help us understand the unexplained scatter we continue to see in the data.

\section{Conclusion}
\label{sec_con}
The re-analysis of the system HD 20794, a bright G6V star (V=4.34 mag) brought on the confirmation of the presence of the two planets with orbital periods of 18-\si{\day} and 89-\si{\day}, combining the RV measurements taken with HARPS and ESPRESSO, and thanks to the correction of the HARPS spectra with the YARARA-pipeline for extraction of velocities, which has permitted us to clean HARPS time-series from different systematic errors. These two planets were present in all the previous papers from the literature on the system and are found with more than 9-$\sigma$ significance. 

We confirmed the presence of the 650-\si{\day} planet which was reported first in \citet{cretignier_yarara_2023}. This planet is of particular interest because it spends most of its orbit inside the HZ of the star. The closeness of the planetary system, summed with the distance of the star and the planet and the planet-to-star contrast ratio make this planet a good candidate for future atmospheric characterization through direct imaging facilities. We call the three planets HD 20794 b, HD 20794 c, and HD 20794 d for planets at 18-\si{\day}, 89-\si{\day}, and 650-\si{\day,} respectively.

HD20794 b has a period of 18.3140$_{-0.0022}^{+0.0022}$ 0.003 \si{\day} and a minimum mass of M$_{p}$sini of 2.15 $\pm$ 0.17 M$_{\oplus}$ with an eccentricity e$_b$ < 0.13. HD20794 c has an orbital period of 89.68 $\pm$ 0.10 \si{\day}, a minimum mass of 2.98 $\pm$ 0.29 M$_{\oplus}$, and an eccentricity e$_d$ < 0.16. HD 20794 d has an orbital period of 647.6$_{-2.7}^{+2.5}$ \si{\day}, a minimum mass of 5.82 $\pm$ 0.57 M$_{\oplus}$, and an eccentricity e$_d$ of 0.45$_{-0.11}^{+0.10}$. The amplitude related to the signals is by far under 1 \si{\meter\per\second}, 0.614 $\pm$ 0.048 \si{\meter\per\second}, 0.502 $_{-0.049}^{+0.048}$ \si{\meter\per\second}, and 0.567$_{-0.064}^{+0.067}$ \si{\meter\per\second} for planets b, c, and d respectively. 

We did not find any evidence of signals related to the stellar rotation in RVs. For the activity indicators, we ran a parallel analysis with YARARA and not YARARA-corrected spectra. In the case of raw indicators, we found evidence of the presence of a magnetic cycle. We can see the cycle in the FWHM, S-index, BIS, and contrast. The timescale of the cycle is compatible with FWHM and S-index, pointing toward a cycle of $\sim$ 3000 \si{\day}. We can see some hints of the stellar rotation period in the analysis of FWHM with GP, where the rotation period results amount to $\sim$ 40 \si{\day}. In the YARARA corrected dataset, we find an excess of power at $\sim$ 40 \si{\day} in BIS that might be related to rotation, as we have seen from our analysis. 

The low level of activity we find confirms the stability of HD 20794 and its suitability as a standard target for observations carried out by future facilities. The brightness of the target permits us to achieve a photon noise level for ESPRESSO of 10 \si{\centi\meter\per\second}. 
%even if the important intranight scatter % you could try to calculate and put some info about this fact
%bring us to think that a better correction for granulation and oscillation is necessary to achieve the sensitivity necessary for the detection of extremely low amplitude signals, in order to reach the barrier of 10 \si{\centi\meter\per\second}, necessary for the detection of an Earth-like planet around a G-type star.

\begin{acknowledgements}
      NN acknowledges funding from Light Bridges for the Doctoral Thesis "Habitable Earth-like planets with ESPRESSO and NIRPS", in cooperation with the Instituto de Astrofísica de Canarias, and the use of Indefeasible Computer Rights (ICR) being commissioned at the ASTRO POC project in the Island of Tenerife, Canary Islands (Spain). The ICR-ASTRONOMY used for his research was provided by Light Bridges in cooperation with Hewlett Packard Enterprise (HPE). 
      JIGH, AKS, RR, CAP, NN, VMP, and ASM acknowledge financial support from the Spanish Ministry of Science and Innovation (MICINN) project PID2020-117493GB-I00. The project that gave rise to these results received the support of a fellowship from the ”la Caixa” Foundation (ID 100010434). The fellowship code is LCF/BQ/DI23/11990071.
      Co-funded by the European Union (ERC, FIERCE, 101052347). Views and opinions expressed are however those of the author(s) only and do not necessarily reflect those of the European Union or the European Research Council. Neither the European Union nor the granting authority can be held responsible for them. This work was supported by FCT - Fundação para a Ciência e a Tecnologia through national funds and by FEDER through COMPETE2020 - Programa Operacional Competitividade e Internacionalização by these grants: UIDB/04434/2020 (DOI: 10.54499/UIDB/04434/2020); UIDP/04434/2020 (DOI: 10.54499/UIDP/04434/2020)". A.M.S acknowledges support from the Fundação para a Ciência e a Tecnologia (FCT) through the Fellowship 2020.05387.BD (DOI: 10.54499/2020.05387.BD).
      This work has made use of data from the European Space Agency (ESA) mission {\it Gaia} (\url{https://www.cosmos.esa.int/gaia}), processed by the {\it Gaia} Data Processing and Analysis Consortium (DPAC,\url{https://www.cosmos.esa.int/web/gaia/dpac/consortium}). Funding for the DPAC has been provided by national institutions, in particular the institutions participating in the {\it Gaia} Multilateral Agreement.
      This work was financed by Portuguese funds through FCT (Funda\c c\~ao para a Ci\^encia e a Tecnologia) in the framework of the project 2022.04048.PTDC (Phi in the Sky, DOI 10.54499/2022.04048.PTDC). CJM also acknowledges FCT and POCH/FSE (EC) support through Investigador FCT Contract 2021.01214.CEECIND/CP1658/CT0001 (DOI 10.54499/2021.01214.CEECIND/CP1658/CT0001).
      This publication makes use of The Data \& Analysis Center for Exoplanets (DACE), which is a facility based at the University of Geneva (CH) dedicated to extrasolar planets data visualization, exchange, and analysis. DACE is a platform of the Swiss National Centre of Competence in Research (NCCR) PlanetS, federating the Swiss expertise in Exoplanet research. The DACE platform is available at \url{https://dace.unige.ch}.
      XD acknowledges the support from the European Research Council (ERC) under the European Union’s Horizon 2020 research and innovation programme (grant agreement SCORE No 851555) and from the Swiss National Science Foundation under the grant SPECTRE (No $200021\_215200$). This work has been carried out within the framework of the NCCR PlanetS supported by the Swiss National Science Foundation under grants $51NF40\_182901$ and $51NF40\_205606$.
     % \textbf{X.D acknowledges the support from the European Research Council (ERC) under the European Union’s Horizon 2020 research and innovation programme (grant agreement SCORE No 851555) and from the Swiss National Science Foundation under the grant SPECTRE (No 200021_215200). This work has been carried out within the framework of the NCCR PlanetS supported by the Swiss National Science Foundation under grants 51NF40_182901 and 51NF40_20560.}
      FPE and CLO would like to acknowledge the Swiss National Science Foundation (SNSF) for supporting research with ESPRESSO and HARPS through the SNSF grants nr. 140649, 152721, 166227, 184618 and 215190. The ESPRESSO and HARPS Instrument Projects were partially funded through SNSF's FLARE, resp. FINES Programmes for large infrastructures.
      The INAF authors acknowledge financial support of the Italian Ministry of Education, University, and Research
      with PRIN 201278X4FL and the "Progetti Premiali" funding scheme.
      This research has made
      extensive use of the SIMBAD database operated at CDS, Strasbourg, France,
      and NASA’s Astrophysics Data System. This research has made use of NASA
      Exoplanet Archive, which is operated by the California Institute of Technology, under contract with the National Aeronautics and Space Administration
      under the Exoplanet Exploration Program. The manuscript was written using \texttt{Overleaf}. Extensive use of \texttt{numpy} \citep{2011_numpy} and \texttt{scipy} \citep{2020_virtanen_scipy}.
      The main analysis was performed in Python 3 \citep{python3_ref} running on a Ubuntu system \citep{ubuntu_2015}.
\end{acknowledgements}

% WARNING
%-------------------------------------------------------------------
% Please note that we have included the references to the file aa.dem in
% order to compile it, but we ask you to:
%
% - use BibTeX with the regular commands:
%   \bibliographystyle{aa} % style aa.bst
%   \bibliography{Yourfile} % your references Yourfile.bib
%
% - join the .bib files when you upload your source files
%-------------------------------------------------------------------

\bibliographystyle{aa} % 
\bibliography{bibliography}
%\bibliography{bibliography,activity,additional}

\begin{comment}

\end{comment}

\begin{appendix}
\section{Methods}
\label{methods_sec}
For the parameter estimation, we used the nested-sampling tool Dynesty \citep{dynesty_2020}. Dynesty gives an estimate of the natural logarithm of the evidence associated with a model, allowing for an easy model comparison. To reckon the different zero-points of different instruments, we always consider an offset term for each instrument in the analysis. A jitter term is added in quadrature to the nominal error of the different instruments for each time series. The jitter term takes into account all the sources of noise we are not modeling for, and the instrumental noise. To remove outliers from different datasets, we bin   the observations nightly and we apply a cut on the dataset consisting of a 3 $\sigma$ clipping joined with the exclusion of measurements where the error is larger than three times the median error for each dataset. To implement Gaussian processes (Sect. \ref{sec_stel_act}) in our analysis we used S+LEAF \citep{2022_delisle_spleaf}. S+LEAF allows for only a certain kind of semi-separable matrix to be taken into account as a covariance matrix. In this way, the computational cost scales linearly with the dimension of the dataset, instead of with its cube, as it used to be in standard implementations. Before a fit for the jitter was available, exploratory periodograms of our datasets were generated adding in quadrature a white noise term to the error on the measurements equal to the standard deviation of the dataset.
%\clearpage
\section{Planetary system} 

We show in Table \ref{table_literature} the previous results obtained in the literature, compared with the results obtained in our work.

\captionsetup{justification=raggedright, singlelinecheck=false}

\begin{table*}[h!]
  \caption{Planetary parameters from the literature works on HD 20794 and this work.}
  \label{table_literature}
  \begin{tabular}{lllll}
    \hline
    \hline
    \noalign{\smallskip}
    Reference & K [ms$^{-1}$] & P [d] & m$_{p}$·sini [M$_{\oplus}$] & e\\
    \noalign{\smallskip}
    \hline
    \noalign{\smallskip}
    \citet{pepe} &  &  &  &  \\
    18-\si{\day} &  0.83 $\pm$ 0.09 & 18.315 $\pm$ 0.008 & 2.7 $\pm$ 0.3 & 0 \\
    40-\si{\day} & 0.56 $\pm$ 0.10 & 40.114 $\pm$ 0.053 & 2.4 $\pm$ 0.4 & 0 \\
    89-\si{\day} & 0.85 $\pm$ 0.10 & 90.309 $\pm$ 0.184 & 4.8 $\pm$ 0.6 & 0 \\
    \citet{feng_hd20794} &  & &  &  \\
    18-\si{\day} &  0.81$_{-0.24}^{+0.00}$ & 18.33$_{-0.02}^{+0.01}$ & 2.82 $_{-0.80}^{+0.10}$ & 0.27 $_{-0.22}^{+0.04}$ \\
    89-\si{\day} &  0.60$_{-0.18}^{+0.00}$ & 88.90$_{-0.41}^{+0.37}$ & 3.52 $_{-0.91}^{+0.48}$ & 0.25 $_{-0.20}^{+0.16}$ \\
    147-\si{\day} &  0.69$_{-0.13}^{+0.15}$ & 147.02$_{-0.91}^{+1.43}$ & 4.77 $_{-0.86}^{+0.96}$ & 0.29 $_{-0.18}^{+0.14}$ \\
    \citet{cretignier_yarara_2023} &  & &  &  \\
    18-\si{\day} &  0.56$_{-0.05}^{+0.05}$ & 18.32$_{-0.01}^{+0.01}$ & 2.0 $_{-0.2}^{+0.2}$ & 0.09 $_{-0.06}^{+0.08}$ \\
    89-\si{\day} &  0.78$_{-0.05}^{+0.05}$ & 89.56$_{-0.10}^{+0.09}$ & 4.7 $_{-0.4}^{+0.4}$ & 0.13 $_{-0.07}^{+0.07}$ \\
    640-\si{\day} &  0.61$_{-0.06}^{+0.06}$ & 644.6$_{-7.7}^{+9.9}$ & 6.6 $_{-0.7}^{+0.6}$ & 0.40 $_{-0.07}^{+0.07}$ \\
    This work &  & &  &  \\
    18-\si{\day} &  0.614 $\pm$ 0.048 & 18.3142 $\pm$ 0.0022 & 2.15 $\pm$ 0.17 &  0.065$_{-0.046}^{+0.065}$\\
    89-\si{\day} &  0.502$_{-0.049}^{+0.048}$ & 89.68 $\pm$ 0.10 & 2.98 $\pm$ 0.29 & 0.077$_{-0.055}^{+0.084}$ \\
    650-\si{\day} &  0.567$_{+0.067}^{-0.064}$ & 647.6$_{-2.7}^{+2.5}$ & 5.82 $\pm$ 0.57 & 0.45$_{-0.11}^{+0.10}$ \\
    
    \noalign{\smallskip}
    \hline
  \end{tabular}
  \medskip
  \begin{flushleft}
  Notes: We can see planets at 18 \si{\day} and 89 \si{\day} are recovered in all literature works on the system 
  
  and by the present work. Planet at 40 \si{\day} is recovered only in \citet{pepe}, the planet 
  
  at 147 \si{\day} 
  is recovered only in \citet{feng_hd20794}, and the planet at 640 \si{\day} is recovered only
  in 
  
  \citet{cretignier_yarara_2023} and this work.
  \end{flushleft}

\end{table*}
%clearpage
\section{L1 periodogram}

\begin{figure}[!h]
    \includegraphics[width=\linewidth]{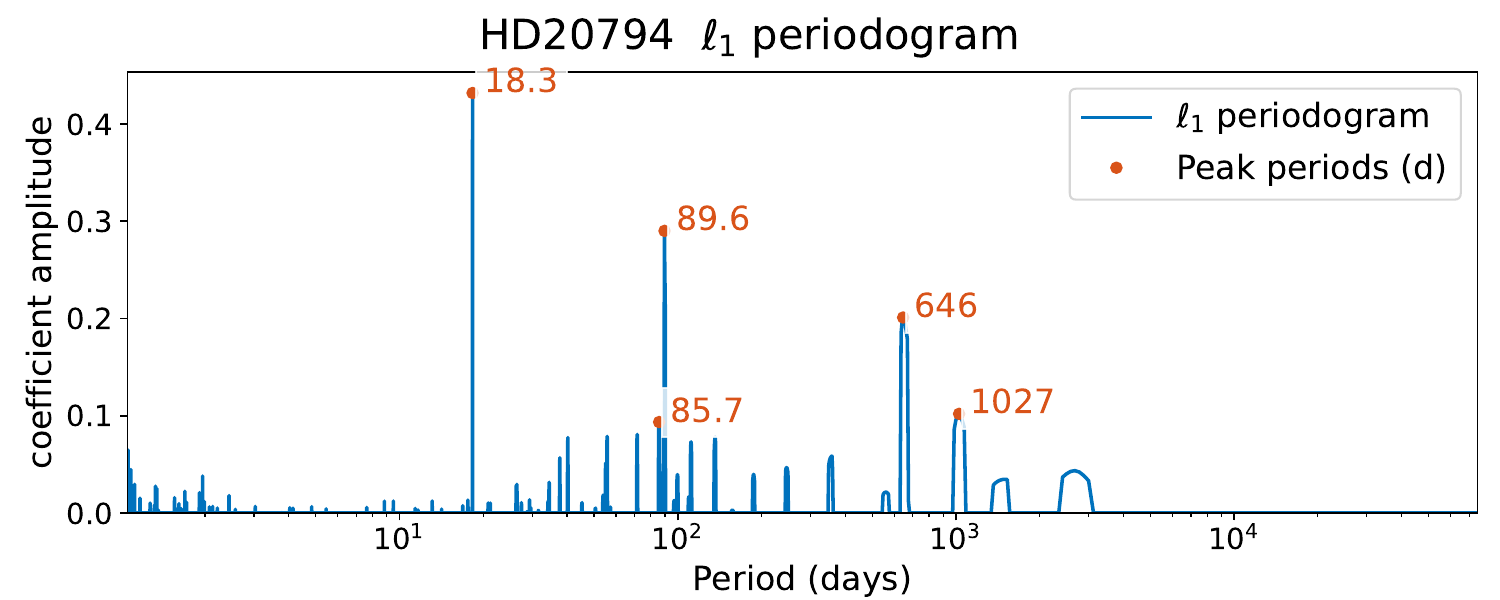}
    \caption{L1 periodogram of the full dataset of RVs for HD20794. We see the most significant peaks at 18.3 d, 89.6 d, and 646 d. These peaks correspond to the planets b, c, and d we are seeing in our analysis. We see other peaks at $\sim$ 85 \si{\day} and 1027 \si{\day}. These peaks appear in our analysis but we cannot claim those as new detections.}
    \label{l1_periodogram}
\end{figure}

A different way to assess the presence of significant signals in the dataset, instead of a GLS periodogram, is the L1 periodogram \citep{2020_hara_l1}. GLS periodogram assesses the significance of each frequency of a grid of frequencies alone. L1 periodogram attempts to fit simultaneously for all the periodicities present in the dataset. We show in Fig. \ref{l1_periodogram} the result of the L1 periodogram applied to HD 20794. We are calculated the L1 periodogram with the LARS algorithm \citep{2004_efron_lars}. We can see the most prominent peaks at 18.3 d, 89.5 d, and 646 \si{\day}. These peaks correspond to the peaks we find in our blind-search analysis. Less significant peaks are found at $\sim$ 85 \si{\day} and 1026 \si{\day}.

\section{111-day signal}
\label{111_day}
The periodogram of the residuals of the 3-Keplerian model shows a double peak in the residuals at 85.4 and 111.5 \si{\day}. The blind search we conducted and the FIP analysis did not permit us to claim this new signal as a safe planet detection. Here, we discuss a dedicated analysis to investigate the characteristics of this signal. We tried a model with normal priors on the periods for the claimed planets and the 111-\si{\day} signal. We considered both a circular and a Keplerian model. In the circular model, we obtained a solution for the 111-\si{\day} signal with K = 0.276 $_{-0.050}^{+0.049}$ \si{\meter\per\second}, and P = 111.51 $_{-0.18}^{+0.19}$ \si{\day}. This corresponds to a planet of mass M$_p$sini = 1.77 $\pm$ 0.31 M$\oplus$ orbiting at a distance of 0.4192 $_{-0.18}^{+0.19}$. In the Keplerian model, we obtained a solution with a significant eccentricity e = 0.40$_{-0.12}^{+0.13}$, K = 0.38 $\pm$ 0.07 \si{\meter\per\second}, and P = 111.52 $_{-0.16}^{+0.15}$ \si{\day}. This corresponds to a planet mass of M$_p$sini = 2.23 $_{-0.34}^{+0.35}$ M$\oplus$ and an orbital semi-major axis of 0.4192 $\pm$ 0.0018. The presence of such an eccentric signal at a period close to the period of planet c would imply a crossing orbit for the two planets, making the system likely unstable. Furthermore, the amplitude of the hypothetical Keplerian signal should be detectable in our dataset considering our detection limit. The high level of eccentricity could be related to the tendency of the fitting algorithm to include all the remaining structures in the data inside the 111-\si{\day} signal. We lack the sensitivity necessary for a proper characterization of this structure.
Additional observations are needed to solve the planetary system of HD 20794 in the lower mass regimes. 
\section{Stellar activity}
\label{stellar_activity_appendix}
The advancement in the development of ultra-stable spectrographs opened the possibility of achieving precision on the single measurement of RV well below 1 \si{\meter\per\second}.
At this level of precision, the change in the stellar RV due to stellar physics-related phenomena becomes detectable so we need to consider it in our models because it can hide or mimic planetary signals in RV time series \citep{1997_saar_stellar_activity,2021_meunier_stellar_activity}. Stars are affected by a plethora of non-equilibrium processes. We can find evolving oscillation patterns, granulation, rotation, flares, plages, and long-term magnetic cycles. These phenomena can affect the RV in timescales varying from minutes to years and with amplitudes ranging from \si{\centi\meter\per\second} to tenths of \si{\meter\per\second}. In this section, we first outline the analysis of the stellar activity.

For oscillation and granulation, some strategies have been developed to mitigate their effect \citep{dumusque_2011_oscillation,2019_chaplin}, including: (i) appropriate exposure times to average out the p-mode oscillations,(ii) multiple observations per night to mitigate granulation signal. 

The stellar rotation period can differ significantly, depending on the stellar type and the stellar age \citep{2015_rotation_alejandro}. For solar-type stars, we do not have examples in the literature of rotation periods longer than 50 \si{\day}. 
The most common tools in the field of exoplanets to deal with rotation related-stellar activity are Gaussian Processes (GP) \citep{rasmussen_gp_2006,haywood_gp_2014,aigrain_gp_2023}.
GPs offer a framework to represent signals that present an excess of power at certain frequencies, without being strictly periodic.
Because of their flexibility, GP can adapt very well to the peculiar features of activity signals. Activity signals are not strictly periodic and are usually unstable over multiples of their characteristic timescale. GPs are flexible enough to capture this variability over time.
We directly sample for the parameters of the covariance matrix between samples of the process under analysis, which correspond to our dataset. In this way, we can model the correlated noise in a Bayesian and mathematically defined framework.  The largest drawback of GP is their tendency to overfit, which can be mitigated by deciding on appropriate kernels. Kernels describe the shape of the covariance matrix between observations. Different kernels have been proposed and used in the analysis of exoplanets. 
 We tried different kernels in our analysis, which we briefly discuss here. The stochastic harmonic oscillator (SHO) is a kernel described in \citet{2017_celerite_gp}. It requires three parameters: amplitude, rotation period, and a quality factor. The Matérn 3/2 exponential period (MEP) kernel and the exponential-sine periodic (ESP) kernel are approximations of the quasi-periodic kernel (Eq. \ref{quasi_periodic_equation}) as defined in \citet{aigrain_gp_2023}:

\begin{equation}
k(\Delta t) = A \exp \left( -\frac{(\Delta t)^2}{2\rho^2} - \frac{2 \sin^2 \left( \frac{\pi \Delta t}{P} \right)}{\eta^2} \right),
\label{quasi_periodic_equation}
\end{equation}

\noindent where we can relate the parameters to physical quantities: the amplitude, $A$, the rotation period, $P$, the timescale of evolution, $\rho$, and the harmonic complexity, $\eta$. 
 
Stars can also show in the RV time series signatures of magnetic cycles, which usually have periods of thousands of days for G-type stars \citep{lovis_2011_activity}. The magnetic long-term cycle can also be seen in many activity indicators \citep{lovis_2011_activity,2011_gomes_da_silva,2016_cycle_alejandro}. 
% scrivi cosa sono gli indicatori in questa sezione
Usually, this kind of noise is modeled with a sinusoid and a few harmonics at rational fractions of the cycle period. Over the years many diagnostic methods have been developed as proxies for activity, and nowadays the stellar activity analysis is a central part of the analysis performed in almost every work on extrasolar planetary systems. Activity indicators are sensitive to the change in the spectral profile stellar activity can give rise to (FWHM, bisector, contrast), or to the behavior of spectral lines, which are good proxies for chromospheric activity (S$_{MW}$, H-$\alpha$, and Na doublet).

 The FWHM is one of the most reliable indicators of stellar activity. The WHM measures the width of the CCF profile. A temporal variation of the broadening of the spectral lines is due to the change in the activity pattern of the star \citep{2007_desort_stellar_activity,2014_dumusque_soap}. This activity indicator has been used in many analyses as a proxy for the rotational period of the star, or alongside RV to better constrain the RV model \citep{suarez_proxima_2020,faria_proxima_2022,suarez_gj1002}.

 The bisector span metric (BIS) measures the asymmetry in spectral lines. The variation in BIS is related to a change in the stellar profile due to the stellar activity happening on the star. The presence of spots on the stellar surface can create some unbalances in the flux from the star with a periodicity related to the rotational period of this. This results in a modification of the shape of the spectral lines \citep{queloz_bis_2001,1988_toner_gray}. The pipelines of HARPS and ESPRESSO automatically compute this indicator. 

The contrast of the CCF measures the ratio between the core of the CCF and the continuum. This indicator is sensitive to the magnetic level of the star, which can change due to phenomena as rotation and long-term cycles.

The H-$\alpha$ index is an indicator of the chromospheric activity of a star. It measures the strength of the core of the line of the H-$\alpha$ emission, or absorption for weakly active stars.  This line is mostly effective for M-dwarfs, even if it can also be used for G-type stars. On G-type stars the indicator is usually sensitive to the contamination of filaments and telluric lines. 
The calculation of this indicator is based on the comparison between the flux in the H$\alpha$ line and the continuum, as explained in \citet{gomes_2011_halpha}. The exact method to extract this indicator can slightly change from work to work. In our analysis, we average the flux in a window with a width of 1.6 Å centered at the center of the line, at 6562.808 Å ($H\alpha_{core}$) and divide it by the sum of the averaged flux in two bands, centered at 6550.87 Å (H$\alpha_{V}$) and 6580.31 Å (H$\alpha_{R}$), with a width of 10.75 Å and 8.75 Å respectively. The equation
for the extraction of the indicator is
\begin{equation}
    \text{H}\alpha = \frac{H\alpha_{core}}{H\alpha_{R} + H\alpha_{V}}.
\end{equation}

% guarda come estrarre H alfa con serval

The core of the emission of the H \& K lines of calcium is an important indicator of chromospheric activity. We used the same approach of \citet{lovis_2011_activity} to derive the Mount Wilson S-index (S$_{MW}$). The calculation is done by defining two triangular passbands centered in the core of the emission lines, respectively at 3968.469 Å for the H line and at 3933.663 Å for the K line, with an FWHM of 1.09 Å. We normalized this for the continuum using two rectangular bandpasses of 20 Å centered at V = 3901.070 Å and R = 4001.070 Å respectively. The equation we used to derive the indicator is 
\begin{equation}
    S_{MW} = \alpha\frac{\Tilde{N_{H}} + \Tilde{N_{K}}}{\Tilde{N_{R}} + \Tilde{N_{V}}} + \beta,
\end{equation}where $\Tilde{N_{H}}$, $\Tilde{N_{K}}$, $\Tilde{N_{R}}$, $\Tilde{N_{V}}$ are the mean fluxes in each band and $\alpha$ and $\beta$ are constants for the calibration, namely, $\alpha$ = 1.111 and $\beta$ = 0.153. For a description of the effect of different phenomena such as spots or plages on the profile of Ca H \& K lines, see \citet{cretignier_hk_2024}. 

Na I D resonance lines (D1: 5895.92 Å; D2: 5889.95 Å) are visible in all kinds of stars as an important absorption feature of the spectra. In cool stars as late G, K, and M dwarfs the Sodium absorption lines are usually present as strong absorption wings and have been demonstrated as a useful tool for studying stellar atmospheres \citep{1997_andretta_na,2007_diaz_na}. A method to compute this activity indicator is described in \citep{2007_diaz_na}. 

\begin{figure*}[!h]
    \includegraphics[width=\textwidth]{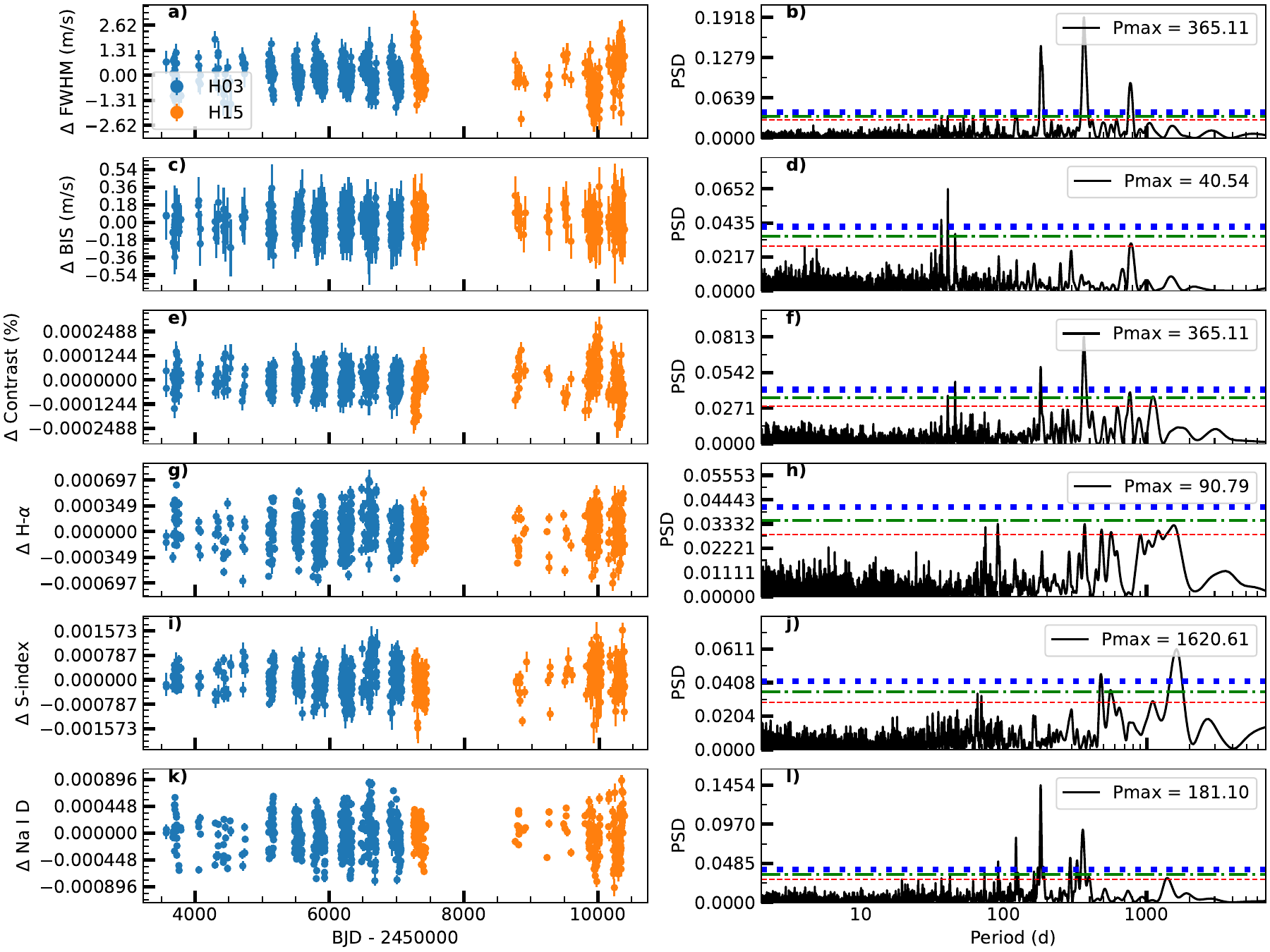}
    \caption{Activity indicators time series derived after the YARARA correction. Panel (a): FWHM time series. Panel (b): GLS periodogram for the FWHM, a signature of the year is present. Panel (c): BIS time series. Panel (d): GLS periodogram for BIS, peak at 40.53 \si{\day}. Panel (e): Time series of Contrast. Panel (f): GLS periodogram of the Contrast, 1-year signature is visible. Panel (g): H$\alpha$ time series. Panel (h): GLS periodogram of H$\alpha$, a peak corresponding to the revolution of the Earth is visible. Panel (i): S-index time series. Panel (j): GLS periodogram of S-index.
    Panel (k): Na I D time series. Panel (l): GLS periodogram of Na I D. 
    The red/green/blue horizontal dashed lines indicate FIP levels of 10/1/0.1 \% respectively.}
    \label{stellar_activity_full_yarara}
\end{figure*}

In Fig. \ref{stellar_activity_full} we show the time series of the activity indicators together with their GLS periodograms \citep{zechmeister_gls_2009}. 
For observations taken before BJD 2453500, we see a higher dispersion in the H03 dataset than for observations made after that date. For this reason, we discard observations made previous to this date. We can see a linear trend present in FWHM and contrast in the H03 dataset. This effect is due to a change in the focus of the instrument
through the years and can be modeled with a linear trend for the H03 dataset alone. 
We desire to model stellar activity on itself. For this reason, we first analyze the activity indicators derived without taking into account the YARARA correction. YARARA corrects spectra also for stellar activity, hiding its signature in the indicators. The trade-off is the presence in our time series of the systematics YARARA corrects for. The analysis of activity indicators corrected for YARARA will be reported in Sec. \ref{yarara_activity_indicators} and YARARA activity indicators and their GLS periodograms are shown in Fig. \ref{stellar_activity_full_yarara}.

\subsection{Temperature of the echelle gratings}

\begin{figure*}[!h]
    \centering
    \includegraphics[width=0.90\textwidth]{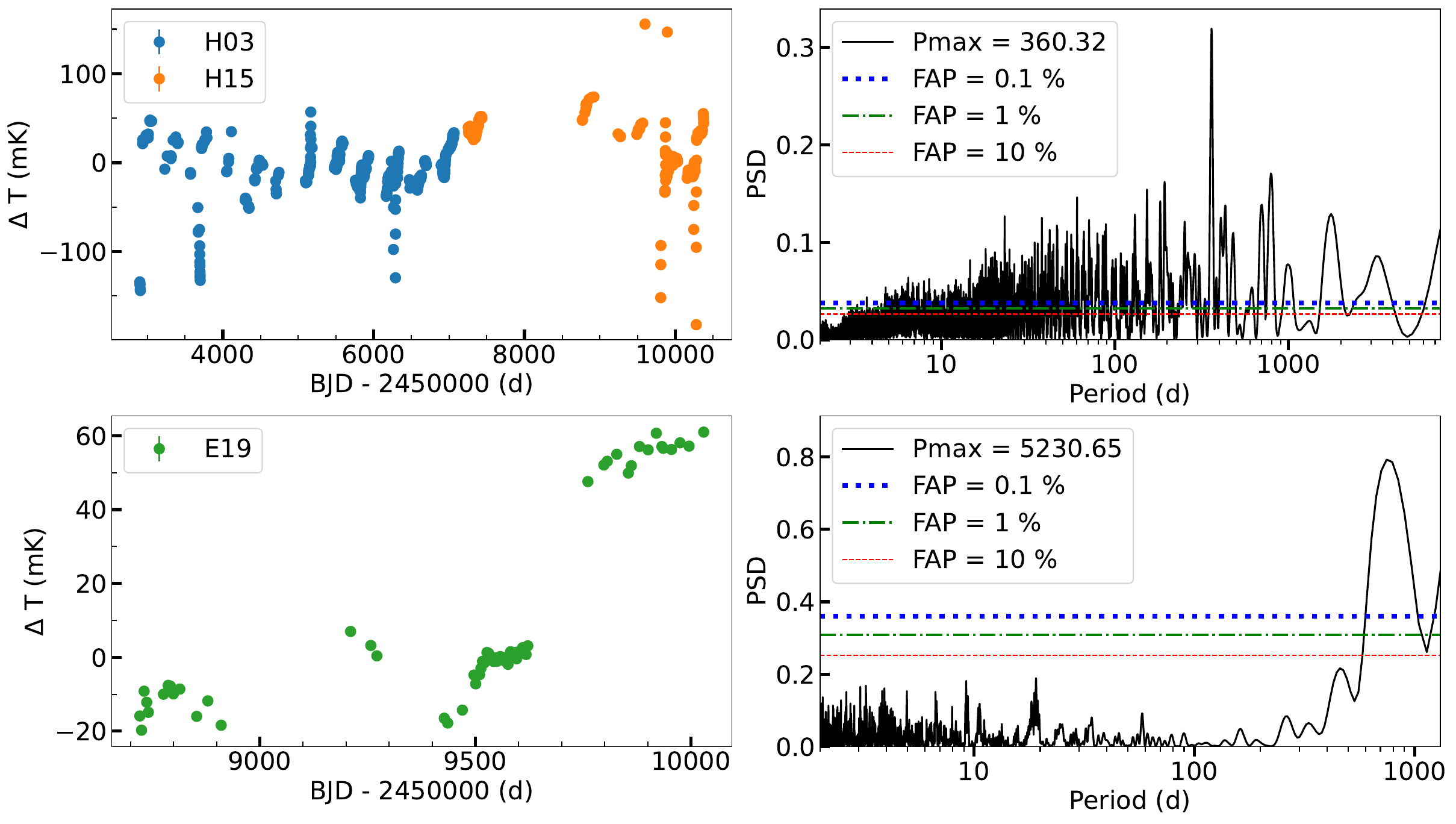}
    \caption{Variation in the temperature of echelle gratings for HARPS (upper left panel) and ESPRESSO (lower left panel) with their respective GLS periodograms. We can see a strong peak in the GLS periodogram of the HARPS dataset close to 1 year, related to the seasonal variation of the instrument.}
    \label{echelle_fig}
\end{figure*}

\begin{figure}[!h]
    \centering
    \includegraphics[width=0.45\textwidth]{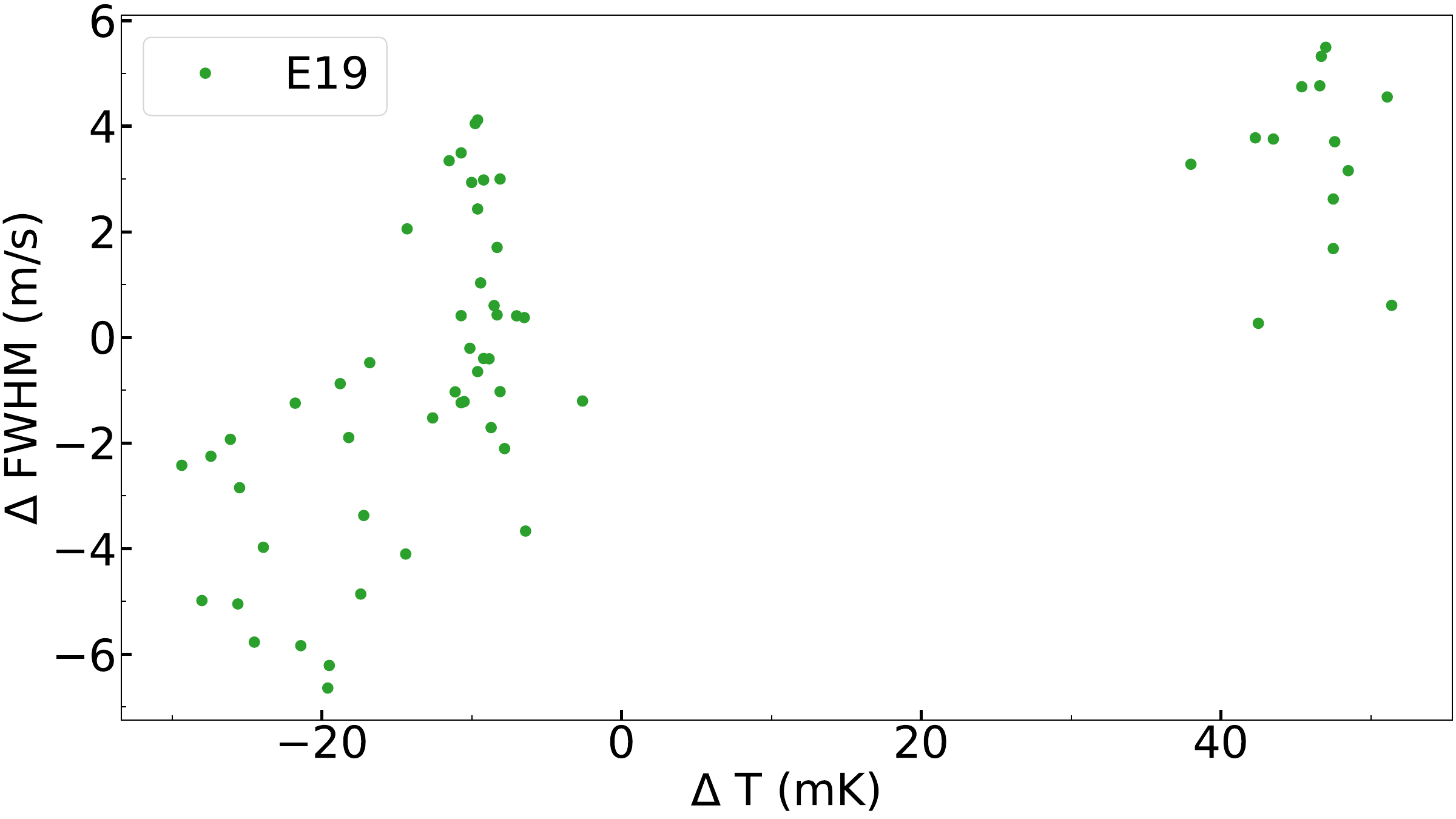}
    \caption{Correlation between variation of the temperature of the echelle grating and the FWHM.}
    \label{echelle_fwhm}
\end{figure}

Modern spectrographs need a stabilized environment in pressure and temperature to achieve the necessary precision in RV measurements. Even if the environmental conditions are strictly controlled and monitored, some noise is still present. This variability in principle can affect RVs and activity indicators time series \citep{suarez_gj1002}. We controlled the variations of pressure and temperatures for different components of HARPS and ESPRESSO. We checked for correlations with activity indicators and RVs. We found a strong linear correlation (Pearson correlation = 0.67, p-value = 2.48·10$^{-9}$) between FWHM and the temperature of echelle gratings for ESPRESSO. We saw a similar correlation for BIS. Also, the HARPS dataset shows similar trends for BIS and FWHM compared to the temperature of the echelle gratings. In the following analysis, we corrected the correlation with temperature for all the activity indicators with linear trends. Figure \ref{echelle_fig} shows the time series of echelle gratings temperature for HARPS and ESPRESSO together with their GLS periodogram. Figure \ref{echelle_fwhm} shows the correlation between FWHM and variation in temperature of echelle gratings. 

\subsection{FWHM}
We analyzed the FWHM as it is extracted by the ESPRESSO pipeline \citep{pepe_espresso_2021}. %instead of the indicator as extracted after the YARARA correction.  this comes from a comment from Nuno, where he asks why we do not use the indicator, but it is written before 

\begin{figure}[!h]
    \includegraphics[width=\linewidth]{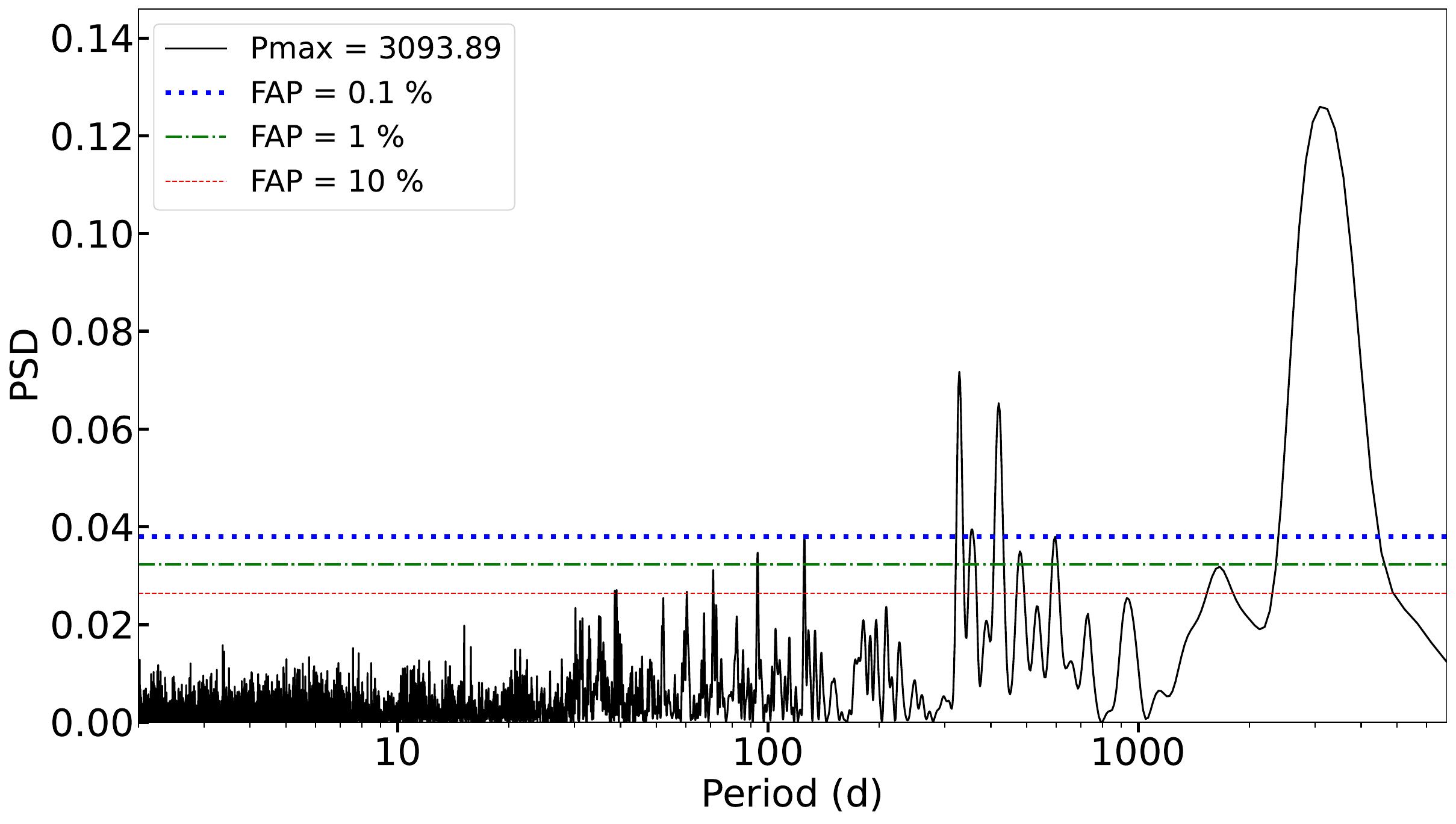}
    \caption{GLS periodogram of the FWHM time series after linear detrending for the temperature of echelle gratings and the change of focus in the H03 dataset. We can see the presence of a long-term magnetic cycle at more than 3000 \si{\day}.}
    \label{fwhm_res_temp}
\end{figure}

The first step in the analysis of FWHM is done by detrending the time series for the change of focus HARPS suffered before fiber intervention. Secondly, we modeled the correlation with the change in temperature of echelle gratings. Once we linearly detrend the FWHM for focus and temperature, we can see a peak in the GLS periodogram of the residuals at 3093.89 \si{\day} as shown in Fig. \ref{fwhm_res_temp}.

\begin{figure}[!h]
    \includegraphics[width=\linewidth]{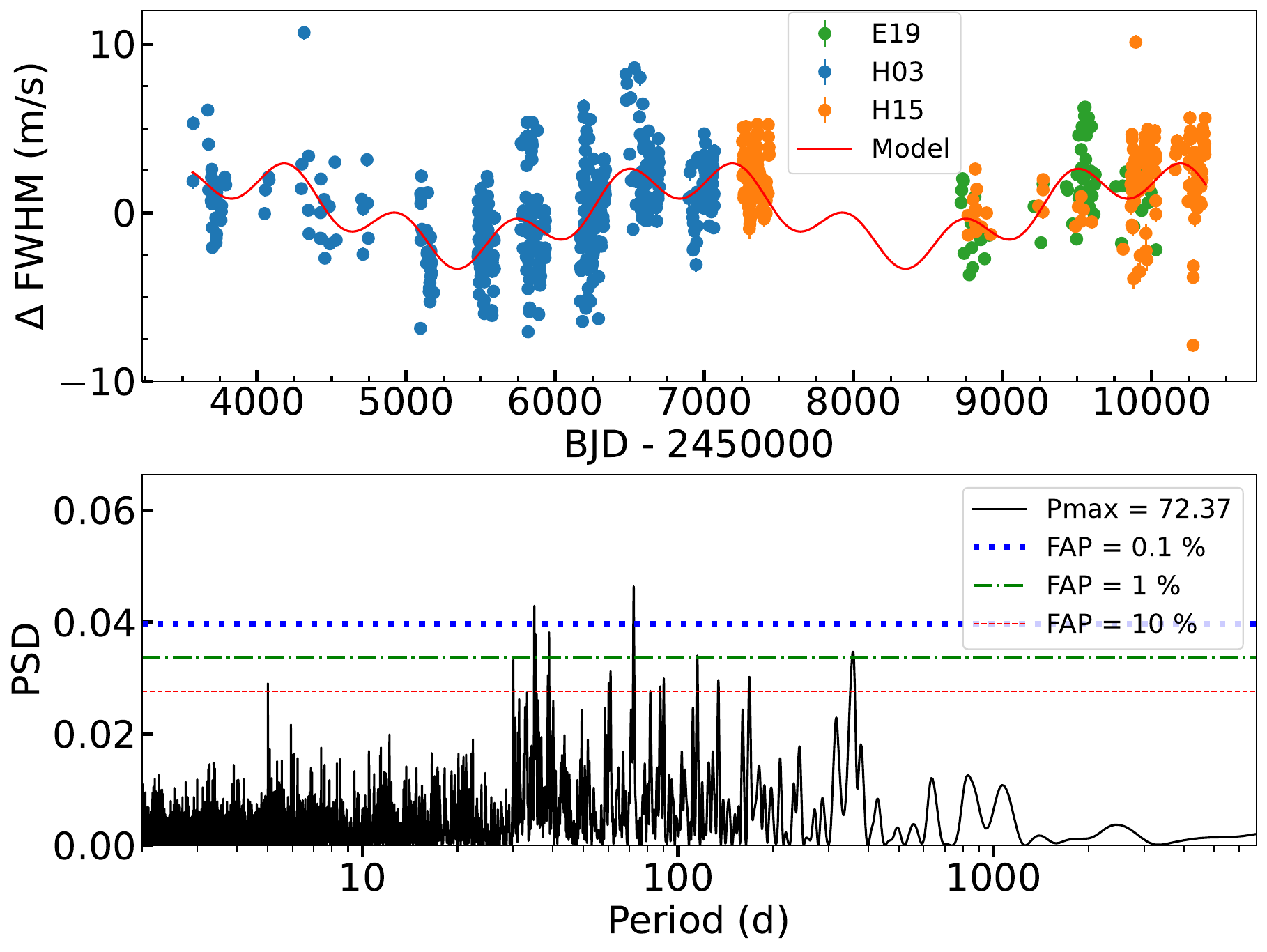}
    \caption{ Model of the cycle in FWHM for HD 20794 (top). The time series is detrended for correlation with temperature and focus.
    GLS periodogram of the residuals once the model is subtracted (bottom).}
    \label{fwhm_model}
\end{figure}

This periodicity is typical for the magnetic cycle of a G-type star. We modeled the cycle with a sinusoidal and a few of its harmonics. We found the best solution to be a model with a sinusoidal and its fourth harmonic, with a period of 1/4 of the main term period. For the cycle, we found a period of 3005$_{-20}^{+25}$ \si{\day}.
In Fig. \ref{fwhm_model} we show a plot of the cycle model with the GLS periodogram of the residuals. The plotted time series is already detrended for temperature changes and focus, while these parameters were fitted simultaneously to the long-term cycle. We searched for the rotation period of the star with GP. We tried different GP kernels to model the suspected rotation signal. We found as the best kernel the MEP kernel. We found a rotation period of 35.0$_{-2.5}^{+3.2}$ \si{\day}. 

\begin{figure}[htbp]
  \centering
  \begin{subfigure}[b]{0.45\textwidth}
    \centering
    \begin{tikzpicture}
            \node[inner sep=0] (image) at (0,0) {\includegraphics[width=\textwidth]{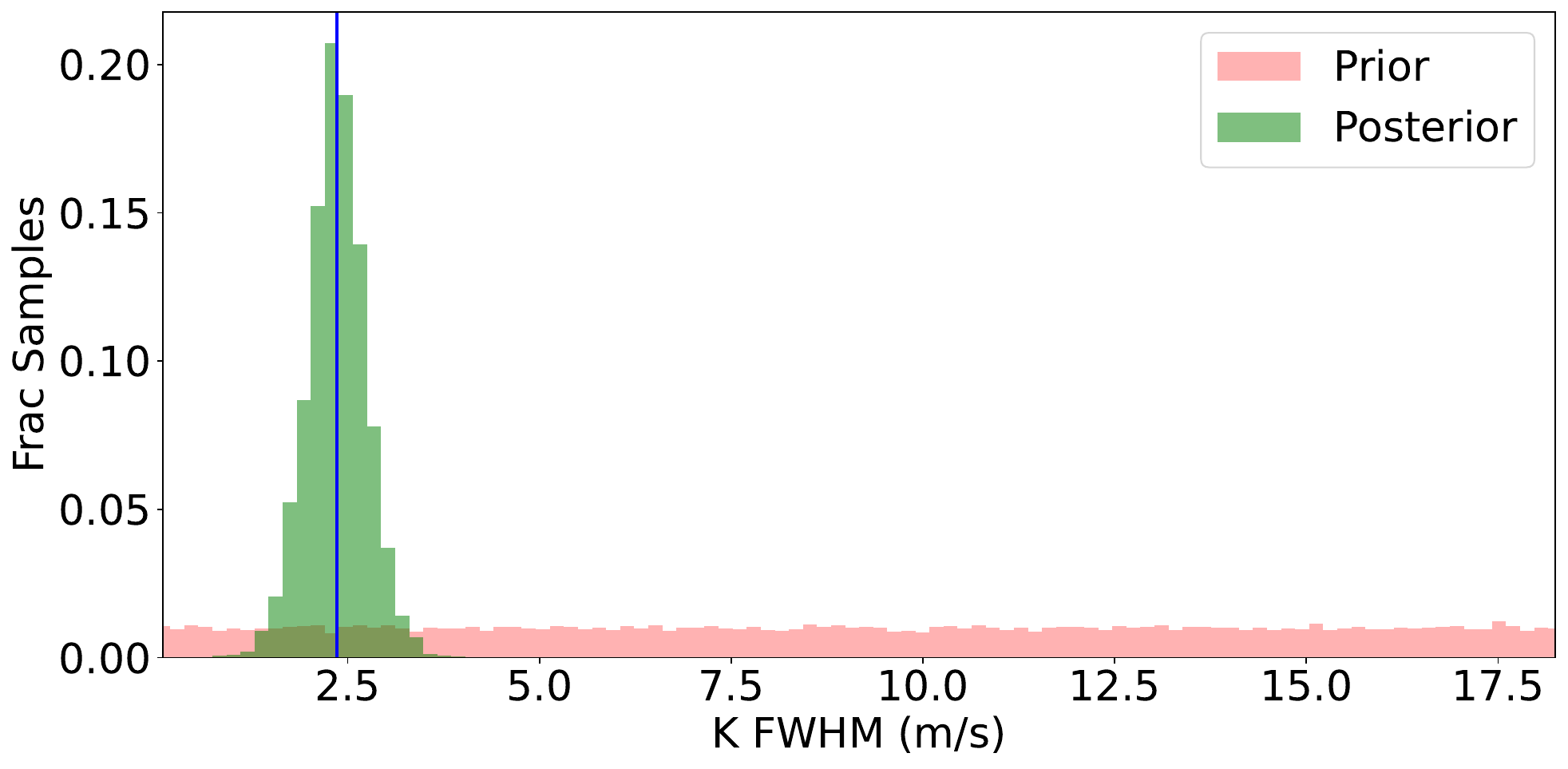}};
            \node[anchor=north west] at ([xshift=0.9cm, yshift=-0.32cm]image.north west) {\textbf{a)}};
    \end{tikzpicture}
    \label{fig:subfig1}
  \end{subfigure}
  \hspace{0.05\textwidth}
  \begin{subfigure}[b]{0.45\textwidth}
    \centering
    \begin{tikzpicture}
            \node[inner sep=0] (image) at (0,0) {\includegraphics[width=\textwidth]{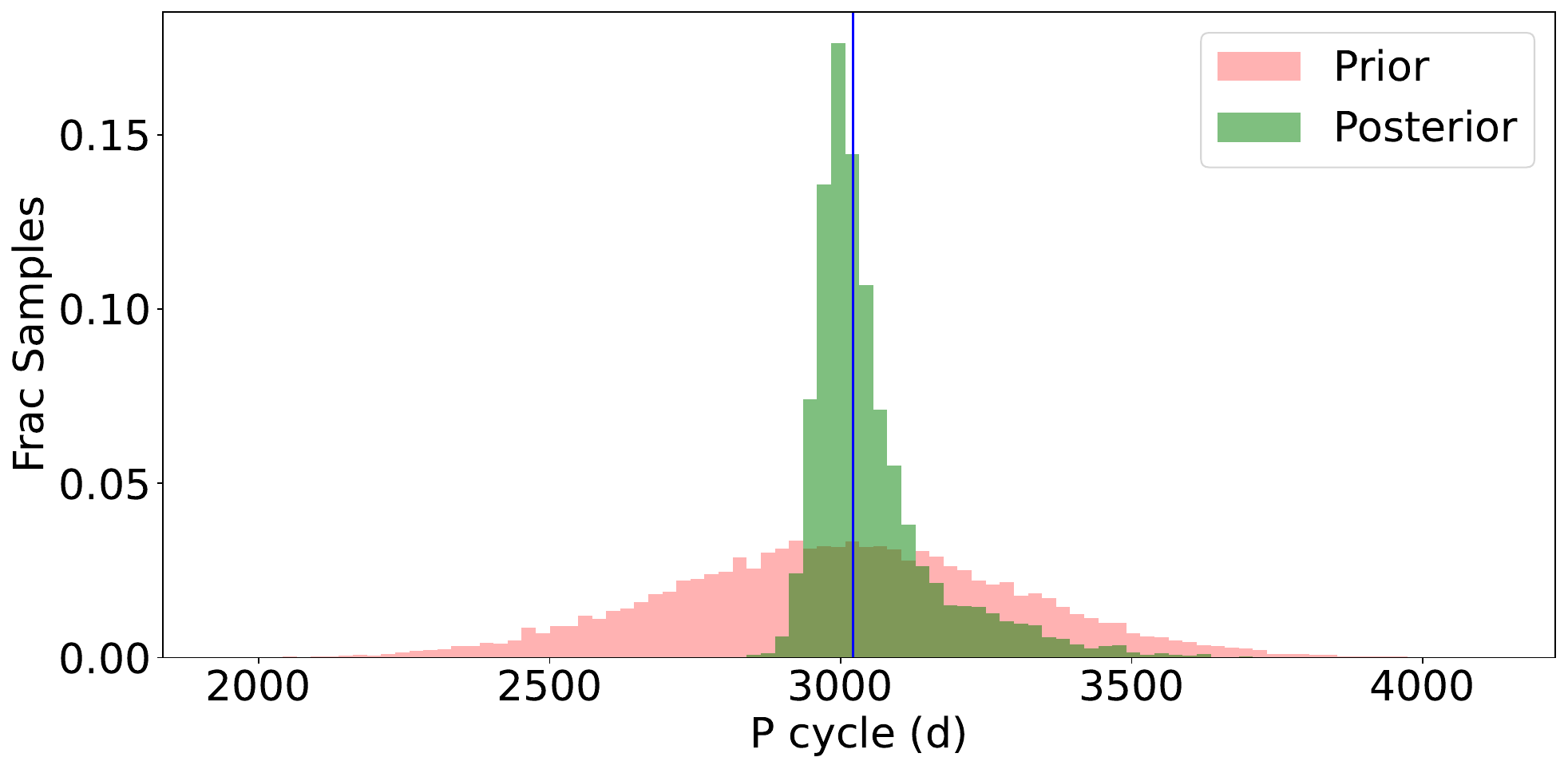}};
            \node[anchor=north west] at ([xshift=0.9cm, yshift=-0.32cm]image.north west) {\textbf{b)}};
    \end{tikzpicture}
    \label{fig:subfig1}
  \end{subfigure}
  \begin{subfigure}[b]{0.45\textwidth}
    \centering
    \begin{tikzpicture}
            \node[inner sep=0] (image) at (0,0) {\includegraphics[width=\textwidth]{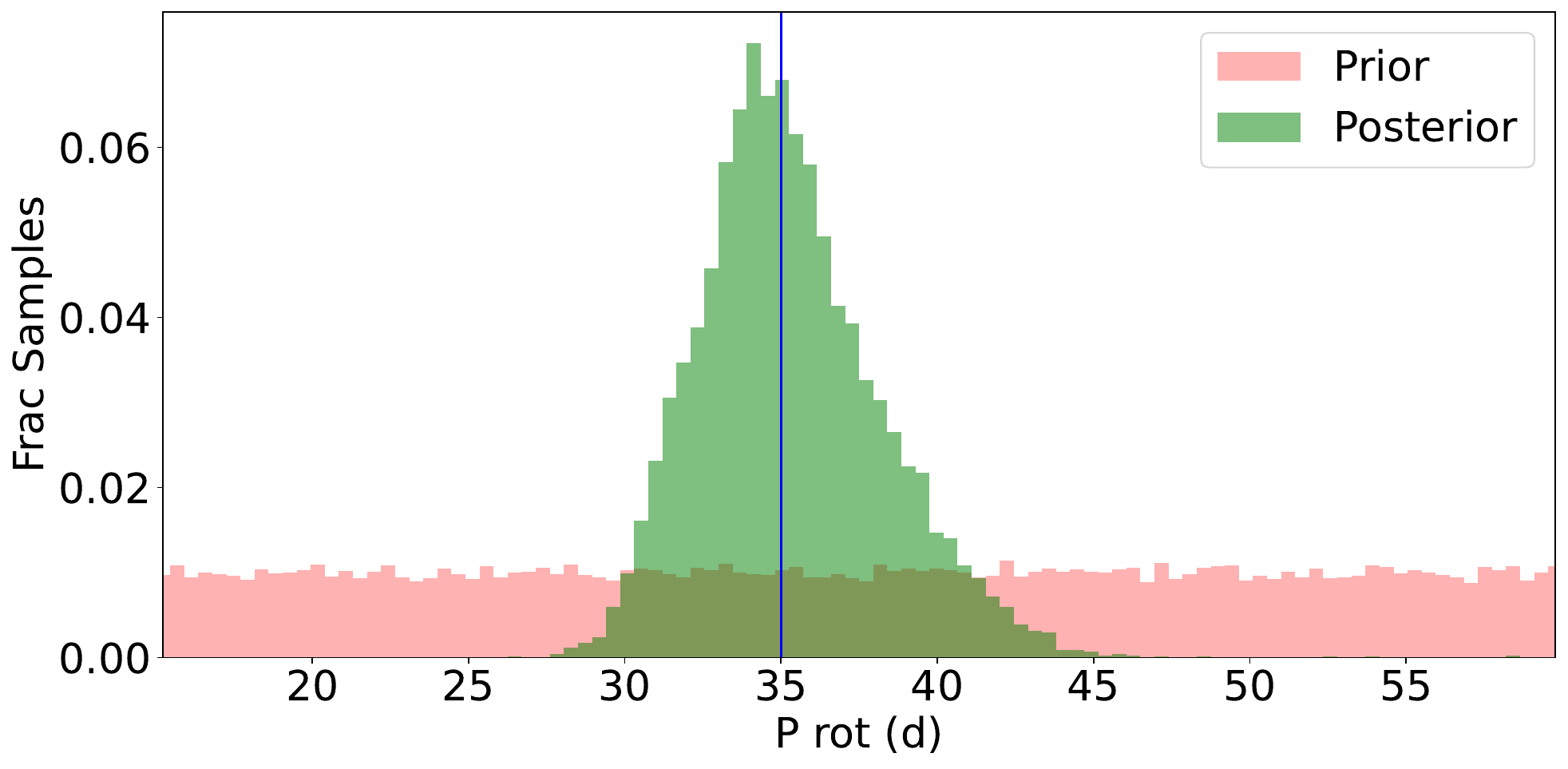}};
            \node[anchor=north west] at ([xshift=0.9cm, yshift=-0.32cm]image.north west) {\textbf{c)}};
    \end{tikzpicture}
    \label{fig:subfig1}
  \end{subfigure}
  \caption{Posterior distributions for parameters of the magnetic cycle and rotation period we can observe in HD 20794 for FWHM. Panel (a): Posterior distribution of the amplitude of the cycle found in FWHM. Panel (b): Posterior distribution of the period of the cycle found in FWHM. Panel (c): Posterior distribution of the rotation period found in FWHM.}
  \label{fwhm_posterior}
\end{figure}

%The timescale of evolution is the most difficult parameter to determine. We can force this parameter to be longer than the rotation period, to avoid this parameter from converging to unphysical solutions, creating a clear overfitting. 
In the GP model, we found a period for the cycle of 3020$_{-50}^{+111}$ \si{\day}. The period of the cycle is compatible with the one we found in the cycle-only model, even if with a larger uncertainty. The amplitude of the cycle remains different from zero at 6.6 $\sigma$. We show the posterior distribution of the amplitude of the cycle, the period of the cycle, and the rotation period in Fig. \ref{fwhm_posterior}

\subsection{BIS}

\begin{figure}[!h]
    \includegraphics[width=\linewidth]{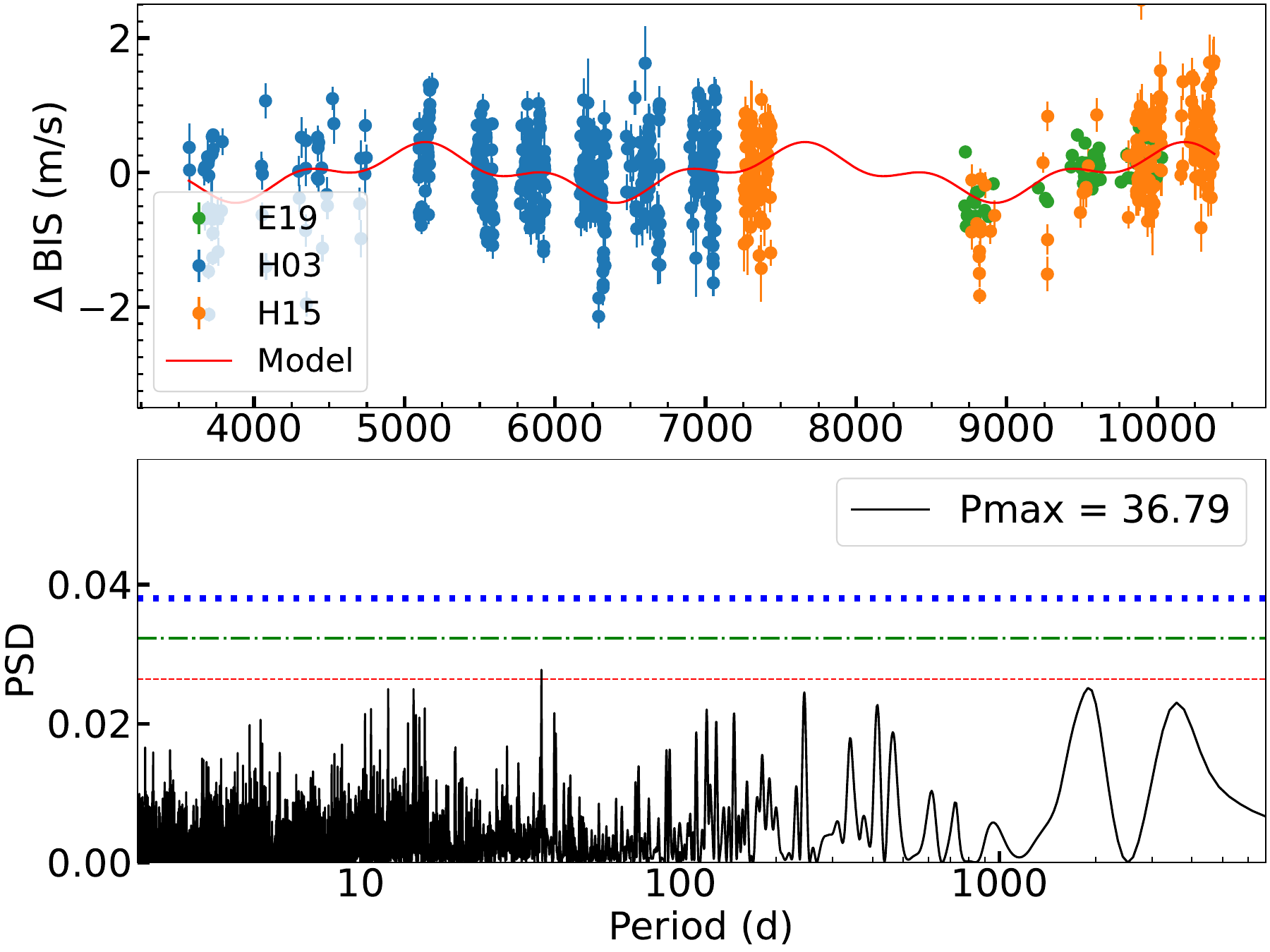}
    \caption{ Model of the cycle in BIS for HD 20794 (top). The time series is detrended for correlation with temperature.
     GLS periodogram of the residuals once the model is subtracted (bottom).}
    \label{model_bis}
\end{figure}

For the analysis of BIS, we followed a procedure similar to that of the analysis of the FWHM. We considered a model with a linear detrending of the BIS against the temperature of the echelle gratings. Once we detrend the indicator for temperature of echelle gratings we can still see strong peaks at 2530 \si{\day} and 850 \si{\day}. These two peaks suggest the presence of a magnetic cycle for HD 20794. The ratio between the two periods is very close to 1/3, pointing toward the presence of a cycle with its harmonic at $P/3$. We modeled the cycle with a sinusoid with a norm prior for the period centered at 3000 \si{\day}, with a width of 250 \si{\day}, and a second sinusoid, with free amplitude and phase but with the period constrained to be 1/3 of the period of the main component. We found a period for the cycle of 2521$_{-34}^{+40}$ \si{\day}. Figure \ref{model_bis} shows the model for BIS over-imposed to the dataset once we have subtracted the correlation with the temperature of echelle gratings. 

\begin{figure}[htbp]
  \centering
  \begin{subfigure}[b]{0.45\textwidth}
    \centering
    \begin{tikzpicture}
            \node[inner sep=0] (image) at (0,0) {\includegraphics[width=\textwidth]{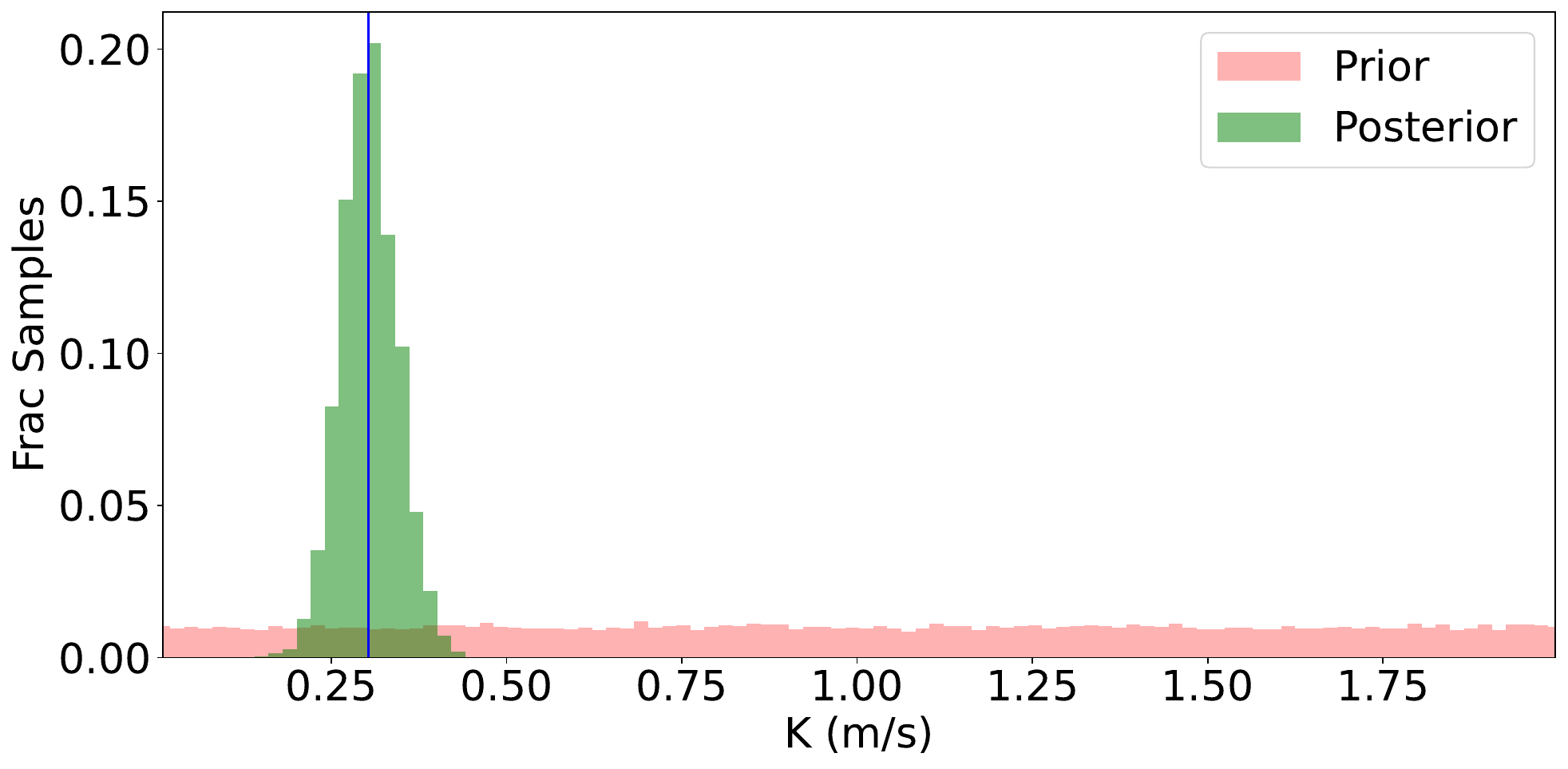}};
            \node[anchor=north west] at ([xshift=0.9cm, yshift=-0.32cm]image.north west) {\textbf{a)}};
    \end{tikzpicture}
    \label{fig:subfig1}
  \end{subfigure}
  \hspace{0.05\textwidth}
  \begin{subfigure}[b]{0.45\textwidth}
    \centering
    \begin{tikzpicture}
            \node[inner sep=0] (image) at (0,0) {\includegraphics[width=\textwidth]{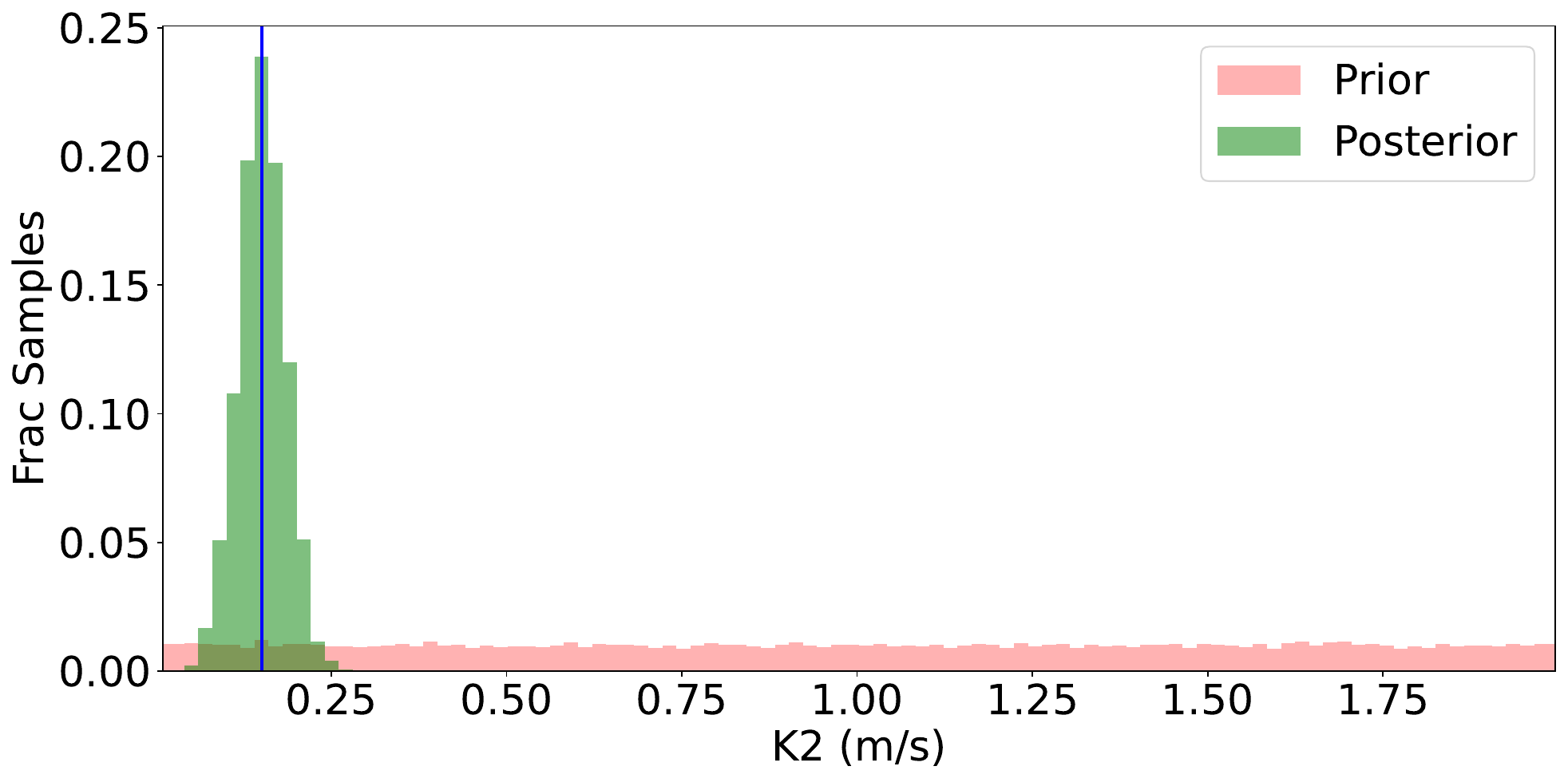}};
            \node[anchor=north west] at ([xshift=0.9cm, yshift=-0.32cm]image.north west) {\textbf{b)}};
    \end{tikzpicture}
    \label{fig:subfig1}
  \end{subfigure}
  \begin{subfigure}[b]{0.45\textwidth}
    \centering
    \begin{tikzpicture}
            \node[inner sep=0] (image) at (0,0) {\includegraphics[width=\textwidth]{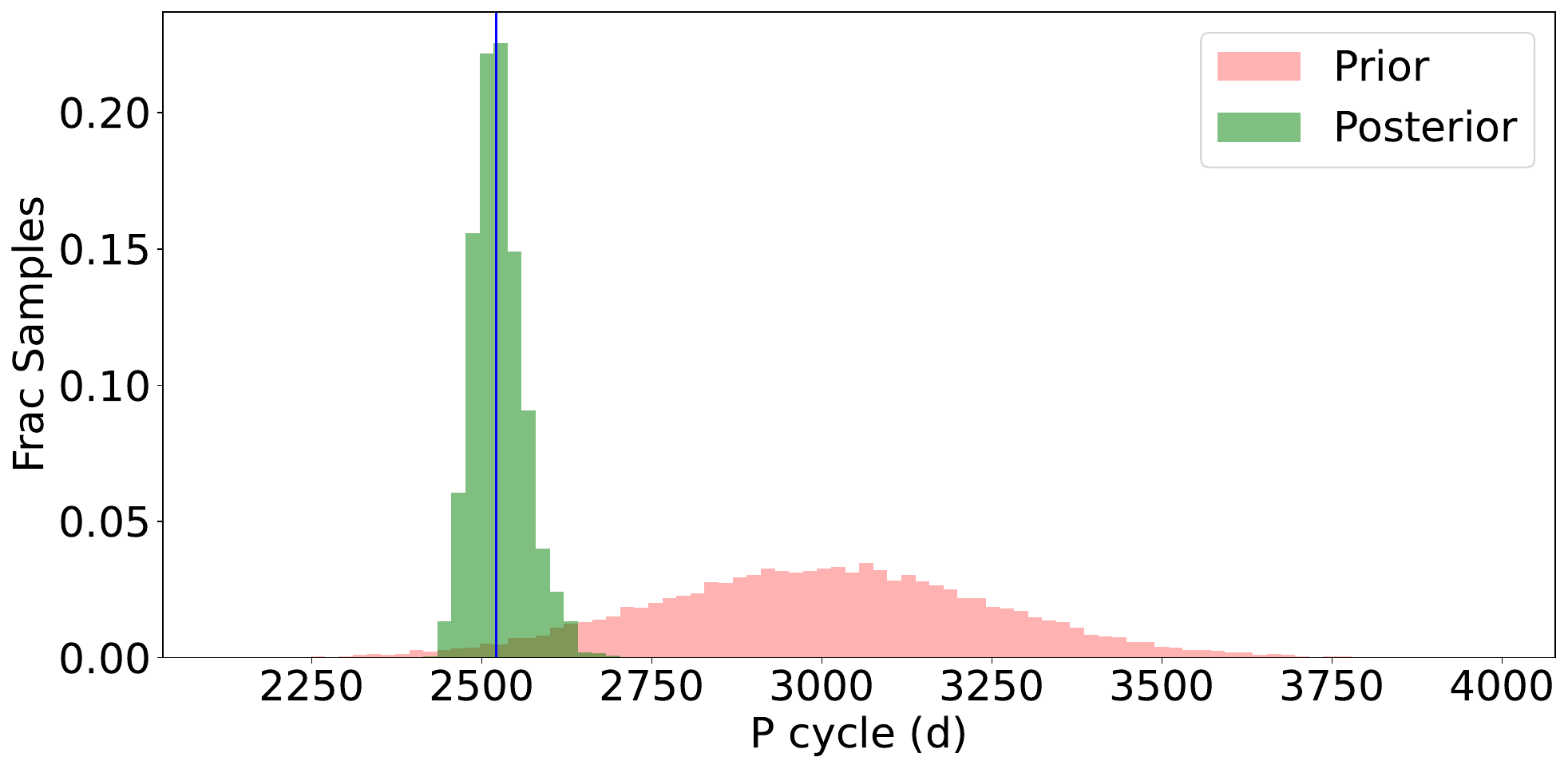}};
            \node[anchor=north west] at ([xshift=0.9cm, yshift=-0.32cm]image.north west) {\textbf{c)}};
    \end{tikzpicture}
    \label{fig:subfig1}
  \end{subfigure}
  \caption{Posterior distributions for parameters of the magnetic cycle we can observe in HD 20794 for BIS. Panel (a): Posterior distribution of the amplitude of the cycle found in BIS. Panel (b): Posterior distribution of the amplitude of the harmonic at P/3 of the period of the cycle found in BIS. Panel (c): Posterior distribution of the period of the cycle found in BIS.}
  \label{bis_posterior}
\end{figure}

Once we subtract the model for BIS we only see peaks with FAP > 1 \%. The most prominent peak in the residuals is at 36.79 \si{\day}. This could be a signature of the rotation period of the star. We tried a fit with a GP but the result did not converge to a defined solution. 
In Fig. \ref{bis_posterior} we show the posterior distribution for the amplitude and the period of the BIS cycle.

\subsection{S-index}

\begin{figure}[!h]
    \includegraphics[width=\linewidth]{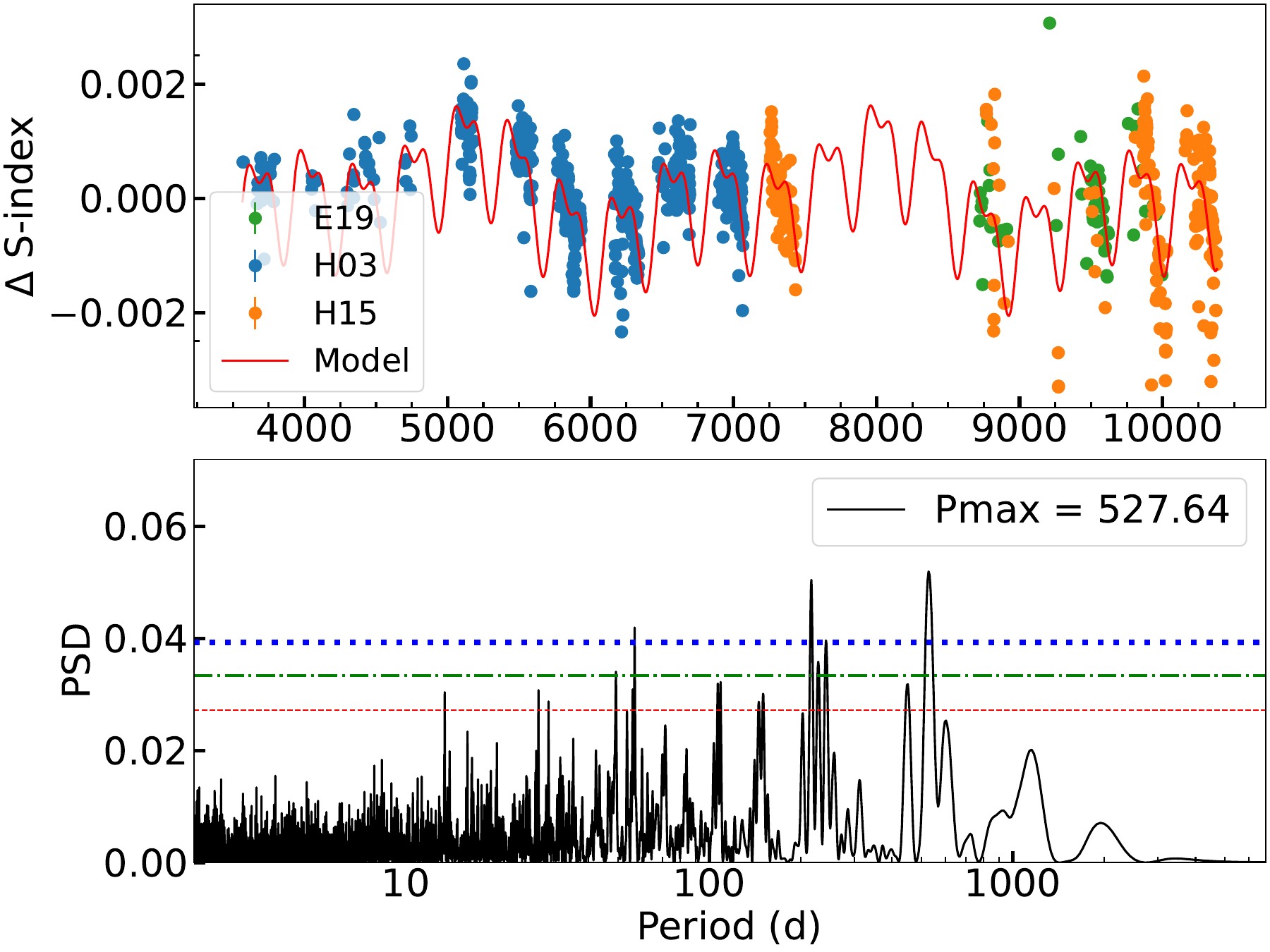}
    \caption{Model of the cycle in S-index for HD 20794 (top). The time series is detrended for correlation with temperature.
    GLS periodogram of the residuals once the model is subtracted (bottom).}
    \label{model_sindex}
\end{figure}

\begin{figure}[htbp]
  \centering
  \begin{subfigure}[b]{0.45\textwidth}
    \centering
    \begin{tikzpicture}
            \node[inner sep=0] (image) at (0,0) {\includegraphics[width=\textwidth]{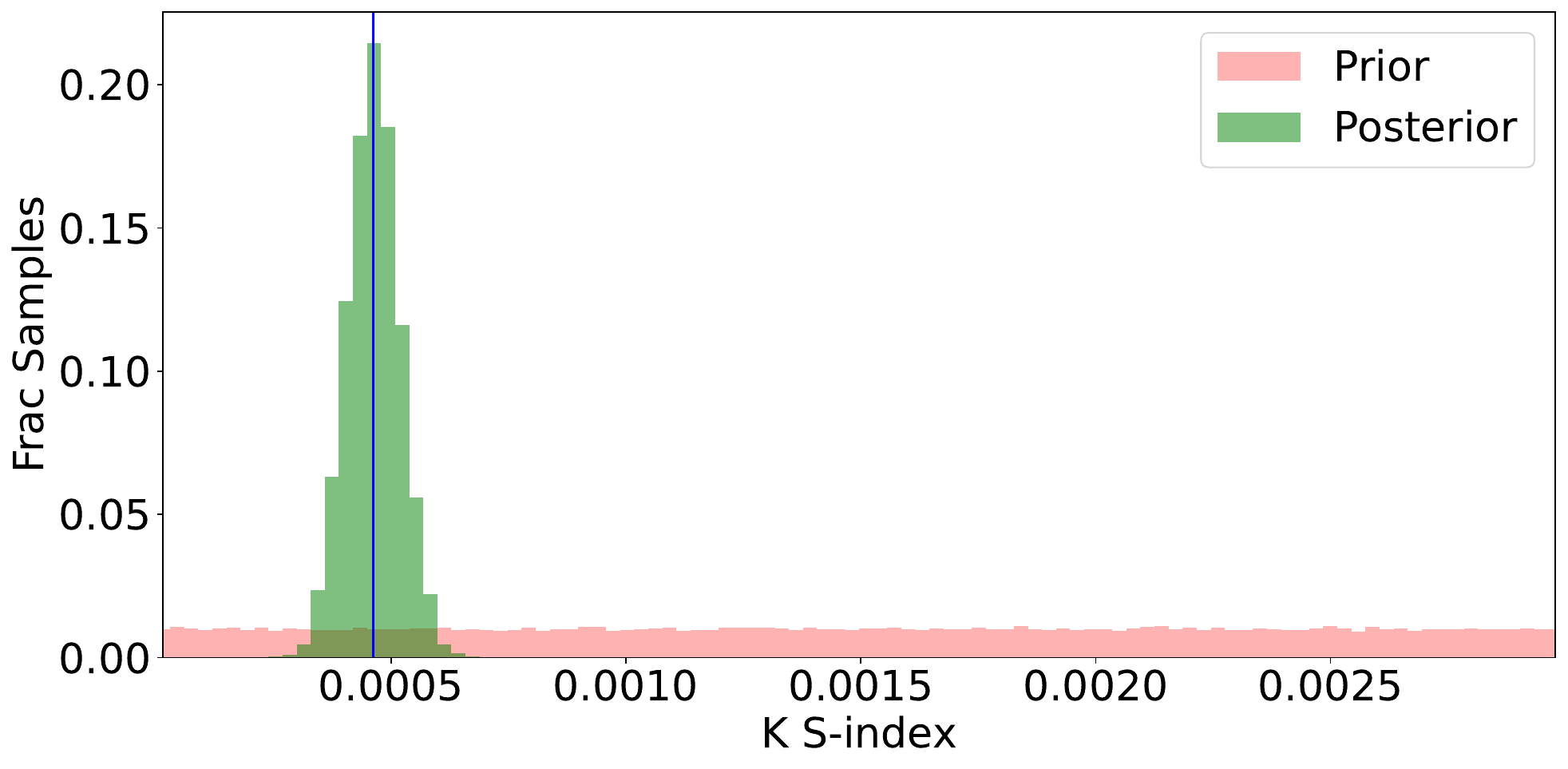}};
            \node[anchor=north west] at ([xshift=0.9cm, yshift=-0.32cm]image.north west) {\textbf{a)}};
    \end{tikzpicture}
    \label{fig:subfig1}
  \end{subfigure}
  \hspace{0.05\textwidth}
  \begin{subfigure}[b]{0.45\textwidth}
    \centering
    \begin{tikzpicture}
            \node[inner sep=0] (image) at (0,0) {\includegraphics[width=\textwidth]{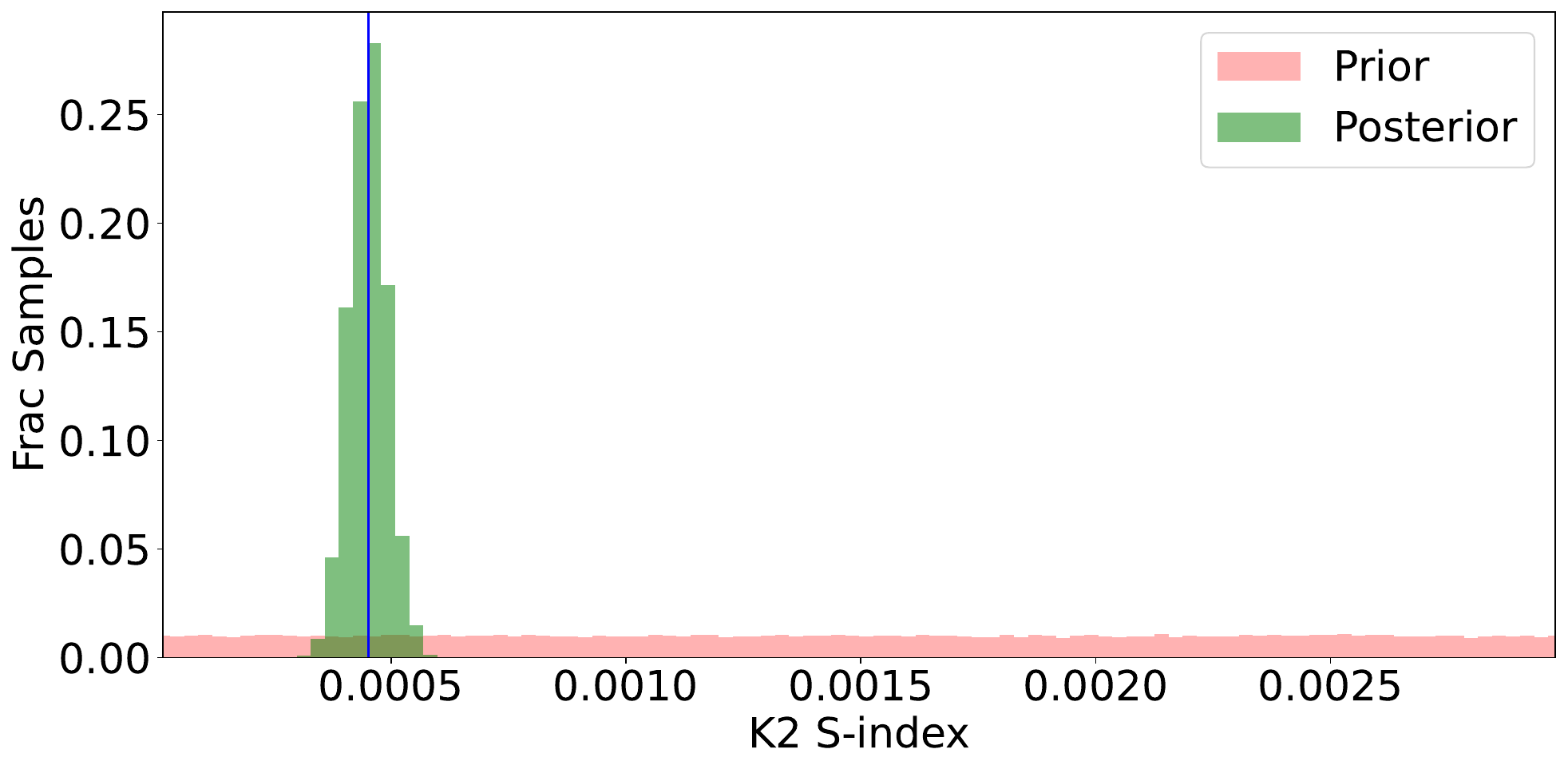}};
            \node[anchor=north west] at ([xshift=0.9cm, yshift=-0.32cm]image.north west) {\textbf{b)}};
    \end{tikzpicture}
    \label{fig:subfig1}
  \end{subfigure}
  \begin{subfigure}[b]{0.45\textwidth}
    \centering
    \begin{tikzpicture}
            \node[inner sep=0] (image) at (0,0) {\includegraphics[width=\textwidth]{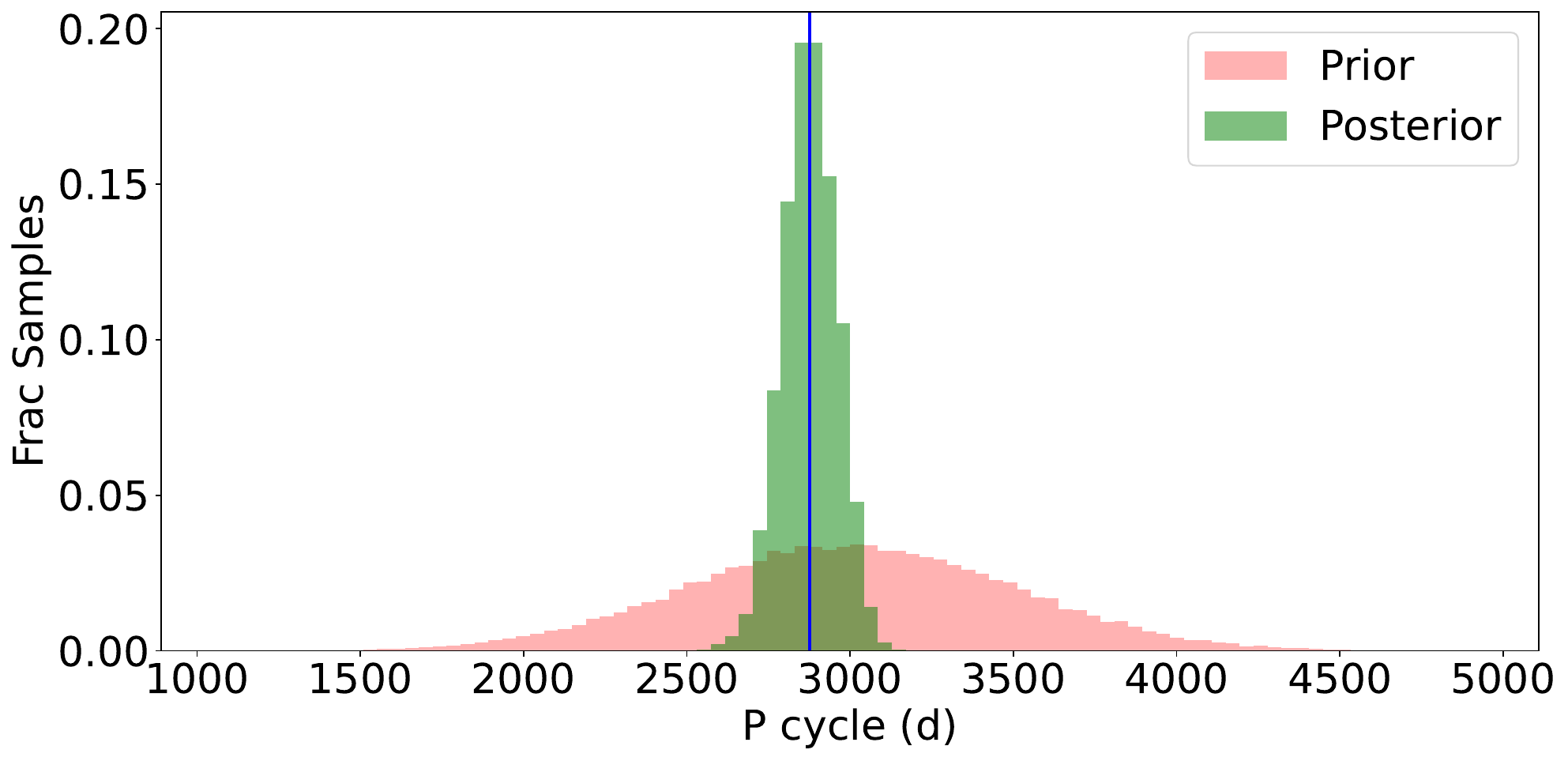}};
            \node[anchor=north west] at ([xshift=0.9cm, yshift=-0.32cm]image.north west) {\textbf{c)}};
    \end{tikzpicture}
    \label{fig:subfig1}
  \end{subfigure}
  \caption{Posterior distributions for parameters of the magnetic cycle we can observe in HD 20794 for S-index. Panel (a): Posterior distribution of the amplitude of the cycle. Panel (b): Posterior distribution of the amplitude of the harmonic at P/2. Panel (c): Posterior distribution of the period of the cycle.}
  \label{sindex_posterior}
\end{figure}

\begin{figure*}[!h]
    \includegraphics[width=\textwidth]{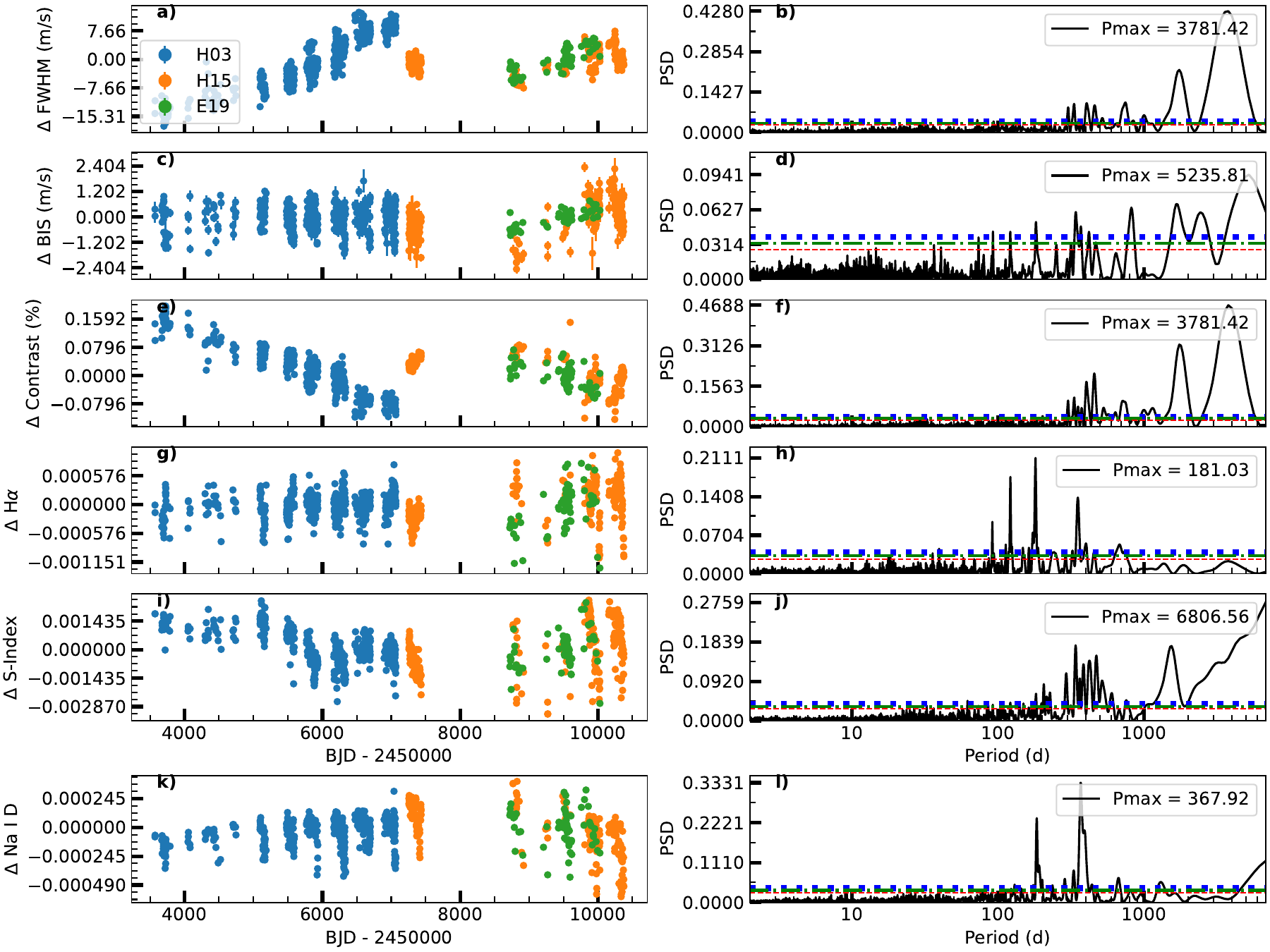}
    \caption{Activity indicators time series derived before the YARARA correction. Panel (a): FWHM time series. It is possible to see trends in the different datasets. For H03 we have a trend with the change of focus of the instrument, while for H15 and E19 we have a linear correlation with the temperature of the echelle gratings. Panel (b): GLS periodogram for FWHM.  Panel (c): BIS time series. It is possible to see a linear trend in E19 and H15. Panel (d): GLS periodogram for BIS. Panel (e): Time series of Contrast. Panel (f): GLS periodogram of Contrast. Panel (g): H$\alpha$ time series. Panel (h): GLS periodogram of H$\alpha$. Panel (i): S-index time series. Panel (j): GLS periodogram of S-index.
    Panel (k): Na I D time series. Panel (l): GLS periodogram of Na I D. 
    The red/green/blue horizontal dashed lines indicate FIP levels of 10/1/0.1 \% respectively.}
    \label{stellar_activity_full}
\end{figure*}

From the dataset and the GLS periodogram of the residuals of Fig. \ref{stellar_activity_full}, we see a quadratic trend in the S-index time series. We tried to model this indicator with a quadratic trend together with a cycle, its first harmonic, a sinusoidal to model 1-year variability, and the first harmonic of that year. For the cycle period, we considered a normal prior centered at 3000 \si{\day} with a width of 500 \si{\day}. The period of the cycle derived for the S-index is P = 2884$_{-90}^{+86}$ \si{\day}. Once we subtracted the model for the cycle, we could see some additional structure in the residuals, resulting in a peak at 527 \si{\day}, but no peaks in correspondence with the orbital periods of the planets we have found in our analysis. We can see some excess in power at lower periods, so we tried a blind search for the rotation period through a GP analysis. The GP analysis did not converge to a satisfying solution, probably due to the higher level of noise in this indicator compared to the FWHM.
In Fig. \ref{model_sindex} we show the model of the cycle for S-index together with the GLS periodogram of the residuals.
In Fig. \ref{sindex_posterior} we show the posterior distribution for the amplitude and the period of the S-index cycle. 

\subsection{Contrast}
In the contrast, we can see a structure similar to the one we find in FWHM. We see a strong linear trend in the H03 dataset, related to the change in focus of the instrument through the years. We can see some trends in the E19 dataset too, related to the change in temperature of the echelle gratings. We started the analysis by detrending the different time series with a linear trend to the temperature of echelle gratings, with an additional term for the H03 dataset to correct the change in focus. Once we subtracted this model we could see peaks in the GLS periodogram of the residuals related {to the 1-year variation} and probably related to the magnetic cycle of the star. We saw peaks at $\sim$ 1800 \si{\day} and $\sim$ 3200 \si{\day}. We modeled this signal with a sinusoid and its first and second harmonic and a sinusoid with normal prior on the period centered around 1-year. We found a solution with a period for the cycle equal to 4065$_{-30}^{+32}$ \si{\day}.  Once we subtracted this signal we still found some structure in the residuals but we did not find any periodicity related to the period of the detected planet or stellar rotation.

\subsection{H$\alpha$}
In H$\alpha,$ we can see the presence of a signal related to the year with the first three harmonics, respectively at 1/2,1/3, and 1/4 of the year-signal. We found the best model a model with three sinusoids, one with a period of 1 year and the other two at half and one-third of this period. When we subtracted this model, we did not see any peak in the GLS periodogram in correspondence with any interesting signal in RVs, nor a region of the parameter space suitable for a signature of stellar rotation. The modeling with a GP did not converge to a result on the rotation period. 
\subsection{Na I D}
In Na I D, we only find a peak in the GLS periodogram at 367 \si{\day} and a second peak at the first harmonic of the year signal. We tried to model the time series with a sinusoidal and a harmonic at half the period. Once we subtracted this model, we can see a quadratic trend in the full dataset. We modeled it and we did not find any significant signal in the GLS periodogram of the residuals. We did not go further in the analysis of this indicator. 

\subsection{YARARA activity indicators}
\label{yarara_activity_indicators}
In our analysis of activity, we did not only consider the activity indicators extracted from the spectra not corrected from YARARA. We also tried to analyze the HARPS dataset alone for activity indicators extracted from spectra corrected for YARARA. Here we report the analysis of BIS. BIS is the only indicator showing a prominent signal in the GLS periodogram that could be related to the rotation of the star, while most of the other indicators just show a signal related to the 1-year variation or its harmonics.

\begin{figure}[!h]
    \includegraphics[width=\linewidth]{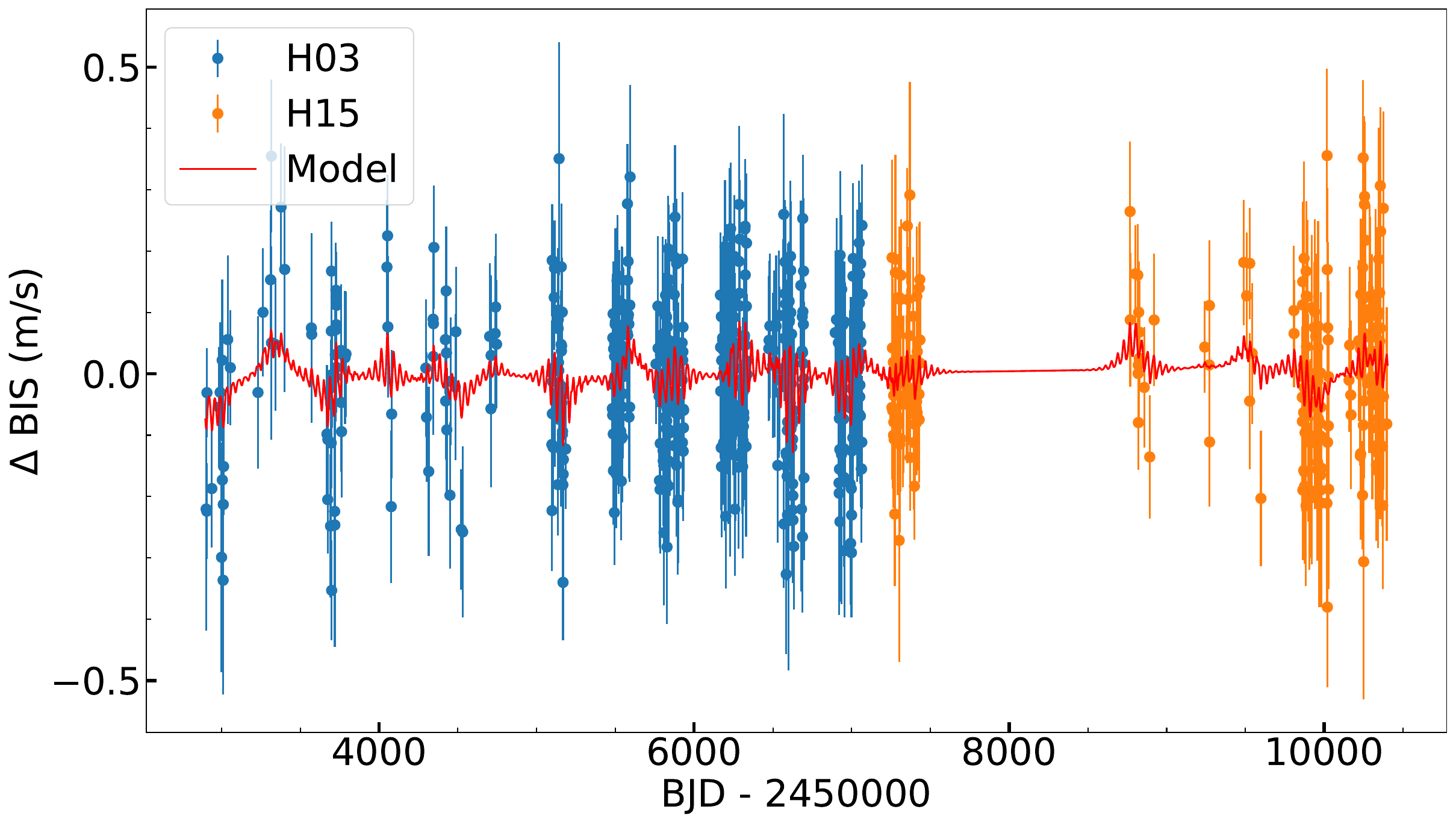}
    \caption{Model for BIS dataset in case of YARARA correction. We can retrieve clues on the rotation period of the star around 40 \si{\day}. The amplitude of the signal is very low as expected from the low level of activity we can observe for HD 20794 and the fact spectra were previously corrected with YARARA}
    \label{bis_model}
\end{figure}

We can see a prominent peak in BIS at 40.53 \si{\day}. This period is of particular interest because it is similar to the period of the 40-d planet (40.114 $\pm$ 0.053) detected in \citet{pepe}. This planet was not confirmed in the following analyses on the target. This points toward a stellar origin for this signal. 
 This periodicity could be related to the rotation period of the star. To test this hypothesis, we tried to model this indicator with a GP, using a MEP kernel. In our analysis, we found a non-converging result. We noticed the mean error on the measurement BIS provided by YARARA is larger than the RMS of the measurement. Assuming the error on the BIS is overestimated we considered as error on the BIS the error we had for RV. With this assumption, 
 we found a convergent solution for the rotation period at 38.8 $_{-2.6}^{+2.4}$ \si{\day}, with a low amplitude, as expected for a quiet star such as HD 20794. We overplotted the model on the HARPS dataset in Fig. \ref{bis_model}. 

\clearpage

\end{appendix}
\end{document}